\documentclass[11pt]{article}
\usepackage{epsfig}\parskip 5pt plus 1pt
\usepackage{eurosym}
\usepackage{amssymb}
\usepackage[utf8]{inputenc}
\usepackage{graphicx}
\usepackage[labelformat=simple]{subcaption}

\usepackage{booktabs,lipsum,tabularx}
\usepackage{lineno}
\usepackage{multirow}
\usepackage{tabularx}
\usepackage{longtable}
\usepackage{multicol}
\usepackage{hyperref}
\usepackage{amsmath}
\usepackage{authblk}
\usepackage{cprotect}
\usepackage[dvipsnames]{xcolor}

\def\nue{\ensuremath{\nu_{e}\ }}

\def\numu{\ensuremath{\nu_{\mu}\ }}
\def\nubarmu{\ensuremath{\overline{\nu}_{\mu}}}

\newcommand{\numunue}{\ensuremath{\nu_\mu \rightarrow \nu_e\,}}

\newcommand{\nubarmunubare}{\ensuremath{\overline{\nu}_\mu \rightarrow \overline{\nu}_e\,}}

\usepackage{geometry}
\newgeometry{vmargin={25mm}, hmargin={25mm,25mm}}
\usepackage{fancyhdr}
\title{Enhanced NeUtrino BEams from kaon Tagging (ENUBET)\vspace{0.3cm} \\  \large{The ENUBET Collaboration} }

\author[a,b]{F.~Acerbi}
\author[x]{I. Angelis}
\author[e]{M.~Bonesini}
\author[e,f]{A.~Branca}
\author[e,f]{C.~Brizzolari}
\author[f]{G.~Brunetti}
\author[r]{M.~Calviani}
\author[e,p]{S.~Capelli}
\author[d,g]{S.~Carturan}
\author[h]{M.G.~Catanesi}
\author[r]{N.~Charitonidis}
\author[i]{S.~Cecchini}
\author[i]{F.~Cindolo}
\author[c,d]{G.~Collazuol}
\author[c]{E.~Conti}
\author[c]{F.~Dal Corso}
\author[c,d]{C.~Delogu}
\author[j,k]{G.~De~Rosa}
\author[e,f]{A.~Falcone}
\author[a]{A.~Gola}
\author[r]{B.~Goddard}
\author[c,d]{F.~Iacob}
\author[l]{C.~Jollet}
\author[r]{V.~Kain}
\author[m]{B.~Kli\u{c}ek}
\author[n,u,v]{Y.~Kudenko}
\author[x]{Ch. Lampoudis}
\author[c,d]{M.~Laveder}
\author[c,d]{A.~Longhin}
\author[o]{L.~Ludovici}
\author[e,p]{E.~Lutsenko}
\author[h,q]{L.~Magaletti}
\author[i]{G.~Mandrioli}
\author[i]{A.~Margotti}
\author[e,p]{V.~Mascagna}
\author[i]{N.~Mauri}
\author[e,f]{L.~Meazza}
\author[l]{A.~Meregaglia}
\author[c]{M.~Mezzetto}
\author[r]{M.~Nessi}
\author[t]{A.~Paoloni}
\author[c,d,r]{M.~Pari}
\author[e,f,r]{E.G.~Parozzi}
\author[i,s]{L.~Pasqualini}
\author[a]{G.~Paternoster}
\author[i]{L.~Patrizii}
\author[i]{M.~Pozzato}
\author[e,p]{M.~Prest}
\author[c,d]{F.~Pupilli}
\author[h]{E.~Radicioni}
\author[j,k]{C.~Riccio}
\author[j,k]{A.C.~Ruggeri}
\author[x]{D. Sampsonidis}
\author[c,d]{C.~Scian}
\author[i]{G.~Sirri}
\author[m]{M.~Stip\u{c}evi\'{c}}
\author[i]{M.~Tenti}
\author[e,f]{F.~Terranova}
\author[e,f,1]{M.~Torti}
\author[x]{S. E. Tzamarias}
\author[e]{E.~Vallazza}
\author[r]{F.M.~Velotti}
\author[t]{L.~Votano}
{\scriptsize{
\affil[a]{Fondazione Bruno Kessler (FBK),  Via Sommarive 18 - 38123 Povo (TN), IT}
\affil[b]{INFN-TIFPA, Università di Trento, Via Sommarive 14 - 38123 Povo (TN), IT}
\affil[c]{INFN Sezione di Padova, via Marzolo 8 - 35131 Padova, IT}
\affil[d]{Università di Padova, via Marzolo 8 - 35131 Padova, IT}
\affil[e]{INFN Sezione di Milano-Bicocca, Piazza della Scienza 3 - 20133 Milano, IT}
\affil[f]{Università di Milano-Bicocca, Piazza della Scienza 3 - 20133 Milano, IT}
\affil[g]{INFN, Laboratori Nazionali di Legnaro, Viale dell'Università 2 - 35020 Legnaro (PD), IT}
\affil[h]{INFN Sezione di Bari, Via Giovanni Amendola 173 - 70126 Bari, IT}
\affil[i]{INFN Sezione di Bologna, viale Berti-Pichat 6/2 - 40127 Bologna, IT}
\affil[j]{INFN, Sezione di Napoli, Strada Comunale Cinthia - 80126 Napoli, IT}
\affil[k]{Università ``Federico II'' di Napoli, Corso Umberto I 40 - 80138 Napoli, IT}
\affil[l]{CENBG, Universitè de Bordeaux, CNRS/IN2P3, 33175 Gradignan, FR}
\affil[m]{Center of Excellence for Advanced Materials and Sensing Devices, Ruder Boskovic Institute, HR-10000 Zagreb, HR}
\affil[n]{
Institute for Nuclear Research of the Russian Academy of Sciences, 117312
Moscow, RU}
\affil[o]{INFN Sezione di Roma 1, Piazzale A. Moro 2, 00185 Rome, IT}
\affil[p]{Università degli Studi dell'Insubria, Via Valleggio 11 - 22100 Como, IT}
\affil[q]{Università degli Studi di Bari, Via Giovanni Amendola 173 - 70126 Bari, IT}
\affil[r]{CERN, Esplanade des particules - 1211 Genève 23, CH}
\affil[s]{Università degli Studi di Bologna, viale Berti-Pichat 6/2 - 40127 Bologna, IT}
\affil[t]{INFN, Laboratori Nazionali di Frascati, via Fermi 40 - 00044 Frascati (Rome), Italy}
\affil[u]{National Research Nuclear University ``MEPhI'', 115409 Moscow, Russia}
\affil[v]{Moscow Institute of Physics and Technology, 141701 Moscow region, Russia}
\affil[x]{Aristotle University of Thessaloniki. Thessaloniki 541 24, Greece.}
}

\date{15 March 2022}
\begin{document}

\begin{figure}[t!]
\centering
\includegraphics[scale=0.45]{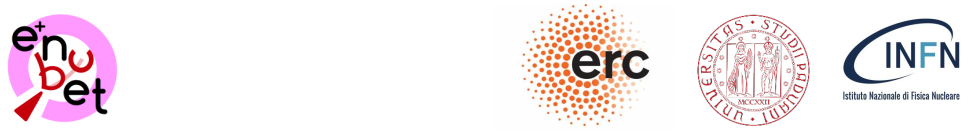}
\end{figure}
    
    \maketitle
    \begin{abstract}
    The ENUBET project aims at demonstrating the feasibility of a ``monitored'' neutrino beam, in which the absolute normalization of the neutrino flux produced by a narrow band meson beam can be constrained at the 1\% level. The electron neutrino component is determined by monitoring large-angle positrons from $K_{e3}$ decays in a 40~m long instrumented decay tunnel. The $\nu_\mu$ flux is provided by the muons produced by two-body decays of kaons and pions and detected along and at the end of the tunnel. Since ENUBET is a very narrow band beam ($p=8.5$~GeV/$c$ $\pm$ 10\%), the transverse position of the neutrino interaction at the detector can be exploited to determine a priori, with significant precision, the neutrino energy spectrum without relying on the final state reconstruction (''narrow band off-axis technique"). Lepton monitoring and narrow band off-axis energy reconstruction can be implemented in a single facility based on standard accelerator technologies for a new generation of high precision $\nu_e$ and $\nu_\mu$ cross section measurements at the GeV scale and for precision searches of physics beyond the standard three neutrino paradigm.
    
%    We will present the optimization and performances of a 20~m long focusing transfer line allowing for a continuous measurement of $K_{e3}$ positrons and $K_{\mu2}$ and $K_{\mu3}$ muons at single-particle level. The (quadrupole-based) focusing system is designed to be operated with a slow extraction proton scheme where protons can be diluted over several seconds. This timing allows for the direct monitoring of muons after the hadron dump and extends the original scope of the project towards a full-fledged ``time-tagged'' neutrino beam: time-coincidences among the lepton at the source and the neutrino at the detector would enable an unprecedented purity and the possibility to reconstruct the neutrino kinematics at source on an event by event basis.

%    Another major milestone is the final design of the ENUBET demonstrator for the instrumented decay tunnel that has been selected on the basis of the results of the 2016-2018 test-beams. This large detector prototype will prove the scalability and performance of the selected detector technology: an iron-scintillator modular sampling calorimeter (for $e$/$\pi$ separation) with a lateral light readout through WLS fibers connected to SiPMs, complemented by a photon veto system (for $e$/$\pi^0$ separation) made by an inner ring of plastic scintillator trackers. It will be finally tested at CERN in fall 2022.
    
    In 2019-2022 ENUBET has devised the first end-to-end simulation of the facility and demonstrated that the precision goals can be achieved in about three years of data taking employing neutrino detectors of moderate mass (ICARUS at FNAL, ProtoDUNE at CERN). The
technology of a monitored neutrino beam has been proven to be feasible and cost-effective, and
the complexity does not exceed significantly the one of a conventional short-baseline beam. The Snowmass 2021 DPF Community Planning Exercise is thus timely for the consideration of monitored neutrino beams hosting the next generation of cross section experiments. The ENUBET results will play an important role in the systematic reduction programme of future long baseline experiments, thus enhancing the physics reach of DUNE and HyperKamiokande. 
    In this document, we summarize the ENUBET design, physics performance and opportunities for its implementation in a timescale comparable with DUNE.
    
%    The ENUBET study offers a sound starting point for the construction of a new facility to host the next generation of cross section experiments. This facility should be aimed at an unprecedented control of the neutrinos at source, their flavor, and energy to sidestep current limitations of short baseline experiments and improve by one order of magnitude the cross section precision.
\end{abstract}
\newpage

\section{Introduction}
%FT
Neutrino oscillations at accelerators~\cite{Mezzetto:2020jka} play a prominent role in the study of the Yukawa sector of the standard model and the mass generation mechanisms. This research field has entered a new phase after the approval of DUNE~\cite{Abi:2020wmh} and HyperKamiokande (HK)~\cite{Abe:2018uyc}. These projects have an unprecedented level of complexity and cost, which resemble the most ambitious collider experiments performed in the last decades as the B-factories or the LHC experiments. Like collider experiments, the progress in the experimental precision must be matched by the progress in theory predictions to fully reap the physics potential of the new facilities. At the time of writing, theory predictions - in particular the study of neutrino cross sections - lag behind what is needed in DUNE and HK~\cite{Katori:2016yel}.

The DUNE and HK Collaborations mitigate this risk with a very careful design of the near \cite{AbedAbud:2021hpb,Abe:2019whr}  and far \cite{Abi:2020loh,Abe:2018uyc} detectors. Still, the traditional near-far cancellation technique is reaching its intrinsic limitations and the systematic uncertainty currently dominates the physics reach. 

This contribution to the Snowmass process addresses this issue in the most direct manner and echoes the outcome of the European Strategy for Particle Physics~\cite{Strategy:2019vxc}, which pointed out such weakness in the global effort to measure neutrino properties. Instead of refining the near-far cancellation technique, we discuss the opportunities and challenges of a new generation of short-baseline cross section experiments,  optimized to reach percent level precision in the inclusive, differential, and doubly-differential cross-sections. These experiments must address the region of interest for DUNE and HK but should not be considered only ancillary facilities of long-baseline experiments. Neutrino interactions at GeV energies \cite{sajjadatar2020} provide a wealth of information on the nuclear structure at moderate $Z$ and can be extended to cover the simplest nuclear system (e.g. hydrogen or deuteron) and explore the high-Z region of relevance for high-density detectors. Their goals are to gain a deep understanding of weak interactions in nuclear physics and provide high-precision data for the study of nuclear structure. These data will inform nuclear effective theories and, in the long term, lattice QCD. We call these facilities a \emph{new} generation of cross section experiments because major advances in this field has been achieved by the previous generation based on dedicated experiments (SciBooNE~\cite{AlcarazAunion:2009ku}, MINER$\nu$A~\cite{Rodrigues:2015hik,Aliaga:2015aqe}, WAGASCI~\cite{Giorgio:2019iuq}, ArgoNeuT~\cite{Acciarri:2020lhp}, etc.), the near detectors of long-baseline (K2K~\cite{Gran:2006jn}, MINOS~\cite{Evans:2013pka}, T2K-ND280~\cite{Abe:2014iza}, NO$\nu$A \cite{Acero:2020eit}) and short-baseline experiments (MiniBooNE~\cite{Aguilar-Arevalo:2013dva}, MicroBooNE~\cite{Abratenko:2020acr}). We expect the next generation of dedicated experiments -- running on a timescale comparable with the data-taking of DUNE and HK -- to achieve major improvements on $\nue$, $\numu$ and $\nubarmu$ cross-sections but also contribute to the short-baseline program for the search of physics beyond the Standard Model. 

In this document we advocate for a new facility to host the next generation of cross section experiments. This facility should be aimed at an unprecedented control of the neutrinos at source, their flavor, and energy to sidestep current limitations of short-baseline experiments and improve by one order of magnitude our knowledge of neutrino cross sections at the GeV scale.

\section{High precision cross section measurements}
%FT
Neutrino cross section measurements are currently dominated by the limited control of the neutrinos produced at accelerators. These limitations affect the knowledge of the flux of $\nu_\mu$, the size of the $\nu_e$ component and the energy spectrum of the neutrinos. An ideal source would hence be a source where the flux is known with a precision ten-times better than current beams ($10\% \rightarrow 1\%$) so that the remaining uncertainties are only detector-related (see below) or - for the case of $\nu_e$, statistics-limited. This is particularly relevant to reduce the systematics of DUNE and HK. 

For DUNE ($L\simeq 1300$~km) and HK ($L\simeq 295$~km), \numunue and \nubarmunubare are of paramount importance to establish CP violation in the neutrino sector and the number of \nue charged-current (CC) interactions observed far from the source is
\begin{equation}
    N^{e}_{FD} = \tilde{M}' \int dE \left[ \phi_{\nu_\mu}^{FD}(E) \ \epsilon_e(E) \ \sigma_{\nu_e}(E) P(\nu_\mu \rightarrow \nu_e) \ + \phi_{\nu_\mu}^{FD}(E) \epsilon_{\mu \rightarrow e} \sigma_{\nu_\mu}(E) \right]
    \label{eq:far_appearance}
\end{equation}
where $\tilde{M}'$ is the number of scattering centers (proportional to the mass and material of the detector), $\phi_{\nu_\mu}^{FD}(E)$ is the total flux integrated during the data taking at $L$ by the \textit{Far Detector} (FD), and  $\epsilon_{e}(E) \sigma_{\nu_e}(E)$ is the visible cross section, i.e., the cross-section corrected by the efficiency of the detector. 
$\epsilon_{\mu \rightarrow e}$ is the mis-identification probability of tagging a \numu as a $\nu_e$, e.g., due to neutral-currents (NC) events or non-reconstructed muons. For low precision experiments ($>5$\% systematic uncertainty), the uncertainties on the flux and the cross-section times detector efficiency can be removed by an identical detector located at short $L$, where oscillation effects are negligible, the \textit{Near Detector} (ND). For the ND, the \numu CC events provide an estimate of the initial flux 
%
%\begin{equation}
%    N_{ND} = \tilde{M}_N \int dE \ \phi^{ND}_{\nu_\mu}(E) \epsilon_{\nu_\mu}(E) \sigma_{\nu_\mu}(E)
%\end{equation}

\begin{equation}  
    N_{ND} = \tilde{M}_N \int dE \ \phi^{ND}_{\nu_\mu}(E) \epsilon_{\mu}(E) \sigma_{\nu_\mu}(E)
    \label{eq:near}
\end{equation}

and the \nue CC events measure the $\nue$ contamination at source:
%
%\begin{equation}
%    \stkout{N^{e}_{ND} = \tilde{M}_N \int dE \left[ \epsilon_e(E) \ \phi(\nu_e) \ \sigma_{\nu_e}(E) \ + %\epsilon_{\mu \rightarrow e} (E) \phi(\nu_\mu) \sigma_{\nu_\mu}(E) \right].}
%\end{equation}

\begin{equation}
    N^{e}_{ND} = \tilde{M}_N \int dE \left[ \phi_{\nu_e}^{ND}(E) \ \epsilon_e(E) \ \sigma_{\nu_e}(E) \ + \phi_{\nu_\mu}^{ND}(E) \epsilon_{\mu \rightarrow e} (E) \sigma_{\nu_\mu}(E) \right].
    \label{eq:near_appearance}
\end{equation}

In this way, the main contribution to the systematic uncertainty should come only from the 
\begin{equation}
 \frac{\epsilon_e (E) \sigma_e (E)}{ \epsilon_\mu (E) \sigma_\mu (E)}    
\end{equation}
ratio~\cite{Huber:2007em} and is strongly mitigated by the lepton universality, i.e. the statement that $\sigma_e (E) = \sigma_\mu (E)$ except for calculable kinematic corrections.
Unfortunately, such an elegant solution does not work at $<5$\% level. The ND and FD are not perfectly identical. Their efficiencies are different ($\epsilon^{ND} \neq \epsilon^{FD}$) both for \nue CC and \numu CC. Even more, $\phi_{\nu_\mu}(E)$ and  $\phi_{\nu_e}(E)$ are different at the near and far location beyond the expected geometrical reduction ($\phi \sim L^{-2}$) because the ND integrates the flux in a much larger neutrino phase-space. The solid angle seen by the ND is much bigger than the FD and this phase mismatch requires corrections both on flux normalization and energy spectrum, which, in turn, increase the systematic uncertainty.  

It is worth noting that flux, efficiencies, oscillation probabilities, and cross sections are all functions of the neutrino energy $E$. The neutrino energy is not known a priori and must be reconstructed event-by-event by the final-state particles of the neutrino interaction in the detector. This reconstruction is not possible in NC events because of the outgoing neutrino and troublesome in CC events every time a final-state particle is missed or mis-reconstructed. Again, this procedure generates a bias affecting the measured oscillation probability.
As a consequence, assuming priors on the neutrino beam (Monte Carlo simulation of the beamline), cross-sections (neutrino-nucleus models), and detector response are mandatory even after the near-far comparison. These priors, clearly, are not free of systematic uncertainties.

Along the same line, there is little hope to predict the Standard Model interactions with matter starting from theory, and all advances in this field require a sophisticated interplay between model-building and high-precision cross section measurements. A detailed account of cross-section limitations is given in \cite{symmetry_cross_section}.

\section{Monitored neutrino beams}
%FT
Monitored neutrino beams are beams where the flux of neutrinos is directly measured and diagnostics does not need to rely on the full simulation of the neutrino beamline. Here, the instrumentation -- mostly located in the decay tunnel --  monitors the production of the lepton associated with the neutrino at the single-particle level. If a muon neutrino is produced by a $ \pi^{+} \rightarrow \mu^{+} \nu_{\mu }$ decay, its appearance is signaled by the observation of the corresponding anti-muon. This observation represents a direct estimate because the number of anti-muons produced by those decays is equal to the number of muon neutrinos. An electron neutrino produced by a kaon decay - for instance  $K^{+}\rightarrow e^{+} \pi^{0} \nu _{e}$ - is signaled by the observation of a positron. Monitoring the production of leptons in the decay tunnel of an accelerator neutrino beam is very challenging due to the rate of secondary and tertiary particles. In the 1980s, monitored neutrino beams were built in the USSR in the framework of the "tagged neutrino beam facility" \cite{tagged}. This facility did not reach a flux sufficient to feed neutrino experiments and was later de-scoped to a tagged kaon beam facility. Current neutrino beams record muons but they have not reached single-particle sensitivity. Their precision on flux (15\%) cannot beat conventional techniques, yet. The most advanced monitored neutrino beam project is ENUBET, which aims at designing a monitored neutrino beam for high precision neutrino cross-section measurements instrumenting most of the decay tunnel with fast and radiation hard detectors, still at a marginal cost (5-10\%) compared with the total cost of the facility.

\section{ENUBET}

The ENUBET Collaboration was formed in 2015 and the feasibility study was supported by the European Research Council in the framework of the ERC Consolidator Grant program (PI A. Longhin). The ERC ENUBET project has delivered the first end-to-end simulation of a monitored neutrino beams where monitoring is performed by the detection of large angle positrons in the decay tunnel from $K^{+}\rightarrow e^{+} \pi^{0} \nu _{e}$. Given the outstanding results obtained in 2016-2019, ENUBET has been approved as a CERN Neutrino Platform experiment (NP06/ENUBET) and extended significantly its physics reach. NP06 monitors not only the kaon positrons but also muons from the leading two body decay of kaons $K^+ \rightarrow \mu^+ \nu_\mu$ and pions ($\pi^+ \rightarrow \mu^+ \nu_\mu$). It became possible by the design of a horn-less focusing system that can be pulsed for several seconds reducing the particle rate at the decay tunnel by nearly two orders of magnitude. Muons from pion decays can then be monitored by additional instrumentation located at the end of the decay tunnel (hadron dump). The NP06 study is site-independent and we have shown that its implementation can be carried out both at CERN and Fermilab. A low energy version, of particular relevance for T2K, could also be implemented in the European Spallation Source (ESS) in a longer timescale and is being investigated in the framework of the INFRADEV EU program. Given the maturity of the design, NP06/ENUBET is currently working in collaboration with CERN (Physics Beyond Collider) to implement this novel facility at the CERN SPS and exploit the existing CERN detectors (ProtoDUNE: NP02 and NP04).  

The ENUBET beam is a conventional narrow-band beam with two differences with respect to the current beams: the decay tunnel is not located along the proton axis of the focusing system and the proton extraction length is slow. Particles produced by proton interactions in the target are focused, momentum selected and transported to the tunnel entrance. Non-interacting protons are stopped in a proton beam dump. Off-momentum particles reaching the decay tunnel are mostly low
energy particles coming from interactions in the collimators and other beamline components together with muons that cross absorbers and collimators. The hadron beam considered has a reference momentum of 8.5 GeV/c with a momentum bite
of 10\% and we expect about 50\% of $K^+$ to decay in a 40 m long tunnel.
A multi-momentum beamline with an optics hosting secondaries down to a few GeV is also under study to increase the neutrino flux below 1 GeV.

Focusing is performed by a set of dipole and quadrupoles (static focusing system) and, in 2019, we demonstrated that the static system can achieve a flux that allows for the measurement of $\nu_e$ cross section at 1\% level, while the statistical error for $\nu_\mu$ interactions is below 0.1\% due to the even  larger $\nu_\mu$ flux.
Such a result was obtained assuming a neutrino detector, located about 100 m from the target of the same size of ProtoDUNE-SP.
A horn based option is now considered a backup option and pursued to further increase the flux.

Special emphasis should be given by the possibility of developing ENUBET as an incremental step toward muon-based neutrino sources with an outstanding physics case on its own. In 2021, the ENUBET and nuSTORM collaborations have started a synergic development, which is described in a dedicated contribution to Snowmass \cite{snowmass_nustorm}.

\section{The ENUBET implementation}

We discuss here the most important outcomes of the ENUBET R\&D: the development of the hadron beamline, simulation and reconstruction algorithms for the lepton tagger, 
the detector prototyping activities and the studies to assess the reduction of systematics on the flux.
%\subsection{Proton driver and target station}
%\label{sec:proton}
% Michelangelo
% Let me know if you'd like more figures. I kept them to the minimum following the guidelines of "executive summary"
\subsection{Burst-mode slow extraction}

ENUBET must employ a slow extraction of the primary protons to mitigate pile-up at the instrumented decay tunnel (i.e.~full intensity continuously extracted in a few seconds). However, in order not to rule out the option of using a magnetic horn for potentially boosting the neutrino flux (Sec.~\ref{sec:horn}), we have studied and developed a novel pulsed slow-extraction method at CERN-SPS, in collaboration with CERN (BE-OP-SPS and TE-ABT-BTP). In this type of extraction, multiple ms-long pulses are slow-extracted during the flat-top at a fixed repetition rate. The main results of this campaign have already been reported in the previous report~\cite{spsc_2020}, and have been confirmed by further off-line analysis. They can be briefly summarized as follows:
\begin{itemize}
    \item A new pulsed slow extraction method (called burst-mode slow extraction) has been successfully implemented and tested at CERN-SPS before the Long Shutdown 2 (LS2). Our measurements showed that the scheme can be readily switched-on and operated with the SPS control tools we developed, and that the most critical parameter is the length of the extracted proton pulses. By upgrading one of the existing SPS control-room applications, we have proven that the new extraction scheme can be successfully optimized down to $10$~ms pulses repeated at $10$~Hz, as first envisaged by ENUBET in~\cite{enubet_eoi}.
    \item During LS2, we developed a full simulation of this extraction scheme to validate the experimental results and explore possible improvements, which could not be tested in the machine before the shutdown. Using two different methods (increase of extraction sextupole strength and amplitude extraction), we found that pulses ranging between $3$ and $10$~ms can be produced without any hardware modification.
\end{itemize}
The burst-mode slow extraction is a general extraction method with application even beyond the horn-based ENUBET layout: in the static focusing option of ENUBET, it can provide a time structure employed at the neutrino detector to reduce cosmic-induced background. Other users of the CERN SPS and PS beamlines can employ this method to increase and control the instantaneous particle rate for pile-up, DAQ and PID studies. 
%\vspace{1cm}

%\subsection{Study of ripples in the slow extraction}
%\noindent 
%%In the context of the collaboration with CERN on the SPS slow extraction, we have recently finalized a study on the spill quality of the nominal slow extraction. 
Ripples appearing on the extracted spill, generally caused by harmonics coming from the magnet power supplies or other electric nuisances, are a well known issue in slow extraction. The suppression of these ripples is the subject of long-standing efforts for both research and medical accelerators~\cite{ripples_sps,autospill,ripples_jparc,ripples_gsi,ripples_cnao}. In the case of high-energy physics experiments, this noise is detrimental because it enhances the instantaneous pile-up and counting uncertainties~\cite{sxw17_johannes}.

\begin{figure}[htb!]
    \hspace{-1cm}~\includegraphics[width=0.6\textwidth]{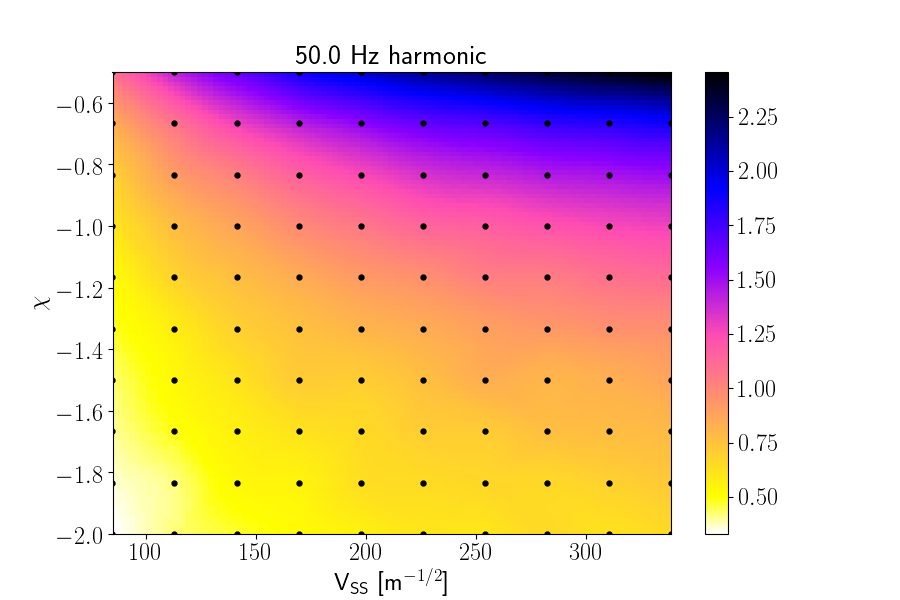}~\hspace{-1cm}~\includegraphics[width=0.6\textwidth]{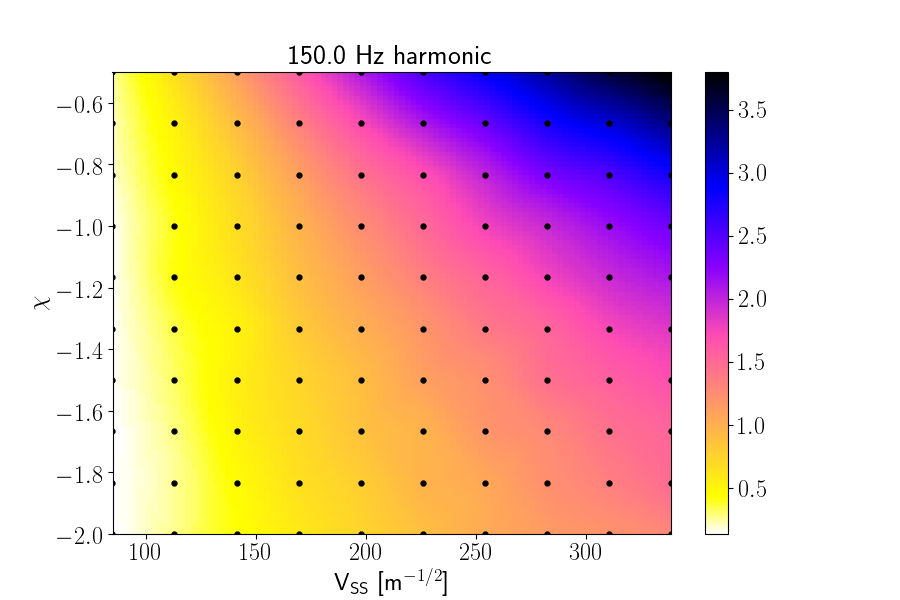}
    \caption{Simulated scan of the relative amplitude of power-supply ripples compared to the nominal SPS case (unit value in the color scale), as a function of the normalized chromaticity ($\chi$) and virtual extraction sextupole strength ($V_{\text{ss}}$).}
    \label{fig:gridHenon}
\end{figure}

We have characterized the frequency response of the slow extraction in terms of its low-pass filter behavior. This has been performed with a full MADX~\cite{madx} model of the process, while we have also developed custom numerical models for fast simulation, which proved to be in good agreement with MADX.
We validated our models both by performing dedicated ripple-injection measurements at the SPS and analyzing operational data. % in order to validate our models.
Ultimately, we could exploit one of the developed fast simulation models to look for possible improvements in the suppression of ripples by scanning two of the main SPS extraction parameters: chromaticity and sextupole strength. The results obtained so far are promising and show that a reduction of a factor $\sim 2$ is within reach, as shown in Fig.~\ref{fig:gridHenon}.

% toglierei questa frase che mi sembra ridondante FT
%This work will be the basis for a renewed effort towards the slow extraction spill quality at CERN, which will be continued in the future years with more dedicated studies and experimental validations (e.g.~of the latter improvements we observed from the parametric scan).
The present work will play an important role for future improvements of the spill quality of the SPS slow extraction, increasing the efficiency of ENUBET and other fixed-target experiments.

\subsection{Target optimization}
\label{sec:target}
% Elisabetta
% A cut and paste from our review is out of context in this report. I reshaped this session linking it directly to the ENUBET needs FT
Neutrino experiments extrapolate and determine the flux of neutrinos from the yield of secondary hadrons produced by the target. This is often the most important source of systematic uncertainties~\cite{Charitonidis:2021qfm}. While it is possible to derive models for secondary production by fitting experimental data~\cite{Bonesini2001}, the uncertainties remain high ($>10$\%), especially for the low energy region, as reported in several surveys~\cite{Chemakin2008,Abgrall2016,Catanesi2008}.

The target of ENUBET does not represent a substantial challenge compared with focusing and instrumentation since we leverage the experience gained by long-baseline experiments and envisage a facility with an average power of about 120~kW \footnote{Assuming a maximal proton rate given by $4.5 \times 10^{13}$ protons over a 24~s super-cycle.}.
On the other hand, a re-optimization of the target has been carried out because the project is in a very advanced state and we are now addressing also engineering issues and costs.
%which do not encourage the use of the beryllium target envisaged in~\cite{Longhin:2014yta}.  

%The neutrino beam energy range is set by selecting and focusing pions and kaons around a given central momentum. Neutrino beam designers have to tune several parameters of the primary proton source, e.g.~the primary proton momentum, extracted intensity, and repetition rate, to reach the sought-for intensity of the facility. Once the primary source is set, target optimization plays the most important role in the optimization of secondary yields. 

Besides the choice of target material, the target geometry determines the re-interaction probability and absorption of the secondary particles coming out of the target after the collision with the primaries. The optimization of the target geometry, however, is not only driven by the produced yields, but also by mechanical constraints and cooling requirements.
%, especially in the cases of high-intensity beams~\cite{Bruno2011}. In CNGS, for instance, the target was an assembly of air-cooled, thin graphite rods, acting as multiple targets that could be hit separately. This configuration allows for the spare targets to be already located on-site in case of necessity.

%Heat management is another parameter that needs consideration, especially in Super beams. The energy deposited in the target material by the primary beam results in tensile stresses radiating out from the central beam spot~\cite{Hurh2012}. The main parameters that determine the heat endurance are the specific heat and the tensile properties of the medium. These properties drive the choice of the heat dissipation system. 
%Heat dissipation may occur through the container that holds the target and the heat transfer from the target material can be either conductive or radiant in solid targets. Conductive internal heat transfer is very effective in metal targets, while radiant heat transfer is better suited for high-emissivity refractory materials and inserts, such as graphite~\cite{Popescu2020}. 

The overall size of the target plays a central role in the cooling design. An increased surface implies a larger radiative heat transfer and the change of the size of a target is driven by the absorption/emission parameters of the material. 
Increasing the size could lead to a lower production yield due to secondary re-interactions inside the target. The optimization is then a trade-off between mechanical robustness against heating and the effective interaction length crossed by the secondaries.

The NP06/ENUBET team has conducted extensive optimization studies based on the FLUKA~\cite{FLUKA1,FLUKA2} and G4beamline~\cite{g4beamline} 
simulation codes, using graphite (density $2.2$~g/cm$^{3}$), beryllium (density $1.81$~g/cm$^{3}$), Inconel (density $8.2$~g/cm$^{3}$) and various high-$Z$ materials such as gold and tungsten. Each target prototype is modeled geometrically as a cylinder with variable radii between $10$ and $30$~mm and lengths extending from $5$ to $140$~cm. 

\begin{figure}[!htb]
\centering
\includegraphics[width=\linewidth]{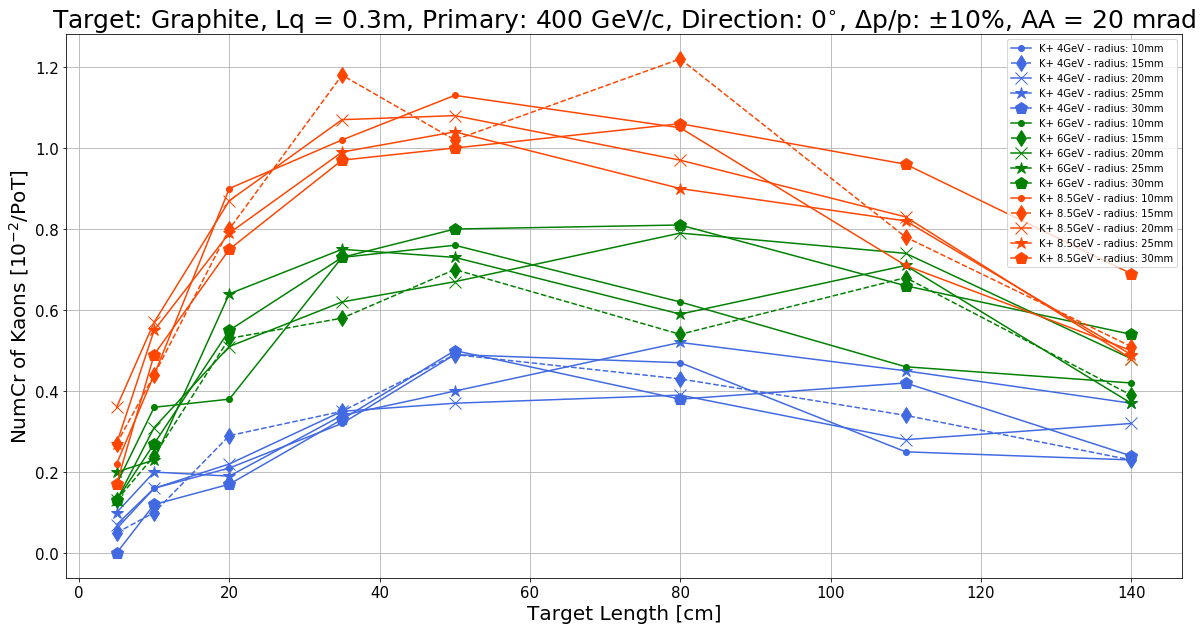}
    \caption{Kaon yields as a function of the graphite target length. The primary beam simulated is a $400$~GeV/c proton beam. The figure of merit for this study is the number of kaons of given energy with $10\%$ momentum bite that enters an ideal beamline with $\pm 20$~mrad angular acceptance in both planes, placed $30$~cm after the target. The error bars are not plotted to ease the reading; statistical errors are negligible ($1$\%), while the Monte-Carlo systematics amounts to $\sim 20$\%.
   Colors refer to different kaon's momenta while the marker style identifies the tagger radius.} 
    \label{fig:K_prod}
\end{figure}

Initially, we analyzed the target secondary particle productions assuming different primary energies. As expected, the number of $\pi^{+}$ per POT linearly grows with the primary proton momenta~\cite{Feynman1969}, which determines the accessible momenta for the secondaries and, hence, the neutrinos. We studied primary protons with momenta of 400, 150, 70, and 50~GeV/c.
FLUKA results confirmed that the nominal energy of the SPS (400~GeV/c) is a good choice for ENUBET, especially for cross section studies in the region of interest for DUNE. This region requires a secondary mean momentum for kaons of about 8.5~GeV. We then proceeded comparing the geometry and material of the target.
We have shown that the best materials for monitored neutrino beams are graphite (2.23 g/cm$^{3}$ density), beryllium and Inconel-718. The kaon yields for graphite are shown in Fig.~\ref{fig:K_prod}. 
%and Fig.~\ref{fig:K_prod_B}. 
Higher-$Z$ materials would prove better candidates for kaons production, however, they pose severe challenge to the cooling system and are not an option we will pursue.  
%\begin{figure}[!htb]
%\centering
%%\includegraphics[scale=0.3]{figs/Kprod_400_AA_20_30_Be_new.png}
%\includegraphics[width=\linewidth]{figs/Kprod_400_AA_20_30_Be_new.png}
%    \caption{Same as Fig.~\ref{fig:K_prod}, but for Beryllium targets.} 
%    \label{fig:K_prod_B}
%\end{figure}
Graphite is a known and well-tested material employed in several neutrino beams thanks to its heat endurance and production yields~\cite{Hurh2012}; Inconel is quite a novel choice that is under consideration for nuSTORM~\cite{Adey:2013pio} and ENUBET, but already at use at CERN in other applications (like the new CERN-PS East Area Beam Stoppers).

The most advanced ENUBET beamline (TRL6 -- see Sec.~\ref{sec:tl}) employs a graphite target with a 3~cm radius and a length of 70 cm. An Inconel target with the same radius and a 50~cm length is also being considered. The Collaboration is in contact with the CERN developers, in particular the CERN SY Department, to address a precise cost estimate of the target station, possibly in synergy with the nuSTORM studies performed at Physics Beyond Collider.

\subsection{Horn optimization}
\label{sec:horn}
% Michelangelo

% Tell me if you'd think I should put more figures

While the baseline option for ENUBET is a fully static beamline, the use of a horn combined with the pulsed slow extraction described above is pursued as an alternative option. In particular, as the magnetic horn is a complex device, a full automatic optimization approach is required in order to find the best geometry for maximizing the flux of focused secondaries. Since the starting point of this work described in the past report~\cite{spsc_2020}, we have made important advances.

\begin{figure}[!htb]
    \centering
    \includegraphics[width=0.8\textwidth]{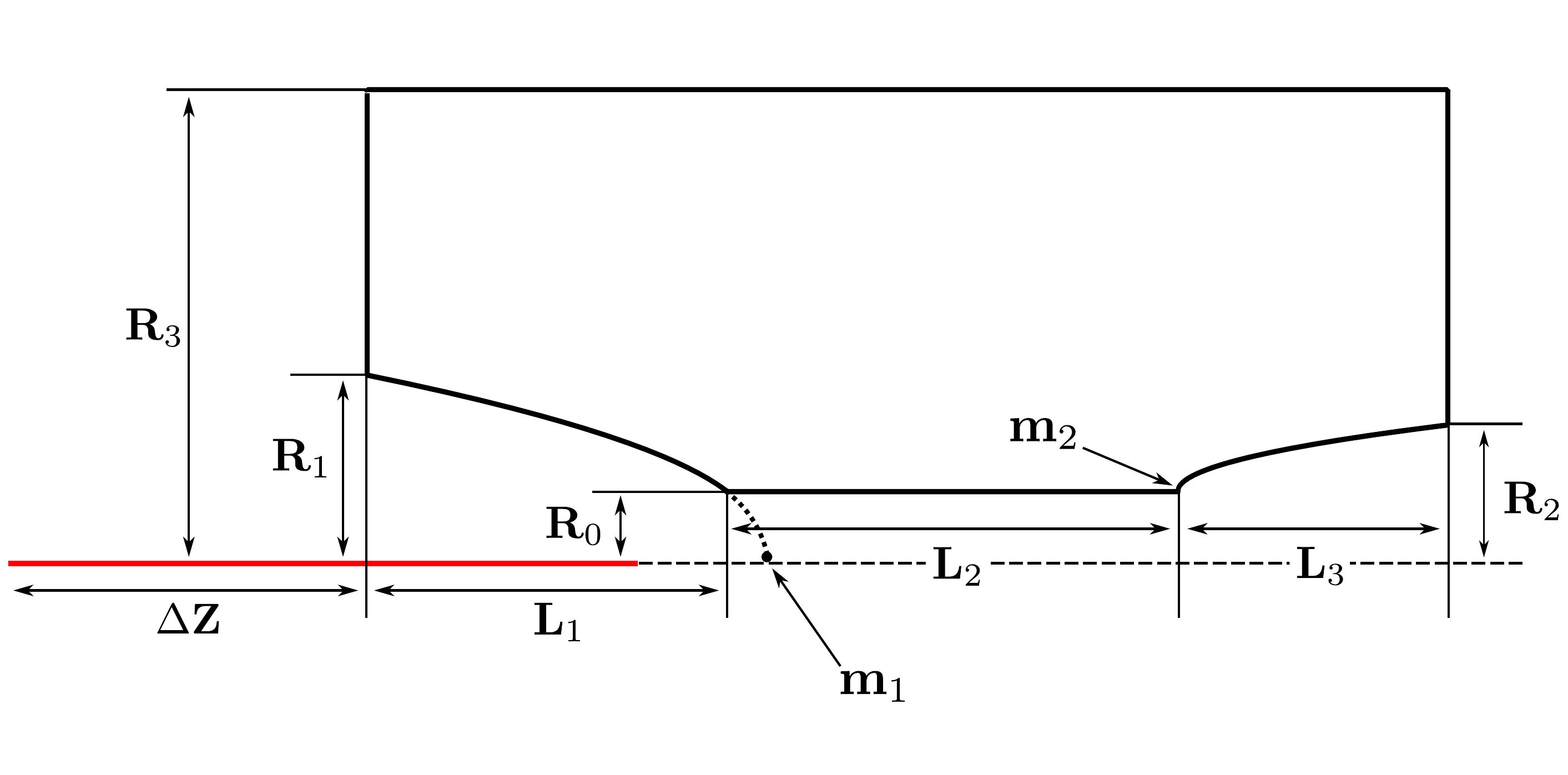}
    \caption{Design of the double parabolic horn available in the developed GEANT4 simulation.}
    \label{fig:horndp}
\end{figure}

First, a full GEANT4~\cite{geant4} simulation model for the magnetic horn has been successfully developed. Other than the MiniBooNE-like geometry previously reported, a new double parabolic geometry has been implemented (Fig.~\ref{fig:horndp}).

We have then completed the implementation of a dedicated framework for optimizing the magnetic horn configuration: it is composed of different software elements and it is interfacing with a job scheduling system (Grid Engine from the CC-IN2P3 cluster), which is required given the computational intensive problem. 
%A schematic of the developed framework is illustrated in Fig.~\ref{fig:hornsw}.
%\begin{figure}[!htb]
%    \centering
%    \includegraphics[width=0.6\textwidth]{figs/sw_struct.pdf}
%    \cprotect\caption{Structure of the optimization framework for the magnetic horn. A Python script manages the full optimization process by interfacing with the Genetic Algorithm (GA) libraries, job-scheduling system, configuration of the GEANT4 model of the horn, and dedicated \verb!C++! analysis routines.}
%    \label{fig:hornsw}
%\end{figure}
Since the only way to obtain a quantitative estimation of the horn fluxes is via a full numerical simulation, and no other crucial information is obtained otherwise, using a meta-heuristic and stochastic-based optimization algorithm is a good way to scan the parameter space for global optima. For this reason, we have based the horn optimization on a Genetic Algorithm (GA) implementation. We have inserted the handling of external constraints on the horn parameters, in order to take into account hardware limitations and requirements (e.g.~total horn length, current, radii, etc). The Figure of Merit (FoM) of the optimization is based on the number of focused $K^+$ in the designed momentum bite, computed at two possible locations:
\begin{itemize}
    \item At the first quadrupole after the horn. This particular FoM has the advantage of being independent from the downstream beamline: it is used to maximize the flux at the output of the horn and inside the acceptance of the first quadrupole.
    \item At the entrance of the decay tunnel. This FoM requires a further computational intensive tracking of the particles, but it provides the quantitative gain for the specific beamline under consideration.
\end{itemize}
In order to test the optimization framework, we have decided to first optimize the horn in a standalone way using the beamline-independent FoM, and then to compare the result with a full beamline tracking using the static configuration. The main results can be summarized as follows:
\begin{itemize}
    \item The optimization framework successfully reaches an optimum in $\sim 100$ iterations (taking typically from a few hours to a day of computation time), the constraints are also enforced correctly.
    \item For different horn geometries and constraints, we were able to reach values of the standalone FoM of a factor $3$ higher than the static case. These results confirm an improvement with respect to the horn configurations used in early studies initially performed in $2016$.
    \item When further tracking the particles along the full static transfer line in the current configuration, we observed an overall reduction in the gain of kaons with respect to the standalone FoM value, up to about a factor $2$. 
    %\item When further tracking the particles along the full static transfer line, we observed a strong dependence of the FoM at the decay tunnel to the particular beamline configuration employed, as shown in Tab.~\ref{tab:hornColl}. %occhio che questo discorso è misleading, la simulazione per studiare le aperture/chiusure dei collimatori l'avevo girata con i files da horn perchè stavamo usando quelli ma il gain si ottiene anche per una configurazione statica (perchè aprendo i collimatori va da sè...), ed era stato fatto per capire il "contributo" delle varie sezioni della TL, in più il motivo per cui non si può comunque adottare nessuna delle configurazioni in tabella (che verrebbe da chiedersi visto il gain why not...) è che cresce molto il background. Bisogna dire che un'intera transferline come quella statica piazzata dopo l'horn probabilmente non permette di sfruttarlo al meglio per cui ci vuole un disegno ad-hoc ma non legherei questa tabella alla FOM dell'horn... GB
%\begin{table}[htb!]
%\centering
%\begin{tabular}{lc}
%\toprule
%& $\mathbf{K^+}$ \textbf{gain at tunnel}\\
%\midrule
%\textbf{Nominal static beamline} & 1.6 \\
%\textbf{Collimators opened to quad. apertures} & 6.8 \\
%\textbf{All collimators opened but last one} & 4.7  \\
%\textbf{All collimators opened but triplet ones} & 3.3 \\
%\bottomrule
%\end{tabular}
%\caption{Gain in the number of $K^+$ at the entrance of the decay tunnel for different collimation configurations of the static version of the ENUBET beamline, compared to the no-horn case.}
%\label{tab:hornColl}
%\end{table}

\end{itemize}

The most important consideration that can be drawn from the previous results is that further studies on a dedicated beamline configuration specific for the horn-focused beam are necessary to really determine the total achievable increase of flux. In fact, the current implementation of the static transfer line is not tuned to be paired with a horn: to exploit this device at best, significant modifications could be needed. For this reason, the next step will be to design and optimize such an alternative beamline.
However, the successful convergence of the developed optimization framework to standalone horn optima is positive: these horn candidates satisfy basic hardware constraints and can be used as starting point for the further beamline optimization. On top of this, the developed optimization framework will be upgraded to become a generic tool available for the optimization of other components of ENUBET (e.g.~collimation).

\subsection{Transfer line}

\label{sec:tl}
\begin{figure}[ht!]
\centering
%   \includegraphics[height=2.5cm]{figs/tlr5.png}
%  ~\hspace{0.1cm}
%  \includegraphics[height=2.5cm]{figs/tlr6.png}
%  \includegraphics[width=\linewidth]{figs/tlr5.png}\\~\\
  \includegraphics[width=\linewidth]{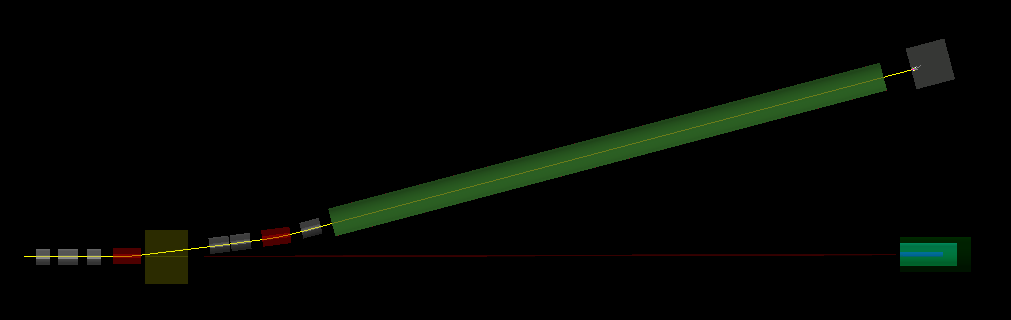}
  \caption[]{Bending elements: dipoles are shown in red while quadrupoles are shown in grey. Both transfer lines have the same design for the proton dump. The hadron dump is placed downstream the tagger exit. At the tagger entrance is placed a INERMET180 block with a conical aperture.}
  \label{fig:tl}
\end{figure}

% Giulia
The overall concept of the ENUBET transfer line and the factors guiding our optimization have been described in~\cite{spsc_2020}. Here we present the latest design for a static transfer line optimized for a 8.5~GeV/c reference particle. We consider normal-conducting bending dipoles with 1.8~T field, providing a total bending of the beam w.r.t the primary proton line of 14.8°. The quadrupoles have an aperture radius of 15~cm.
The optics is optimized considering a 10\% momentum bite and an improved target (described in ~\ref{sec:target}) made by graphite, 70~cm long and with a radius of 3~cm. 
The layout is presented in Fig~\ref{fig:tl}, where we only show the focusing elements, the dumps, the copper block downstream the first dipole with two pipes (one for the primary protons and a tilted one for the ENUBET hadron beam), and the tagger. The absorbers placed all along the transfer line and the rock volumes are nevertheless used in the complete simulation that have been studied and optimized using a complete FLUKA and GEANT4 simulation (Sec.~\ref{sec:simupackages}). The design implements an INERMET180 tungsten alloy block at the tagger entrance with a conical aperture.

Downstream the target and before the first quadrupole of the transfer line we also make use of a 5~mm thick tungsten foil to eliminate the positrons generated by the interactions in the target. Its thickness has been chosen after a series of simulations to optimize it by checking the ratio of positron coming from the target and hitting the tunnel walls over the $K^{+}$ flux. The target has been completely surrounded by concrete.

\begin{figure}[ht!]
\centering
   \includegraphics[width=\textwidth]{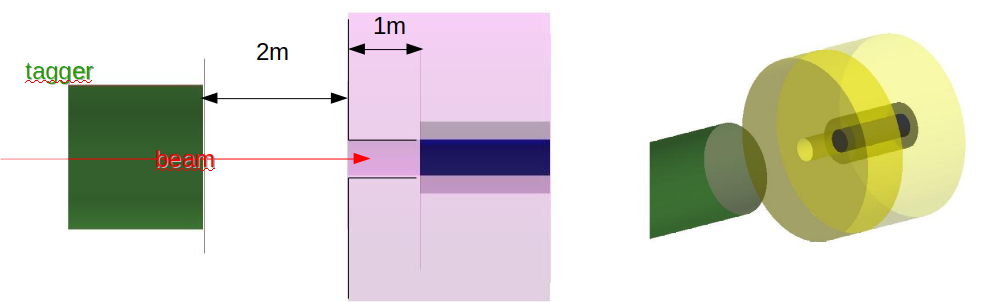}
  \caption[]{Preliminary new design of the hadron dump that helps reducing considerably the contribution to the background in the tagger due to the backscattering. There is 1 meter of borated concrete with an opening for the beam in front of the dump. The dump itself is composed by three cylindrical layers: the core is made by graphite, the second layer is made by iron and the last one is made by borated concrete.}
  \label{fig:hDump}
\end{figure}

The proton dump is composed by three cylindrical layers: a 3~m long graphite core, surrounded by aluminum, which in turn is covered by iron.
The hadron dump is placed 2~m after the tagger exit and a new design has also been studied in order to reduce the backscattering reaching the last part of the decay tunnel that would result in additional particles hitting the instrumentation. A preliminary design that follows the same structure of the proton dump and that allows to reduce the flux in the tunnel due to backscattering is presented in Fig.~\ref{fig:hDump}: a graphite core of 50~cm diameter is placed inside a layer of Iron (1~m diameter), covered by borated concrete (4~m diameter), 1~m of additional borated concrete is placed in front of the hadron dump leaving the opening for the beam. A more complete simulation performed with FLUKA and GEANT4 will determine the final design but the one presented here has already the capability to reduce by a significant amount the neutron flux all along the tagger. In particular in the last meters of the tunnel where the neutron fluence is more significant the ratio between neutrons from the hadron dump hitting the tagger w.r.t those with the old design is $\sim$~0.2.

\begin{table}[ht!]
\centering
\begin{tabular}{ |l || c | c || c| }
\hline
Transfer line & $\pi^{+}$ [$10^{-3}$/POT]& $K^{+}$  [$10^{-3}$/POT]  \\
label & [8.5 $\pm$ 10\%] GeV/c & [8.5 $\pm$ 10\%] GeV/c  \\
\hline
%previous TL & 2.05 & 0.185 &  \\
%\hline
%TLR5 & 3.4 & 0.28 & 1.5\\
TLR6 & 4.2 & 0.4  \\
\hline
\end{tabular}
\caption{Fluxes obtained with G4Beamline at the tunnel entrance for $\pi^{+}$ and $K^{+}$ in the [8.5 $\pm$ 5\% GeV/c] range in units [$10^{-3}$/POT]. In the last column we report the ratio to the previous result on $K^{+}$ flux.}\label{tab:flux}
\end{table}
\begin{figure}[ht!]
\centering
   \includegraphics[width=\textwidth]{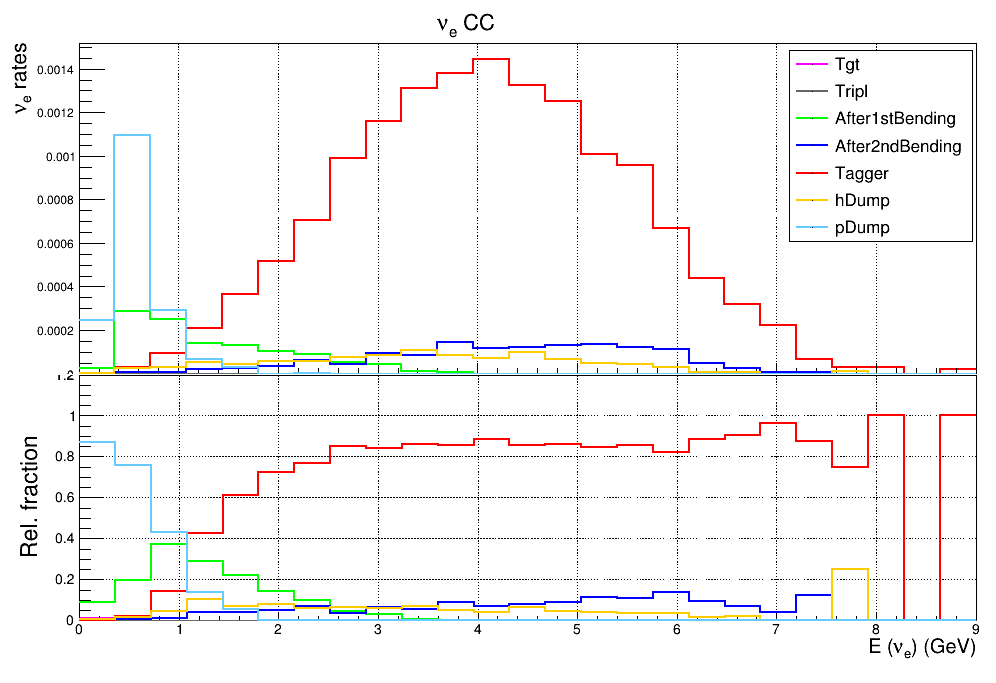}
\caption[]{Top: $\nu_e^{CC}$ interactions considering a 6~$\times$~6~m$^{2}$ front face of a neutrino detector located 50~m downstream the tunnel exit divided in categories corresponding to the position along the transfer line where the neutrino was generated. The red spectrum corresponds to neutrinos generated inside the tagger. Bottom: relative fraction of each category to the total $\nu_e^{CC}$ rate.}
  \label{fig:nuecc}
\end{figure}

In Tab.~\ref{tab:flux} the flux obtained at the tunnel entrance for 8.5~GeV/c $\pi^{+}$ and $K^{+}$ is reported.
%, considering 5\% momentum bite for comparison, we present the increase in kaon flux obtained with respect to the previous estimate as well.
%The fluxes were computed again with a complete G4Beamline simulation for the TLR5 with the new optimized target as it results in a flux increase.
In Fig.~\ref{fig:nuecc} we present the spectra for the $\nu_{e}^{CC}$ interactions seen in a far detector placed 50~m downstream of the tunnel end with the TLR6, divided in categories depending on the origin of the neutrino along the transfer line. The neutrinos coming from the decay tunnel are clearly separated in energy from those generated in the proton dump and in the first section of the beamline. Moreover, 73.5\% of the total $\nu_{e}$ flux is generated inside the tunnel, and above 1~GeV this component represent more than 80\% of the total flux. Below 1~GeV the main component is given by neutrinos produced in the proton-dump region, which will allow us to further improve the separation of the signal against this component by optimizing the proton dump position.
The contribution of $\nu_{e}$ coming from the straight section in front of the tagger (i.e. from the second dipole to the tagger entrance) amounts to $\sim$~7\% and another $\sim$5~\% is given by the section after the tunnel exit and the hadron dump. Even though indistinguishable from the $K_{e3}$ neutrinos produced inside the decay tunnel this 12~\% can be corrected for by relying on the simulation.

%Comparing with previous results we have an increase factor of 2.5 in the number of $\nu_{e}^{CC}$ seen at the far detector with the TLR6 , and 1.4 with the TLR5. 
By assuming 4.5$\times$10$^{19}$ POT/year at the SPS the TLR6 enables to reach 10$^{4}$~$\nu_{e}^{CC}$ at the far detector about two years. A previous design (TLR5) with the new optimized target would allow to reach the same statistics in about 3 and half years.

\subsection{Doses in the decay tunnel}
\label{sec:irradiation}
% Andrea
%[Andrea]
The FLUKA simulation of the double-dipole beamline (TLR5) has been updated including proper shielding to provide a sensible estimate of the ionizing doses and neutron fluences for all the elements of the beamline. 
We are particularly interested in the decay pipe where scintillators and Silicon Photo Multipliers (SiPM) will be located and the location of the first rad-hard focusing quadrupole. We are also evaluating the activation rates to design engineering solutions for a possible accessibility of the instrumented area for maintenance.

%Such layout of the beamline has been defined by a dedicated study performed
%with the TRANSPORT and G4BeamLine codes. The transfer line from the target to the decay tunnel produces an intense and collimated hadron beam with low levels of stray particles. The optimization performed with G4Beamline minimizes the length of the transfer line to reduce losses from kaon decays occurring before the entrance of the decay tunnel. 

\begin{figure}[htb!]
\centering
  \includegraphics[width=\linewidth]{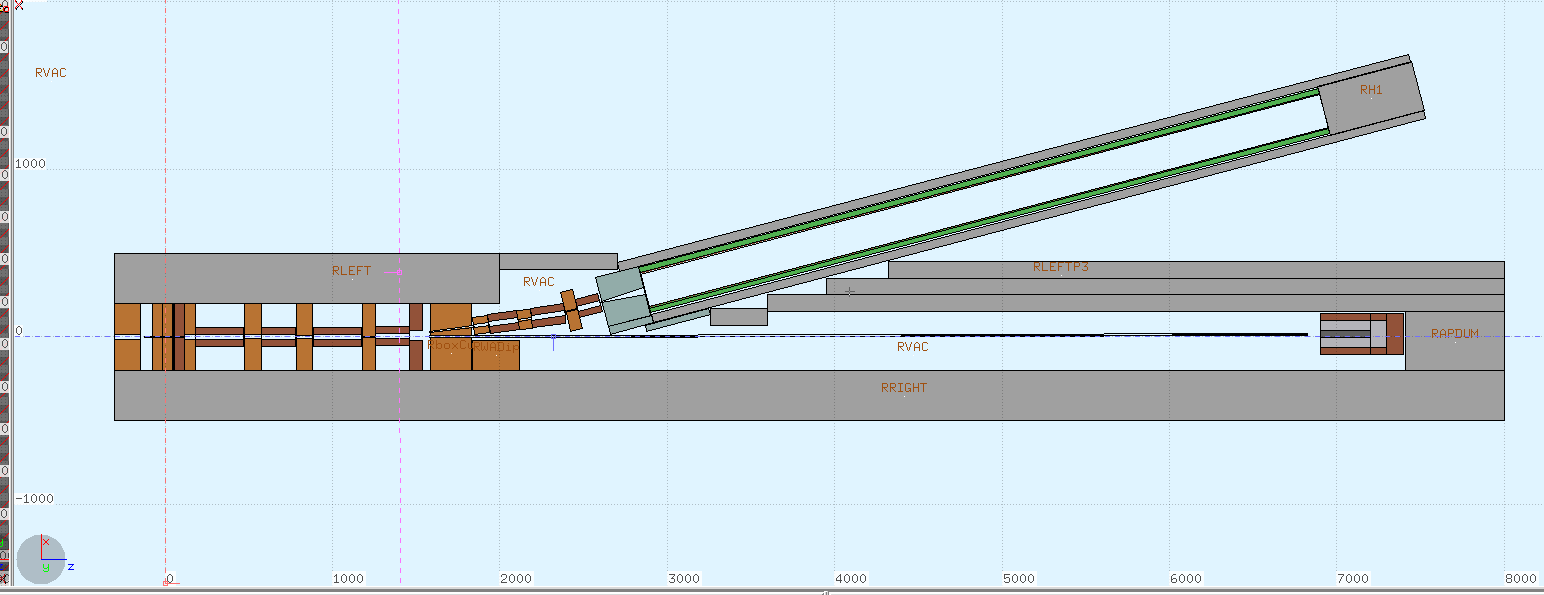}
  \includegraphics[width=\linewidth]{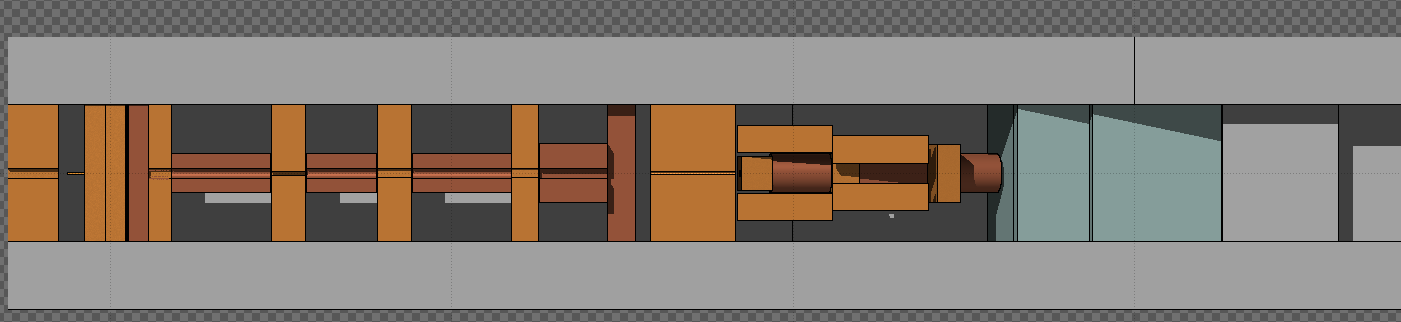}
  \caption[]{Top: FLUKA implementation of the double-dipole transfer line model TLR5. The regions in green are those composed of borated polyethylene, concrete is in gray, tungsten is in greenish while orange and brown represent copper and iron, respectively. Bottom: 3D rendering of the first straight section with a cut along a vertical plane passing through the proton beam.}
  \label{fig:FLUKAlayout}
\end{figure}

The geometry layout of active elements and shielding is shown in Fig.~\ref{fig:FLUKAlayout}.
The FLUKA geometry description is semi-automatically generated within GEANT4. The FLUKA model hence faithfully reproduces the GEANT4/G4Beamline one.
%In this scheme
%\footnote{Several variations - i.e. employing a pair of bending dipoles - are still under scrutiny.} 
%the beamline consists of an ``on-axis'' quadrupole triplet followed by a single dipole and an ``off-axis''
%quadrupole triplet. The overall bending angle of the resulting
%neutrino beam with respect to the proton axis is $\sim$~7.4$^\circ$.
%Quadrupoles and dipoles were dimensioned to achieve a collimated beam of
%pions and kaons at an average momentum of 8.5~GeV/c and a momentum
%bite of 5-10\% with the shortest possible length to avoid losing too
%many neutrinos from early decays of kaons ($\beta\gamma c \tau
%\sim$~63~m at 8.5 GeV/c). 
%While implementing the beam optics in FLUKA several optimizations were
%performed mainly in terms of collimators and shielding. In particular 
%we explored for the effectiveness of a Tungsten plug in front of the decay pipe to
%protect the calorimeter from background particles thanks to the large
%stopping power of this dense material. This design is currently also
%adopted in the latest double-dipole version of the beamline (Sec.~\ref{sec:tl}) and most of these considerations apply to TRL6, as well.
%
%Recently also the double-dipole transfer line has been modeled in FLUKA (Fig.~\ref{fig:tlr5_fluka}).
%
%We expect a further reduction of neutron fluences and we are in the process of tuning 
%the shielding with FLUKA and update the results shown in Fig.~\ref{fig:neutronS} for the single-dipole beamline.

\begin{figure}[htb!]
    \centering
    \includegraphics[width=\columnwidth]{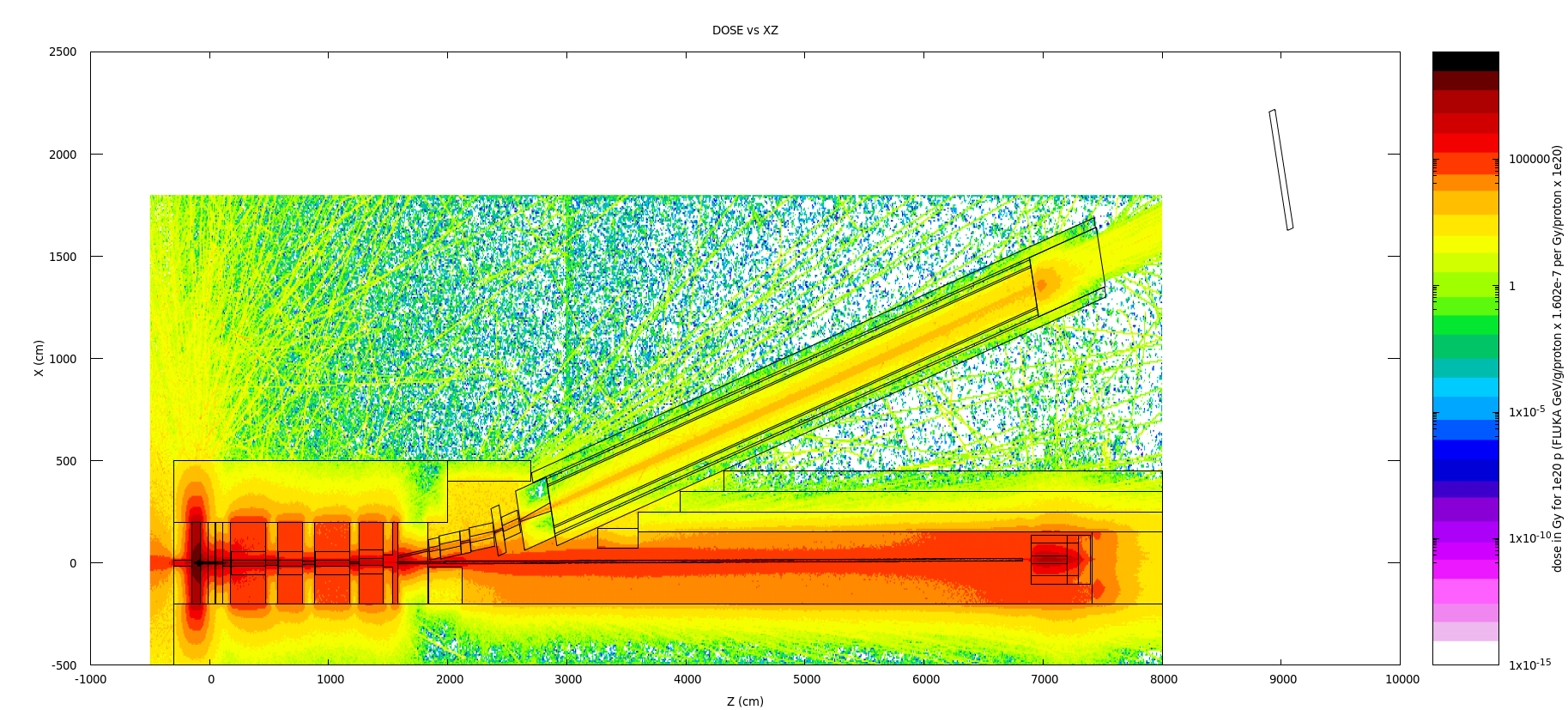}
    \caption{Dose map obtained in Gy for 10$^{20}$~POT (FLUKA). The first quadrupole sits roughly between $z=200$ and 500~cm.}
    \label{fig:mapdose}
\end{figure}

\begin{figure}[htb!]
\centering
  \includegraphics[width=\linewidth]{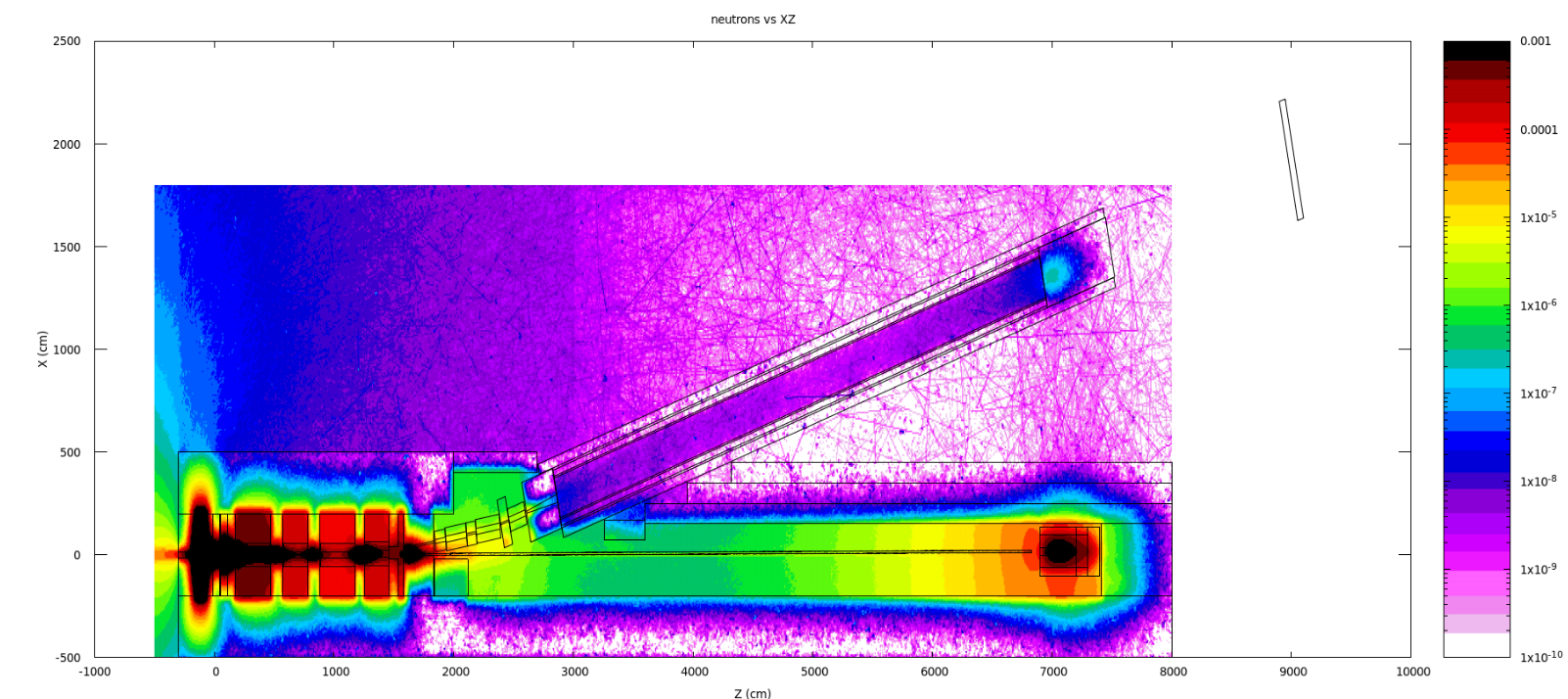}
  \caption[]{Top: neutrons fluences for TLR5.}
  \label{fig:tlr5_fluka_neutrons}
\end{figure}

The map of the accumulated dose obtained with FLUKA in Gy for 10$^{20}$~POT is shown in Fig.~\ref{fig:mapdose}. The dose at
the hottest point of the quadrupole closest to the target is 
of the order of 100-300 kGy for 10$^{20}$~POT.
This is a totally acceptable value in terms of operability of a suited magnet.
Neutron fluences are shown in Fig.~\ref{fig:tlr5_fluka_neutrons} for the same amount of accumulated POT.

\begin{figure}[htb!]
\centering
  \includegraphics[width=0.5\linewidth]{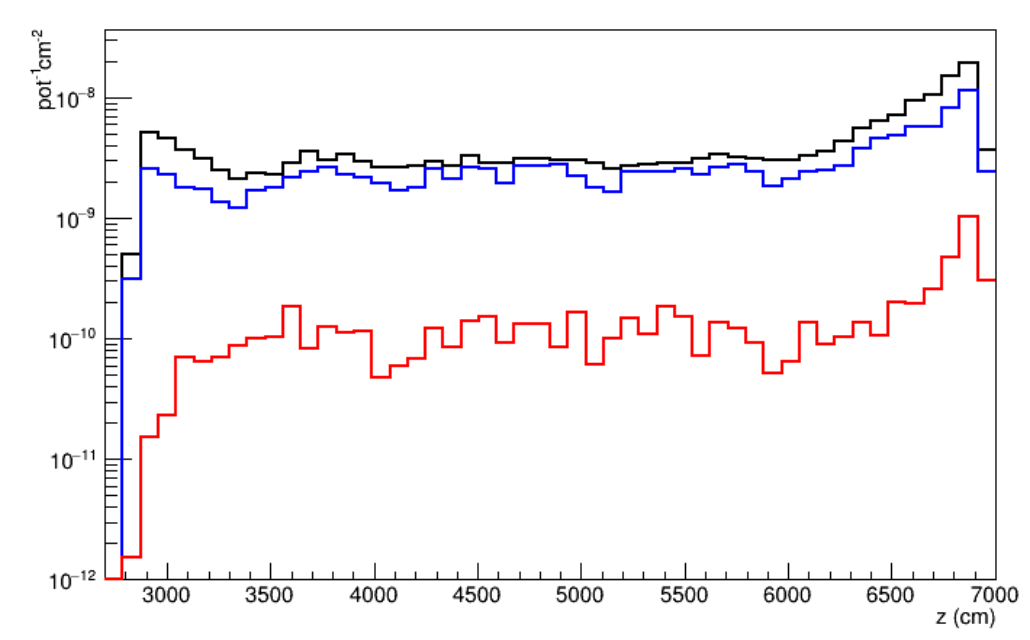}%
 \includegraphics[width=0.5\linewidth]{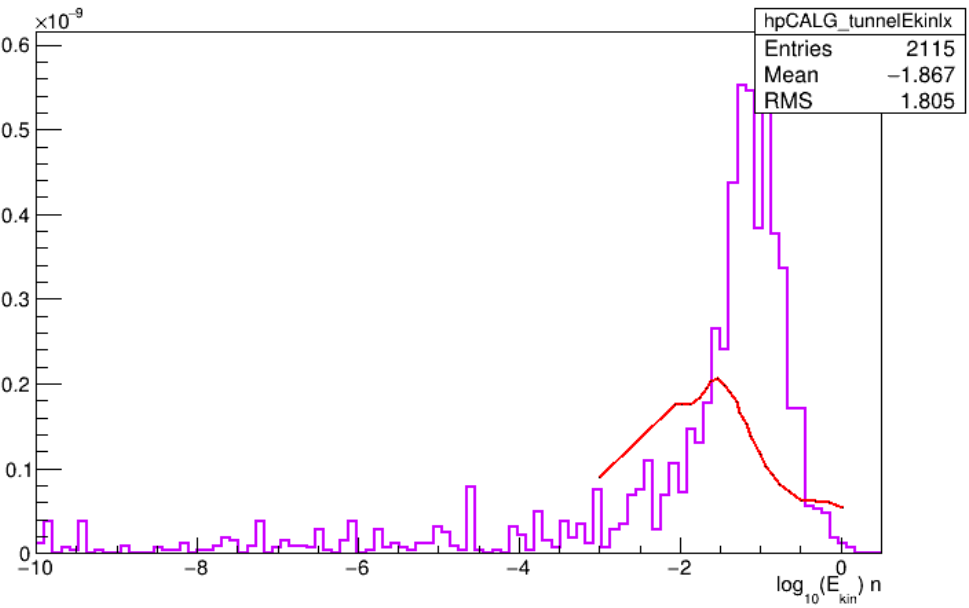}
  \caption[]{Left: FLUKA estimate of neutrons/POT/cm$^2$ as a function of the longitudinal coordinate along the tagger. The black line represents the inner surface of the calorimeter, the blue one the surface after the iron absorbers - 11~cm - 
  and the red one the surface after additional 30~cm of BPE i.e. the region where the SiPM will operate. Right: in Magenta the FLUKA estimate of the kinetic energy spectrum of neutrons (GeV) reaching the SiPM in the - 25~m long - middle section of the tagger. In red the damage function for Silicon.}
  \label{fig:tlr5_fluka_neutronsT}
\end{figure}

The SiPM are protected by a shielding 
of Borated polyethylene (BPE, 5\% Boron concentration) with a thickness of 30~cm.
%as shown in Fig.~\ref{fig:neutronS}, left.
In Fig.~\ref{fig:tlr5_fluka_neutronsT} we show the distribution of neutrons (neutrons/POT/cm$^2$) as a function of the longitudinal coordinate along the tunnel ($z$), at the inner surface of the tagger (black), at the surface between the iron and the BPE (blue) and at the outer surface of
the BPE (red). The neutron reduction induced by adding this layer of material amounts to 
a factor of $\sim$~18, averaging over the expected energy spectrum and it settles at about $7\times 10^{-11}$~n/POT/cm$^2$ in the middle region of the tagger
($3.5\times $~10$^9$~n/cm$^2$ for 5~$\times$~10$^{19}$~POT). This value is about a factor 10 lower than what was estimated with the single dipole beamline. The bulk of neutrons reaching the SiPM have kinetic energies O(10-100)~MeV/c$^2$ (Fig.~\ref{fig:tlr5_fluka_neutronsT}, right).

At the time of writing the latest TLR6 version of the beamline is also being implemented both in GEANT4 and FLUKA\footnote{Our GEANT4 software (Sec.~\ref{sec:simupackages}) implements a feature for semi-automatically producing the geometry input card for FLUKA.}.

\subsection{Multi-Momentum Beamline}
\label{sec:multimom}
%Elisabetta 

Most of the ENUBET studies have been performed considering a beamline with a central momentum of 8.5~GeV/c. This choice is optimal for $\pi^{+}$/e$^{+}$ separation and to cover the energy range of interest for DUNE and HyperK \cite{enubet_proposal}. 
On the other hand, the bulk of the produced neutrinos are in the multi-GeV range ("DUNE optimized beamline") and we aim at achieve a design flexible enough to run lower momentum secondaries, enriching the low-energy region of the ENUBET spectrum ("HyperK optimized beamline").

We present in this section the ongoing R$\&$D side efforts on a beamline design which employs a secondary multi momentum (4, 6, and 8.5~GeV/c). This design was conceived to widen the cross-section energy range available to ENUBET, including the region of interest of T2K/HyperK.

This beamline design is based on existing CERN magnets and on the same principles as other low-energy secondary beamlines at CERN~\cite{nikos}. The conceptual layout can be summarized as follows: downstream the optimized Graphite target, a large-aperture quadrupole triplet defines the initial phase-space of the charged kaons. Large aperture dipoles and iron collimators select the particles, including the sought-for kaons, in a narrow momentum range. In this way the majority of background particles outside the nominal momentum band are not transported by the beamline. The 13.35$^{\circ}$ dipole deflection allows for a proper dump of the 400~GeV/c primary beam without contaminating the decay tunnel, and increase the momentum selection resolution. The proton dump configurations are still under evaluation. The remaining positron background is filtered by a 5~mm tungsten absorber. Finally, another quadrupole triplet performs the final focusing of the beam so that it enters the decay tunnel with a small divergence, and the emitted positrons produced by the kaon decay have similar angles. The overall maximum angular acceptance of this preliminary design is $\pm 20$~mrad in both planes. 

\begin{figure}[htb!]
\centering
  \includegraphics[width=\linewidth]{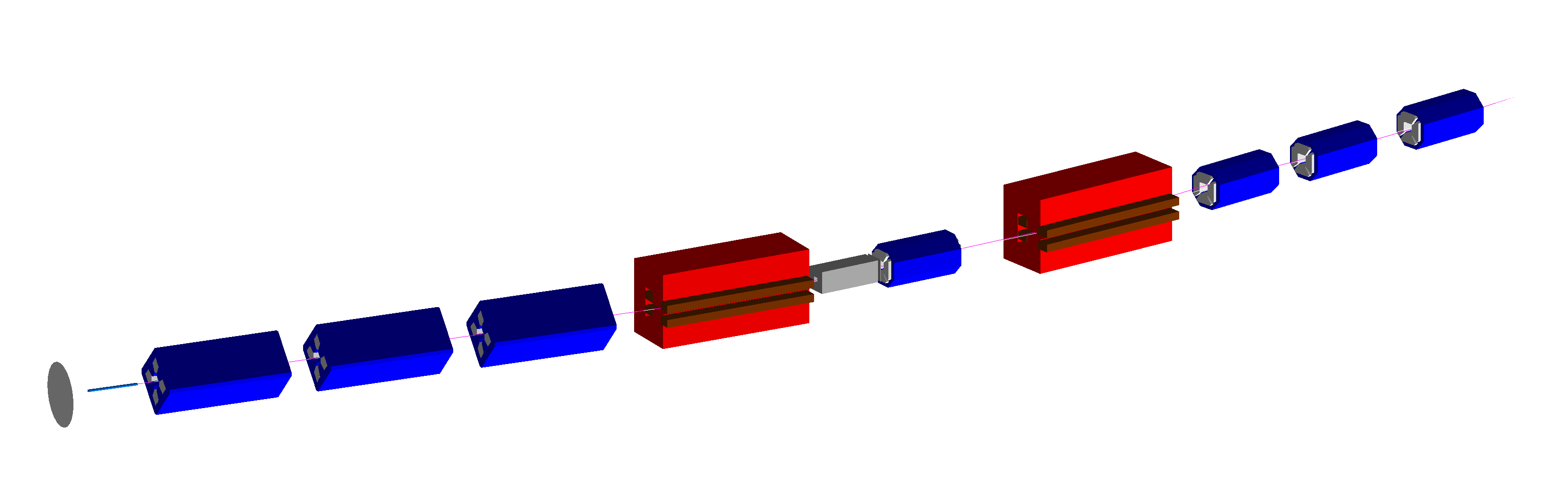}
  \caption[]{Multi Momentum Beamline layout in G4Beamline. 
  Starting with a 80~cm long graphite target, we placed a Q200 quadrupole triplet. The bending section is composed of two large deflection bends, a collimator of $9\times9$ cm$^{2}$ aperture followed by a 5~mm thin tungsten absorber and a QFL focusing magnet. Before the decay tunnel, we find a final QFL quadrupole triplet. }
  \label{fig:bmb}
\end{figure}

The optics optimization for ENUBET's multi-momentum  beamline is performed at first-order using TRANSPORT~\cite{transport} and G4Beamline~\cite{g4beamline}. The results will be then validated with MAD-X/PTC-TRACK~\cite{madx} to estimate higher-order effects, while the background reduction studies will be performed using FLUKA.  
Since positrons dominate the production of secondary particles especially in the lower energies ($<$ 6~GeV/c), we are considering placing the whole beamline tilted with respect to the target. Placing the beamline at a horizontal angle would drop the emission of positrons from the target with a greater factor than the kaons. This effect is important at the desired momenta and allows for further background suppression. However, this configuration requires special care in the design of the dump of non-interacting protons, which is currently under evaluation.

\subsection{Beam background reduction studies}
\label{sec:bbred}

The search for optimal configurations with the genetic algorithm originally developed for the horn has been extended to the full beamline.
The optimization is run with a large number of parallel jobs running on
the Lyon IN2P3 computing farm. In particular the last collimator before the tagger has a conical aperture whose radiuses have been optimized together with the apertures of downstream collimators. The procedure has proven to work and is being fine tuned. A pre-selection of tracks at the target level is being applied to make the execution speed more effective and use a larger statistic for the figure of merit. In a first scan a figure of merit based on an integral S/B ratio in the tagger was defined. 
Currently we are implementing also some information on the shape
of backgrounds to avoid picking up configurations in which the S/B
is good in terms of overall normalization but in which background
is more difficult to be separated from background due to a similar
distribution in energy and in the position along the tagger.

%The last collimator plays an important role in
%determining the background in the tagger. It is
%currently modeled as a box of tungsten with a conical
%aperture determined by the radius at entrance and exit ($r_1$, $r_2$). 

%A systematic optimization of the last collimator geometry has been performed. A standalone GEANT4 simulation has been used simulating only the relevant elements for the sake of CPU-time reduction while using as input the tracks from the full GEANT4 simulation at the collimator upstream plane. We have studied the ratio between the yields of positive pions exiting from the collimator ($y_e$) and that of positive pions impinging on the inner surface of the tagger ($y_b$) as a function of $r_1$ and $r_2$.
%As it is visible in Fig.~\ref{fig:G4CollOpt} the fraction of pions hitting the instrumentation has a minimum at about (55, 140)~mm. This corresponds to a slightly more open and wide cone with respect to the first tentative optimization - (45, 80)~mm. The improvement at the lepton selection level is currently under test with the full  simulation/reconstruction/analysis chain.
%\begin{figure}[htb!]
%\centering
%  %\includegraphics[width=0.9\linewidth]{figs/G4_CollOpt.png}
%    \includegraphics[width=\linewidth]{figs/G4_CollOpt_1.png}
%  \caption[]{Scan of the apertures of the last collimator. Left: ratio of the number of $\pi^+$ entering the collimator and the number of $\pi^+$ impinging on the tagger (in \%, $100 \times y_e/y_b(r_1,r_2)$ using the notation given in the text). Right: geometry of the collimator.}
%  \label{fig:G4CollOpt}
%\end{figure}

\subsection{Instrumented decay tunnel and PID}

In the following Sec.~\ref{subsec:lateral}
and we describe the final, published, analysis of test beam data 
taken during fall 2018 at CERN. With respect to last year a more solid understanding of saturation effects has been obtained which is providing essential guidance in the choice of the final SiPM. It should be noted that, with respect to that design several aspects have evolved (the layout of WLS fibers, the scintillator thickness), as we will explain when discussing the delivery of the final demonstrator (Sec.\ref{sec:demonstrator}).

\subsection{Lateral readout calorimeter}
\label{subsec:lateral}
% Michele and Claudia B.

\begin{figure}[htb!]
\centering
  \centering
  \includegraphics[width=0.7\linewidth]{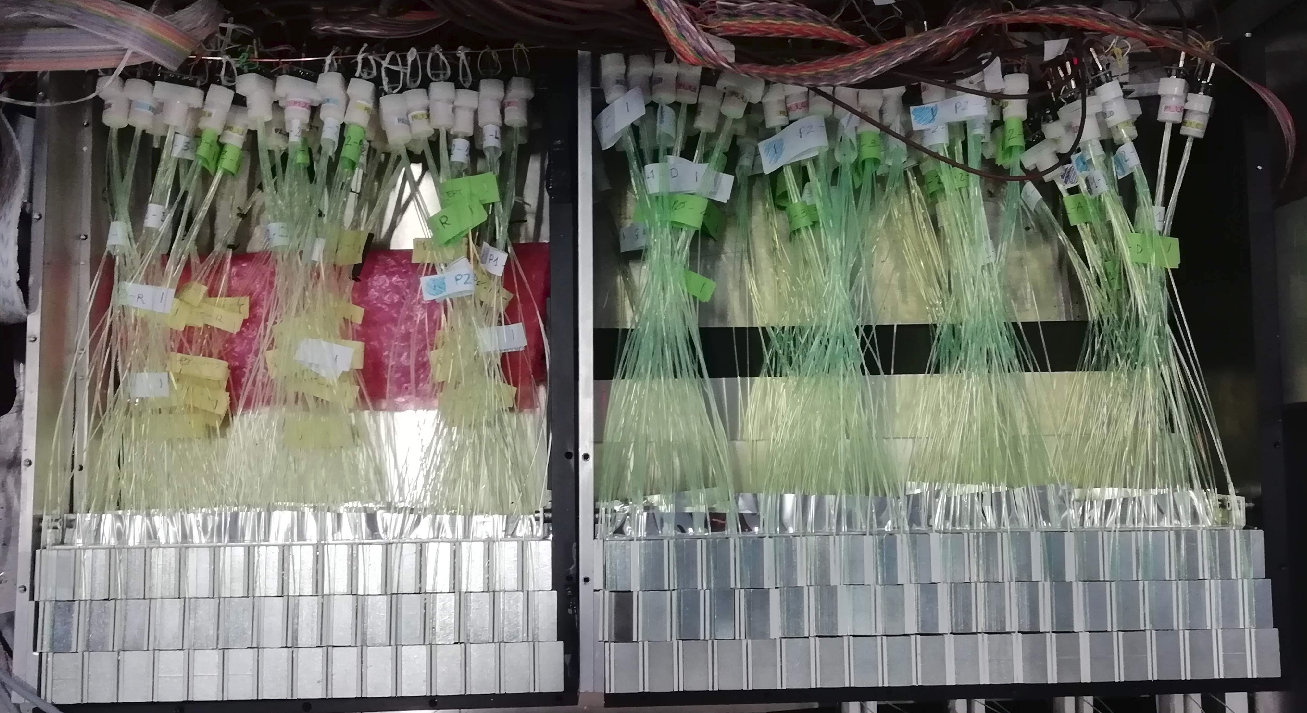}%
   \caption[]{ The full calorimeter tested at CERN PS-T9 beamline in September 2018.}
  \label{fig:prototype_calo}
\end{figure}

Positron tagging and $e^+$/$\pi^+$ separation require a longitudinal segmented calorimeter with an energy resolution of $<25\%/\sqrt{E\mbox{(GeV)}}$\cite{Pupilli:neutel17} in the energy range of interest (1-3~GeV).
The technology of choice is a sampling calorimeter, whose building element is the LCM (Lateral Compact Module), assembled from five $3\times3\times1.5~\mbox{cm}^3$ steel tiles interleaved with five $3\times3\times0.5~\mbox{cm}^3$ plastic scintillator tiles (Eljen EJ-204~\cite{eljen}).
Two WLS fibers of 1~mm diameter are inserted and glued into two grooves placed at two opposite sides of each scintillator tiles. Each LCM has hence ten WLS fibers that are then coupled with a Silicon PhotoMultiplier (SiPM). 
The SiPMs are produced by Advansid~\cite{advansid}, have an active area of $4\times4~\mbox{mm}^2$ and a cell size of 40~$\mu$m.
The prototype tested in fall 2018 was assembled from 84~LCM arranged in 7 planes on a $3\times4$ matrix (Fig.~\ref{fig:prototype_calo}). The first three longitudial planes were instrumented with Kuraray Y11 WLS fibers and the remaining four with Saint Gobain BCF92 WLS fibers. The total length of the calorimeter covers $30.1~X_0$ and $3.15~\lambda_0$.
The transverse dimension of $12\times9~\mbox{cm}^2$ allows for full containment of electromagnetic showers up to 5 GeV.

\begin{figure}[htb!]
\centering
  \centering
  \includegraphics[width=0.7\linewidth]{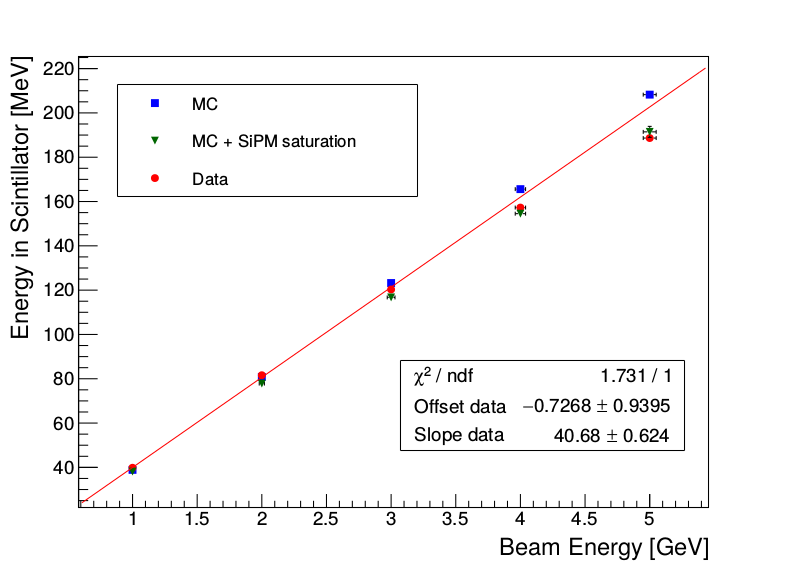}%
   \caption[]{Energy reconstructed in the calorimeter versus beam energy for a 100 mrad run. Testbeam data
(red dots) are compared with Monte Carlo simulation including (green triangles) and not including (blue
squares) the SiPM saturation. The horizontal errors correspond to the momentum bite of the beam. The vertical error bars (not visible in the plot, since of $\mathcal{O}(0.1\%)$ and covered by the marker) in ``MC'' and ``Data''
are given by the standard error of the mean of the gaussian fit performed on the electron peaks. The vertical
error bars in ``MC + SiPM saturation'' are given by the uncertainty on the number of pixels available to the
light collection (the lowest estimate is $\sim4580$, while the highest estimate is $\sim5400$).}
  \label{fig:linearity}
\end{figure}

The extraction of the fibers from the lower planes of the calorimeter is made possible by shifting the 3 longitudinal planes by 3.5~mm with respect to each other and by extending the groove for the WLS fibers from the scintillator to the steel tiles. 
The calorimeter was put under test in September 2018 at CERN PS-T9 beamline, with a mixed beam of $e^-$, $\mu^-$ and $\pi^-$, with beam momementum ranging from 1 to 5~GeV. The response to electromagnetic showers was tested by tilting the prototype at 0, 50, 100 and 200~mrad w.r.t. the beam axis, in order to simulate the impact angle of the positrons inside the decay tunnel.
The SiPM voltage bias was $V_{bias}=31$~V, (4~V Over Voltage - OV).
The results indicate that the performance of the calorimeter is the same for front and inclined runs.
Figure~\ref{fig:linearity} shows the reconstructed energy in the scintillator for data and MC in a 100~mrad run. Above 3~GeV it is possible to observe a deviation from linearity, $\sim3\%$ ($\sim7\%$) at 4~GeV (5~GeV).
This effect is caused by the saturation of the SiPMs, which is enhanced by the rather large cross-talk at $V_{bias}=31$~V, $P_{x-talk}\simeq44\%$, as measured in a dedicated setup at the INFN Bologna labs.
To account for these saturation effects, the number of expected fired cells in the SiPM in the MC has been corrected with the following equation~\cite{sipm_saturation}:
\begin{equation}
\label{eq:sipm_sat}
    N_{\rm{fired}}\simeq N_{\rm{max}}\left(1-e^{-N_{\rm{seed}}/N_{\rm{max}}}\right),
\end{equation}
in which the expected average number of p.e. in an LCM ($N_{\rm{seed}}$) is smeared for Poisson fluctuations ($N_{pe}$) and increased by the aforementioned cross-talk effects ($N_{\rm{seed}}~\equiv(1~+~P_{x-talk})\cdot N_{pe}$). The SiPMs equipping the calorimeter have 9340 cells, but the fibers are put in mechanical contact and illuminate fewer of the available cells, namely $N_{\rm{max}}\simeq 5000 < 9340$. Despite the uncertainty on $N_{\rm{max}}$, Eq.~\ref{eq:sipm_sat} manages to account for non linearities in the detector, as shown in Fig.~\ref{fig:linearity}.

\begin{figure}[htb!]
\centering
  \centering
  \includegraphics[width=0.7\linewidth]{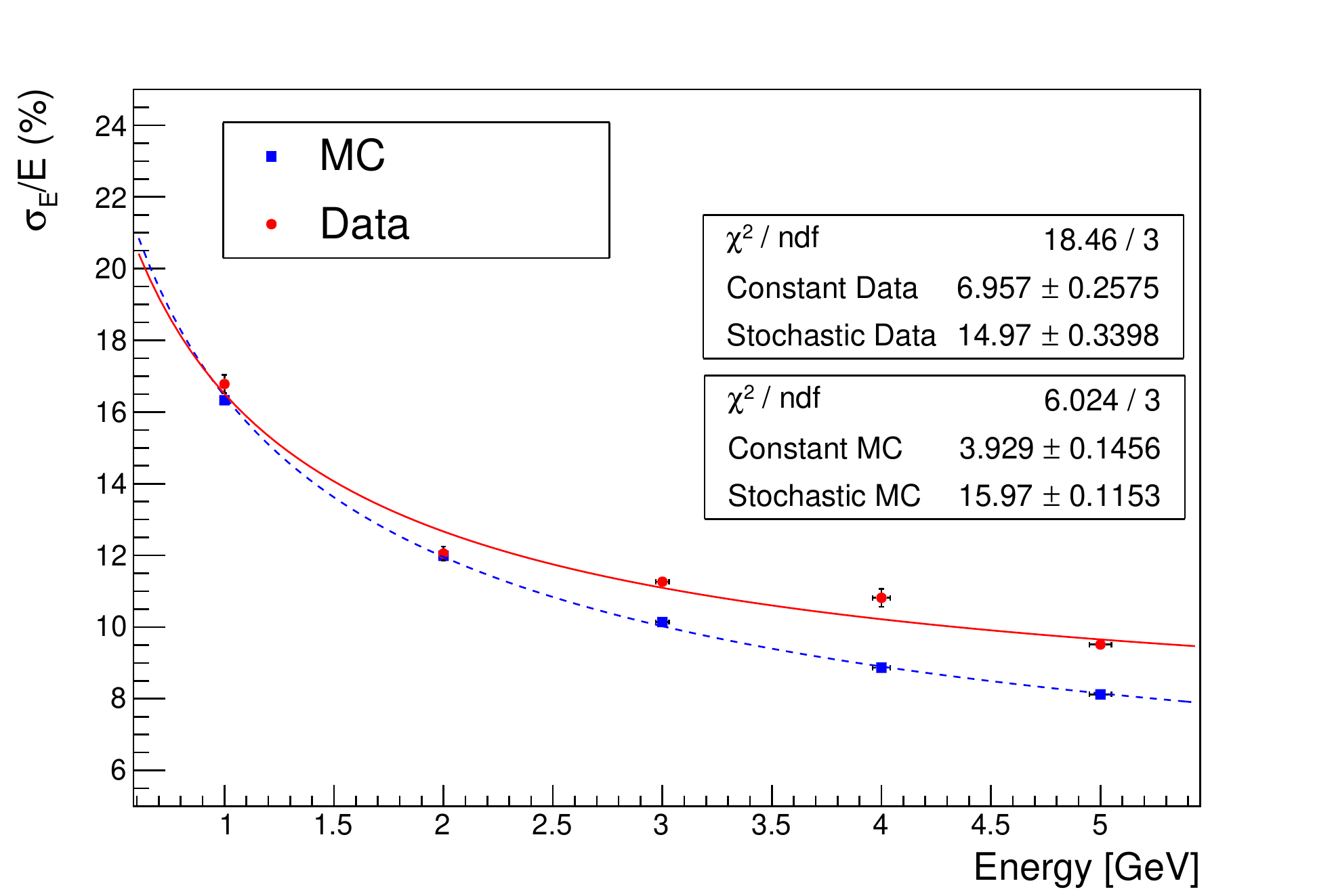}%
   \caption[]{Energy resolution versus beam energy for particles impinging on the front face (0~mrad run) for
data (red dots) and simulation (blue squares). The fit parameters for data and simulation (MC) are shown in
the top and bottom insets.}
  \label{fig:resolution}
\end{figure}

\begin{figure}[!htb]
\centering
\includegraphics[width=0.7\columnwidth]{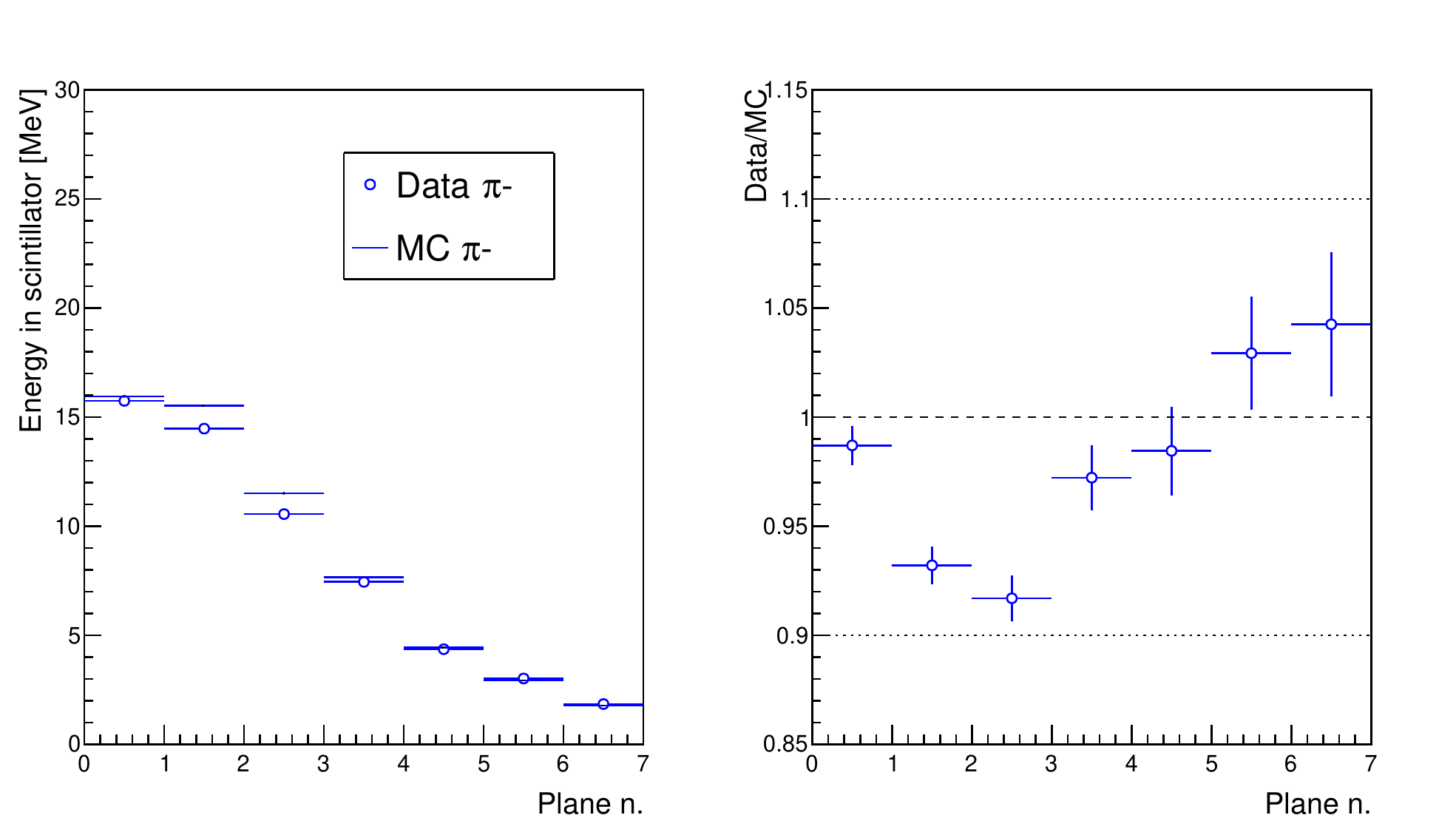}
\caption[]{\label{fig:profile_pi}
Left: average energy deposited in the scintillator as a function of calorimeter planes for 3~GeV pions. Each LCM corresponds to 0.45~$\lambda_0$.
Right: energy ratio between data and MC.}
\end{figure}

The energy resolution for a 0~mrad run is shown in Fig~\ref{fig:resolution}. The points are fitted to $\sigma_E/E = S/\sqrt{E(\rm{GeV})}\oplus C$, in which $S$ is the sampling (stochastic) term and $C$ is the constant one.
At 1~GeV the energy resolution is 17\%. At high energy the simulation deviates from the data, mostly due to saturation effects. In particular, by the time of the publication of these results, it was understood that the impact point of the particle affects its contribution to saturation, which is higher for particles impinging at the center of the module and lower for events near the edge of the tile, as the energy is shared between two adjacent LCMs and two different SiPMs.
Since the separation of signal and background in ENUBET is performed by analyzing the energy deposit pattern in the LCMs, the simulation of this pattern was tested with a $\pi^-$ beam in the same energy range as for the electrons.  
The mean $\pi^-$ energy deposited in each plane of the calorimeter in the data was evaluated and compared to the simulation, showing a rather 
good agreement: the discrepancies do not exceed 10\% and are comparable to the uncertainty due to low-energy hadronic shower simulation 
(Fig.\ref{fig:profile_pi}).
These results were published on Journal of Instrumentation in August 2020 ~\cite{TB_sept18}.

Following the publication, an optical simulation is currently being implemented in the simulation of the calorimeter, in order to replicate the aforementioned SiPM saturation effect associated with the initial impact point of the particle on the calorimeter. The first steps of this simulation, which recreated the signal collected by a single plastic scintillator tile read with two WLS fibers coupled to a single SiPM, have been supported by a set of dedicated tests performed in INFN Bologna; data and simulation showed very good agreement.
Moreover, in collaboration with Fondazione Bruno Kessler (FBK), we are currently identifying and testing different sensors as to find the more suitable one.

\subsection{Photon veto}
\label{sec:t0}
% Michele and Gabriele
% in progress....
In addition to the calorimeter, ENUBET is instrumented with a photon veto detector ($t_0$--\textit{layer}) to provide both photon identification capabilities and precise timing of particles in the instrumented decay tunnel. In order to achieve the main goal, the requirements are i) a photon identification efficiency at 99\%, and ii) a time resolution of  $\sim$1 ns. The $t_0$-layer is composed of doublets of plastic scintillator tiles (3$\times$3$\times$0.5 cm$^3$) mounted below the LCMs and positioned every 7 cm. In this way positrons from kaon decays in the ENUBET working condition cross five doublets on average.

Data taken during beam-test at CERN in 2018 have been fully exploited during 2020 and the results have been published in~\cite{TB_sept18} and can be summarized as follows.
The light yield for a single mip crossing a $t_0$ tile allow for a collection of 25 p.e. at MPV with a time resolution estimated to be $\sim$400 ps, well within the requirements of the project.

\begin{figure}[htb!]
\centering
  \centering
  \includegraphics[width=0.42\linewidth]{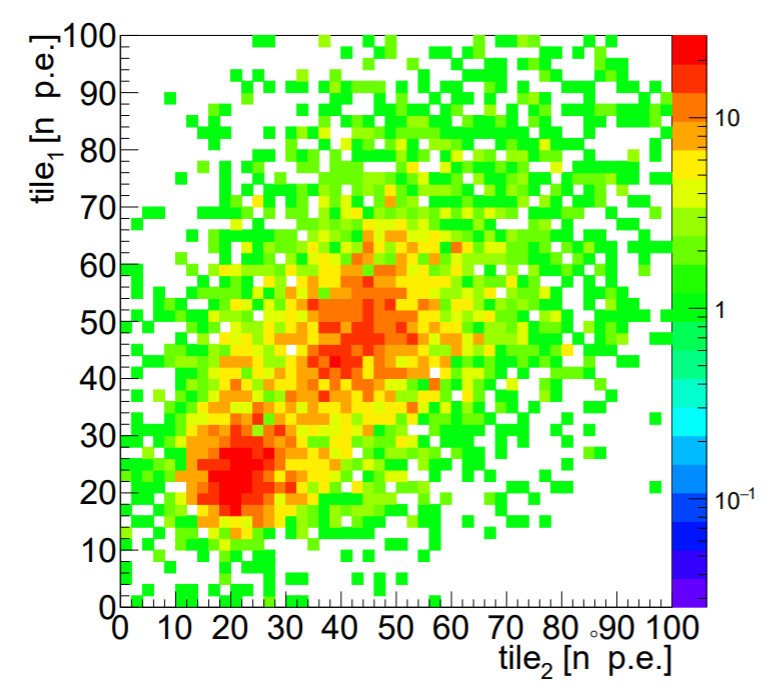}%
 \includegraphics[width=0.58\linewidth]{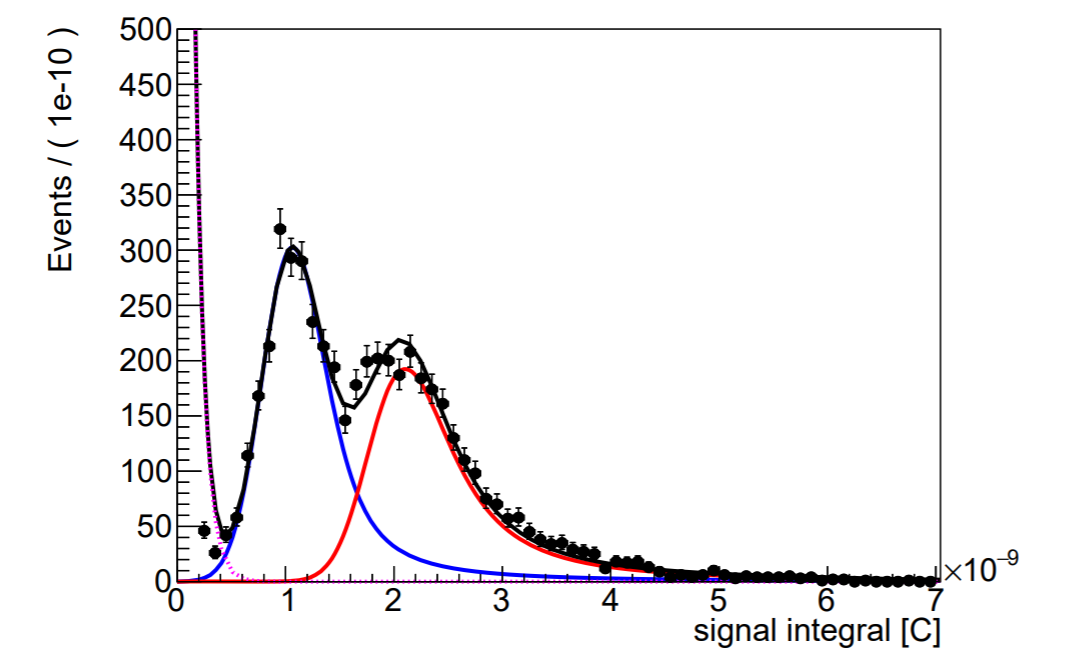}
  \caption[]{(Left: The number of p.e. collected in one tile versus the number of p.e. collected in another tile. Right: Fit on data by using a composite model: black dots represent the data sample, the black line is the composite model. The purple, blue and read lines are respectively dark current contribution, the signal and the background PDF )}
  \label{fig:photon-veto}
\end{figure}

Moreover, the capability of the $t_0$-layer to separate one mip from two
mips was investigated as shown in fig. \ref{fig:photon-veto}
A single $t_0$-tile is capable of selecting the signal due to one mip with an efficiency of 87\% and a background (2-mip like) rejection efficiency of 89\% are obtained with a corresponding value for the purity of about 95\%.

% Valerio, Fabio I.

%\subsubsection{Ongoing studies for the selection of the final SiPM}
%Two different SiPM models, a $4\times 4\;\mathrm{mm^2}$ RGB type (ADVANSID, 40~$\mu$m cells) and a $6 \times 6\;\mathrm{mm^2}$ NUV-High Density 3 type (see Fig.~\ref{fig:sipm_study}), have been coupled to a scintillator tile through WLS fibers in order to be tested in a dark chamber with cosmic ray at INFN-LNL. The latter, also employed by the CTA collaboration~\cite{CTAsipm}, include new production technologies from Fondazione Bruno Kessler (FBK) while the former are the ones used in our previous tests. The amplitudes of the signals induced by cosmic rays have been recorded for different SiPM supply voltages. The peak position values of the resulting Landau distributions are plotted against the supply voltage in Fig.~\ref{fig:sipm_study}.
%\begin{figure}[htb!]
%\centering
%  \includegraphics[width=0.9\linewidth]{figs/sipm_study.pdf}
%  \caption[]{Top Left: RGB SiPM. Bottom Left: NUV SiPM. Right: Landau peak position against supply voltage for RGB SiPM (red points) and NUV SiPM (violet points).}
%  \label{fig:sipm_study}
%\end{figure}
%Other comparisons are being performed with fast laser light inputs. Recently new generation RGB SiPM provided by FBK are also in the evaluation progress. 

\subsubsection{Waveform digitizer design}
% by Erik (+ Valerio)
% Valerio: non metto figure per non appesantire il report, ma se servisse rendering e foto della scheda di cui si parla sono qui https://agenda.infn.it/event/23872/contributions/121330/attachments/74773/95300/digi_sep2020.pdf (slide 5 e 6) 
A prototype digital board to handle the signals from a 4 channel ADC board (250~MS/s, 14~bit) and to provide a USB interface with the PC has been designed around an Altera 5CGXBC3B6F23I7N and manufactured. It is undergoing extensive tests in order to validate the different parts.
%\begin{itemize}
%    % I/O towards the PC 
%    \item Two connectors allow the use of either USB2 or USB3 interface for data I/O operations towards the PC. The board has been fully validated with the USB2 interface for a data rate up to 8~MByte/s and is undergoing tests with the an external USB3 interface board (UMFT601A from FTDI) where we expect to increase by at least a factor 4 this performance.
    % Data communications: the Gigabit transceivers
%    \item For the data communications, six Gigabit (3.125 Gbit/s) transceivers have been implemented: 2 on coaxial cables pairs, 2 transceivers on optical fiber (with Finisar FTFL8524 optical modules) and 2 on the HSMC connector. 
%    A complete test procedure has been deployed and all the links are performing in the correct way.
    % Data communications: ancillary links
%    \item Two RJ-45 connectors with 4 LVDS lines each have been implemented as auxiliary connections to interface with other DAQ parts, e.g. during in a test beam.
    %ADC interface
%    \item All the four channels of the board with ADS4249 ADC have been tested and are working correctly.
%\end{itemize}
In the readout of the final ENUBET prototype this custom electronics will be complemented and cross-checked with commercial digitizers and a set of 64-ch boards by CAEN (A5202) based on the WeeROC CitiROC-1A ASICs which will only readout the signals' amplitude and time.
\begin{figure}[ht!]
\centering
\includegraphics[width=0.47\textwidth]{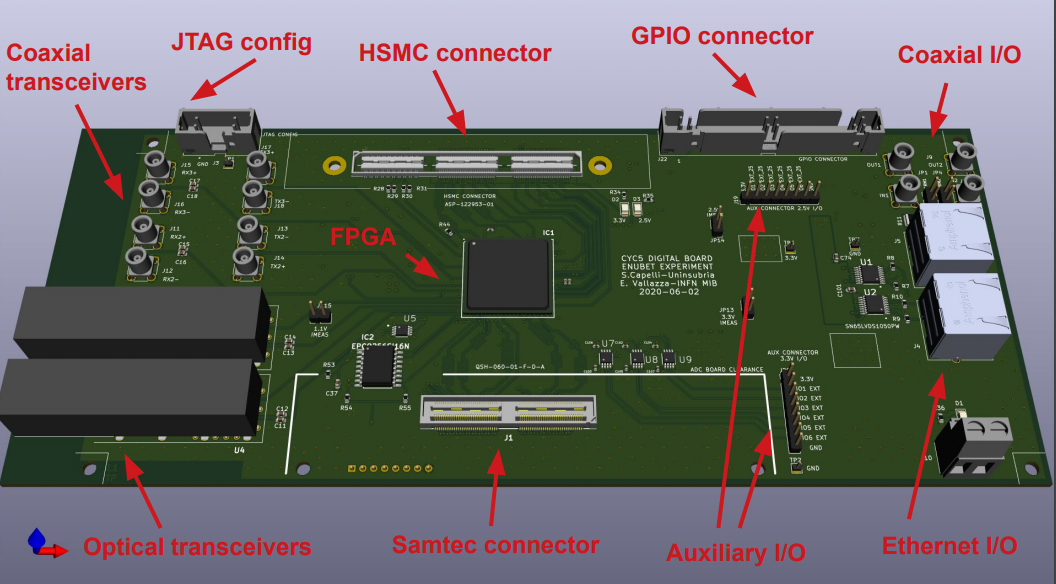}%~
\includegraphics[width=0.5\textwidth]{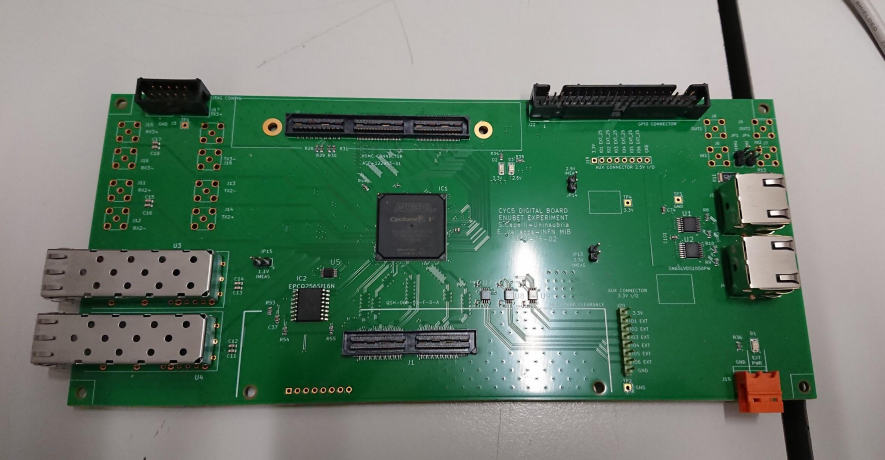}
\caption{Layout and picture of the prototype digital board.}
\label{fig:board}
\end{figure}

\subsection{Simulation}
\label{sec:simulation}
% Fabio, Claudia D.

\subsection{Simulation packages}
\label{sec:simupackages}
\begin{figure}[ht]
\centering
\includegraphics[scale=0.3]{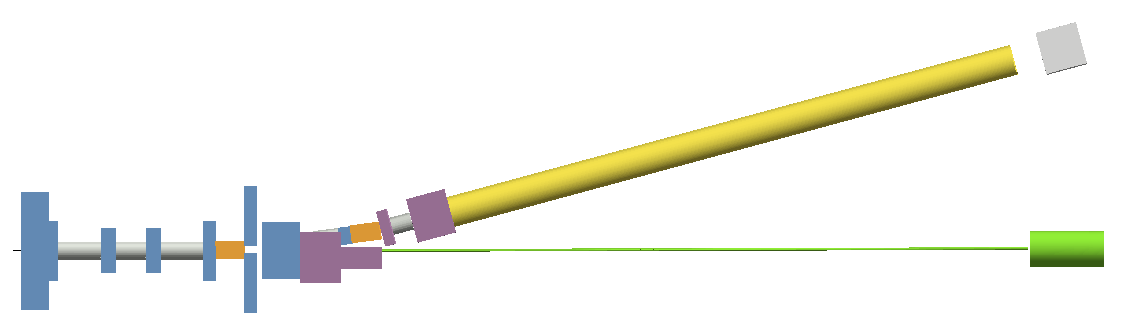}
\caption{GEANT4 simulation of the TLR5 double dipole beamline.}
\label{G4_tlr5v4}
\end{figure}

% Claudia D.
The ENUBET facility has been described in G4Beamline, GEANT4 and FLUKA.
The beam components are implemented in G4Beamline, that fully simulates particle transport and interactions (described in Sec.~\ref{sec:tl}), while FLUKA is used for the fine tuning of collimators and shielding (Sec.~\ref{sec:bbred}) and for doses assessment (Sec.~\ref{sec:irradiation}).
The GEANT4 simulation (G4TL) is redundant, since it contains the same setup implemented in the other software packages, allowing the cross-check of the results. It reproduces the entire facility: the elements of the transfer line, the tagger and the neutrino detector, with the exception of the tagger geometry. 
The simulation has been validated against the G4Beamline results for one of the previous versions of the transfer line (TLR2, single dipole version). This has been achieved by comparing the spectra and the entering angle of particles at the tagger entrance obtained with G4Beamline.
Recently also the double-dipole transfer line (TLR5) has been modeled in GEANT4 (Fig.~\ref{G4_tlr5v4}), and the latest TLR6 version is also being implemented.

One of the strengths of this simulation is the possibility of controlling all the parameters of the setup, from the collimator dimensions and positions to the magnet apertures and fields, using a control card. Moreover, GEANT4 gives access to information on particle decays and their histories. 

\begin{figure}[ht]
\centering
\includegraphics[width=\linewidth]{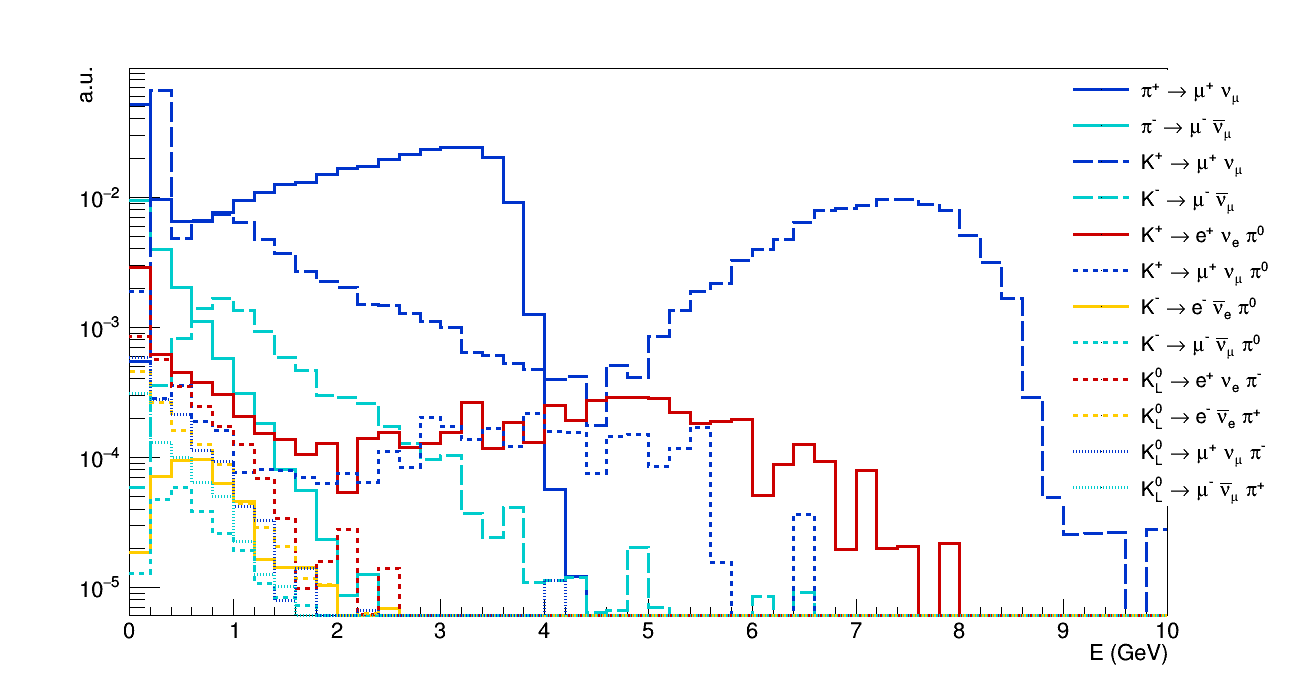}
\caption{Neutrino interaction rates at far detector, breakdown by decay mode. It must be noted that here the implementation of the shielding was not finalized. Hence these fluxes can be considered an upper limit for the neutrino components from the target region. The rates are approximated by weighting the fluxes with energy. Neutrino generators will be used in the future.}
\label{G4_rates}
\end{figure}

Fig.~\ref{G4_rates} shows the output of the TLR5 simulation. The decay information about the particles simulated throughout the beamline, such as the position of the decay and the momenta of the secondaries, allows the reconstruction of neutrino tracks. It is then possible to estimate the flux and the rate of neutrinos at the far detector, and divide it according to the neutrino flavour or to the decay mode.

The instrumentation of the ENUBET facility has been implemented in detail in a GEANT4 standalone package (G4TAG) that reproduces the detectors in the decay tunnel (the photon veto and the calorimeter modules) for the monitoring of large angle leptons from kaon decays (Sec.~\ref{sec:positronsel},  \ref{sec:mutagger}) and the stations after the hadron dump for the monitoring of muons from pion decays (Sec.~\ref{sec:muchambers}).
The simulation includes the propagation and decay up to the hadron dump of particles provided by the G4Beamline simulation of the transfer line (Sec.~\ref{sec:tl}) at the tunnel entrance.
The response of the detectors in the tunnel is simulated at hit level, not considering the scintillation process and light propagation.
The results concerning the lepton monitoring reported in the following are based on the assumption of a 2 s slow extraction with 4.5$\times$10$^{13}$~POT per spill.

Since the G4TAG simulation package can only account for decays in the tunnel, we implemented a shared structure based on a ROOT TTree: by merging the information from the transfer line (G4TL), we can trace all the decays occurring along the entire beamline.  
%We implemented a shared structure with the GEANT4 simulation of the tagger and muon monitoring instrumentation, based on a ROOT TTree.
%This simulation reproduces the detectors in the decay tunnel and the muon stations after the hadron dump. Its starting point is the beam composition at the tunnel entrance, as simulated by G4beamline.
%This means that, using the tagger simulation only, we can take into account only the decays that occur downstream of the tagger. By merging the information from the transfer line simulation, we can take into account decays occurring along the entire beamline.

The assessment of the systematic uncertainties on the neutrino flux will be performed within this GEANT4 framework, exploiting the possibility to tune all the parameters of the simulation, following the procedure described in Sec.~\ref{sec:systematics}.
%As a next step, the systematic uncertainties on the neutrino flux from the beamline parameters will be studied, using the GEANT4 capability of tuning all the parameters of the simulation, following the procedure described in Sec.\ref{sec:systematics}.

\subsection{Pile-up study with waveform simulation}
\label{sec:wfsim}
% Evgenii, Michelangelo, Valerio
In order to assess the effects of the pile-up on the overall detector performances, a software framework has been implemented to simulate the calorimeter response at a single channel level. Each visible energy deposit coming from the GEANT4 simulation of the instrumented tunnel (G4TAG) is initially converted into photons hitting the SiPM using a conversion factor of $\sim$15~photo-electrons/MeV, obtained from past test beams and cosmic rays measurements results. Secondly, the SiPM response is simulated using the GosSiP software tool (Generic framework for the Simulation of Silicon Photomultiplier~\cite{gossip}) which allows a fine control on all the sensor operating parameters. A waveform for each channel is generated and then processed by a pulse detection algorithm whose results are eventually converted back into a time series of energy deposits and returned to the event builder.

\begin{figure}[ht!]
\centering
\includegraphics[width=\textwidth]{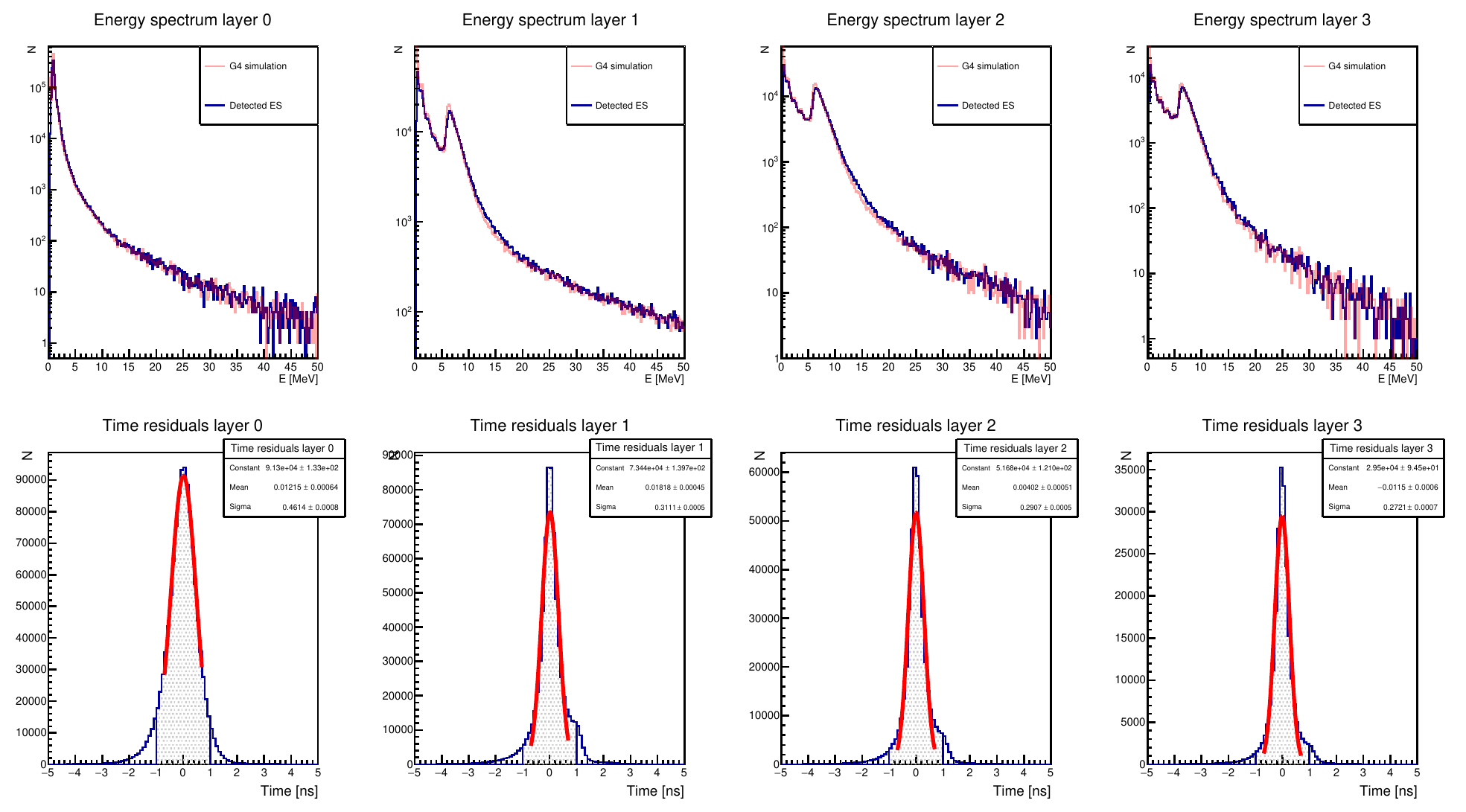}
\caption{Upper plots: energy spectra for the different calorimeter layers as provided by the deposited energy in the GEANT4 simulation (blue) and as measured from peak amplitudes after the waveform simulation of each SiPM (red). Bottom plots: residuals for the different calorimeter layers between the original time of the hits in the GEANT4 simulation and the pulse time as detected by the peak finding algorithm. The highlighted areas represent more stringent time intervals for the efficiency measurements (see Tab.~\ref{tab:hitrate}).}
\label{fig:pileup}
\end{figure}

The algorithm has been optimized aiming to have a good correlation between the measured pulse amplitude and the original visible energy deposits, a high efficiency and an accurate time resolution in the pulse detection, which are key parameters for the event building. The algorithm performance has been measured for different input beam conditions and several electronics and SiPM intrinsic parameters. In the upper part of Fig.~\ref{fig:pileup} the energy spectra measured by the detection algorithm are superimposed to the ones from the G4TAG simulation for the different calorimeter layers; in the bottom part the distribution of the residuals between the original hit time and the measured ones is shown. In Tab.~\ref{tab:hitrate} a summary of the achieved detection efficiency is shown. 
\begin{table}[ht!]
\begin{tabular}{ |p{0.25\textwidth}||p{0.15\textwidth}|p{0.15\textwidth}|p{0.35\textwidth}| }
% |p{3cm}||p{3cm}|p{3cm}|p{3cm}|
%\hline
%\multicolumn{4}{|c|}{Layer 2} \\
\hline
Transfer line and extraction scheme & Hit rate per LCM& detection efficiency & detection efficiency limited to a  $\pm$~1~ns time range\\
\hline
TLR5 slow & 1.1~MHz & 97.4\% & 93.7\%\\
TLR5 fast & 10.4~MHz & 89.7\% & 84.0\%\\
TLR6 slow & 2.2~MHz & 95.3\% & 89.6\%\\
\hline
\end{tabular}
\caption{Achieved detection efficiency as a function of the input beam transfer line and extraction scheme (the listed data refer to the 2$^{\rm{nd}}$ calorimeter layer). In the last column the efficiency defined over a more stringent time interval of $\pm$1~ns with respect to the true hit time is shown (this corresponds to the highlighted area in the bottom plots of Fig.~\ref{fig:pileup}). The slow extraction scheme assumes 4.5$\times$10$^{13}$ POT in 2~s; in the fast scenario, a factor 10 increase in the particle rate is assumed.}
\label{tab:hitrate}
\end{table}
Given the large number of channels, the simulation for the whole tagger is run exploiting the CERN HTCondor batch system and the EOS space is used for the data storage. This allowed us to increase the statistics by more than 100000 times from test files to a complete simulation of the output signal with over 800000~SiPMs.

Once fully debugged and properly integrated, the output of the waveform simulation will be finally included in the lepton reconstruction chain described in the next sections.

\subsection{Positron monitoring}\label{sec:positronsel}

The event building
is the preliminary step for the identification of leptons: it correlates in space and time hits in the tunnel instrumentation belonging to the same decay, dealing with mixing effects induced by beam halo particles and other mesons' decay modes. The algorithm setup for $K_{e3}$
monitoring is oriented towards a preselection of positron candidate events, through the identification of a
visible energy deposit in the LCMs of the innermost layer exceeding 28~MeV\footnote{the Landau fit of the energy released by a mip in a LCM has a most probable value of $\sim$6.5~MeV.} as ``seed'' for the event reconstruction.
LCM and $t_0$ signals correlated to the seed are then clustered taking into account their position and timing, with cuts unchanged with respect to the previous report~\cite{spsc_2020}, and the procedure is iterated over all the signals recorded.

The positron selection and its separation from the hadronic background is performed exploiting the longitudinal, transverse and radial segmentation of the calorimeter, while the suppression of the residual photon background is accomplished employing the $t_0$-layer.
Electromagnetic showers will be more localized than the ones initiated by hadrons from the beam halo or from other kaon decay modes.
Muons and non-interacting hadrons, already suppressed by the request of at least one energy deposition (the event seed) largely exceeding the mip one in the event building process, can be efficiently discarded by their single track topology.

A set of variables describing the energy deposition pattern in the calorimeter is implemented in a multivariate analysis, based on a Multilayer perceptron Neural Network (NN) provided by the TMVA toolkit~\cite{TMVA}. 
Unlike the previous version of the positron selection described in the latest report, the $e^+/\gamma$ separation
is not accomplished anymore with a cut based analysis of the signals in the $t_0$-layer. Indeed, the energy depositions on the four most upstream tiles of the reconstructed event are used as additional variables of the NN.
The Neural Network training is performed defining a signal sample composed by $K_{e3}$ positrons. The background sample, that includes hadrons, muons and photons both from the beam halo and mesons' decays, has been enriched with electrons and positrons originated in the transfer line, in order to exploit differences in angular and longitudinal distributions with respect to signal positrons for their rejection.

%Fig.~\ref{fig:NNvar_ana0} shows the distributions of the 19 variables used in the current implementation of the NN for different particle species entering the tunnel, together with their definition and with the output of the NN classifier.
The event classification is done according to the event seed and the distributions refer to events reconstructed by the event builder.
%\begin{figure}
%\centeringProton dri
%\includegraphics[scale=0.7]{figs/NNvar_ana0_new.pdf}
%\caption{Distribution of variables used in the NN for $K_{e3}$ positron selection, for events reconstructed by the event builder. Particle species entering the decay tunnel are identified by different colors as reported in the legend. Golden yellow represents the positron signal. From left to right and from top to bottom: (1) fraction of the total visible energy of the LCM with the largest energy deposition in the first calorimeter layer; (2) as (1), but in the second layer; (3) as (1) but in the third layer; (4) energy fraction in the first layer; (5) energy fraction in the second layer; (6) energy fraction in the third layer; (7) energy fraction in the first and second layers; (8) fraction of energy released in the LCM with the largest energy deposition and in the downstream LCM of the first layer; (9) as (8) but in the second layer; (10) as (8) but in the third layer; (11) fraction of energy of the LCMs in the same $\phi$ sector as the seed one; (12) total visible energy; (13) total number of LCMs clustered in the event; (14) longitudinal coordinate of the seed LCM; (15) $\phi$ sector of the seed. The left plot in the last row (16) shows the distribution of the NN classifier. The last four plots in that row (17, 18, 19, 20) are the new variables included in the NN to describe the energy deposition in the four most upstream tiles of the $t_0$-layer, in log scale on the $y$-axis to enhance visibility.}
%\label{fig:NNvar_ana0}
%\end{figure}

Focusing on the new variables associated to the $t_0$-layer, the bin with a null energy deposition in the tile is dominated by photons from the beam halo (in orange) and kaon decays producing a $\pi^0$ (in yellow), with a residual contribution from $K_{e3}$ events (golden yellow) in which the $\pi^0$-related part has been reconstructed by the event builder. 
The Landau-like distribution with an approximate 0.85~MeV most probable value is well compatible with minimum ionizing particles crossing the 0.5~cm thick plastic tiles of the $t_0$-layer. Finally the second small structure visible around 1.7~MeV is due to photon conversions in the upstream tiles.

%The NN variables are shown in Fig.~\ref{fig:NNvar_ana1} for a choice of the cut on the classifier ($>$0.91) enhancing the signal content. 
The suppression of photon like events is evident in the $t_0$-related variables where almost all the events with and energy deposition not compatible with a mip one are discarded. The variables encoding information on the calorimeter energy strongly suppress hadronic decays of kaons (yellow) and non-collimated pions (green), while beam induced positrons (black) are left as the main, irreducible, component of the background.

%\begin{figure}
%\centering
%\includegraphics[scale=0.7]{figs/NNvar_ana1_new.pdf}
%\caption{Distributions of the NN variables surviving a signal-enhancing cut on the NN classifier.}
%\label{fig:NNvar_ana1}
%\end{figure}

Fig.~\ref{fig:eff_vs_SN} shows the overall efficiency of the selection, including the geometrical acceptance\footnote{that amounts to $\sim$53\%}, as a function of the signal-to-noise ratio (S/N), obtained for different cuts on the NN classifier.
The previous selection, with the cut-based analysis on the $t_0$-layer (in dashed blue) is compared with the new one here presented (in solid blue).
%the larger efficiency at low purity in the latter one is due to $K_{e3}$ events in which the $\pi^0$-related part has been reconstructed by the event builder. 
The improvement is remarkable: by choosing as an example a cut on the classifier giving an efficiency of $\sim$~24\%, the S/N significantly increases, going from $\sim$1.6 to $\sim$~2.1.
% The optimal cut (0.815) is chosen by maximizing a figure of merit defined as the product of the efficiency and the purity of the selection.
% With this criterion, positrons are selected with an efficiency of 24\% (23.7\% ?) and a S/N of $\sim$1.6.
The improved design of the proton target and of the meson transfer line ensures a larger meson yield at the decay tunnel entrance (Sec.~\ref{sec:tl}), but at the same time determines an increase of the hit rate on the tagger instrumentation of a factor $\sim$2.7.
Despite the larger mixing effects in the event building phase, that spoil the NN discrimination capabilities, the improved selection still allows to monitor positrons with an efficiency of $\sim$~22\% and a S/N of $\sim$~2 (see the red line in Fig.~\ref{fig:eff_vs_SN}, left).

\begin{figure}
\centering
\includegraphics[scale=0.65]{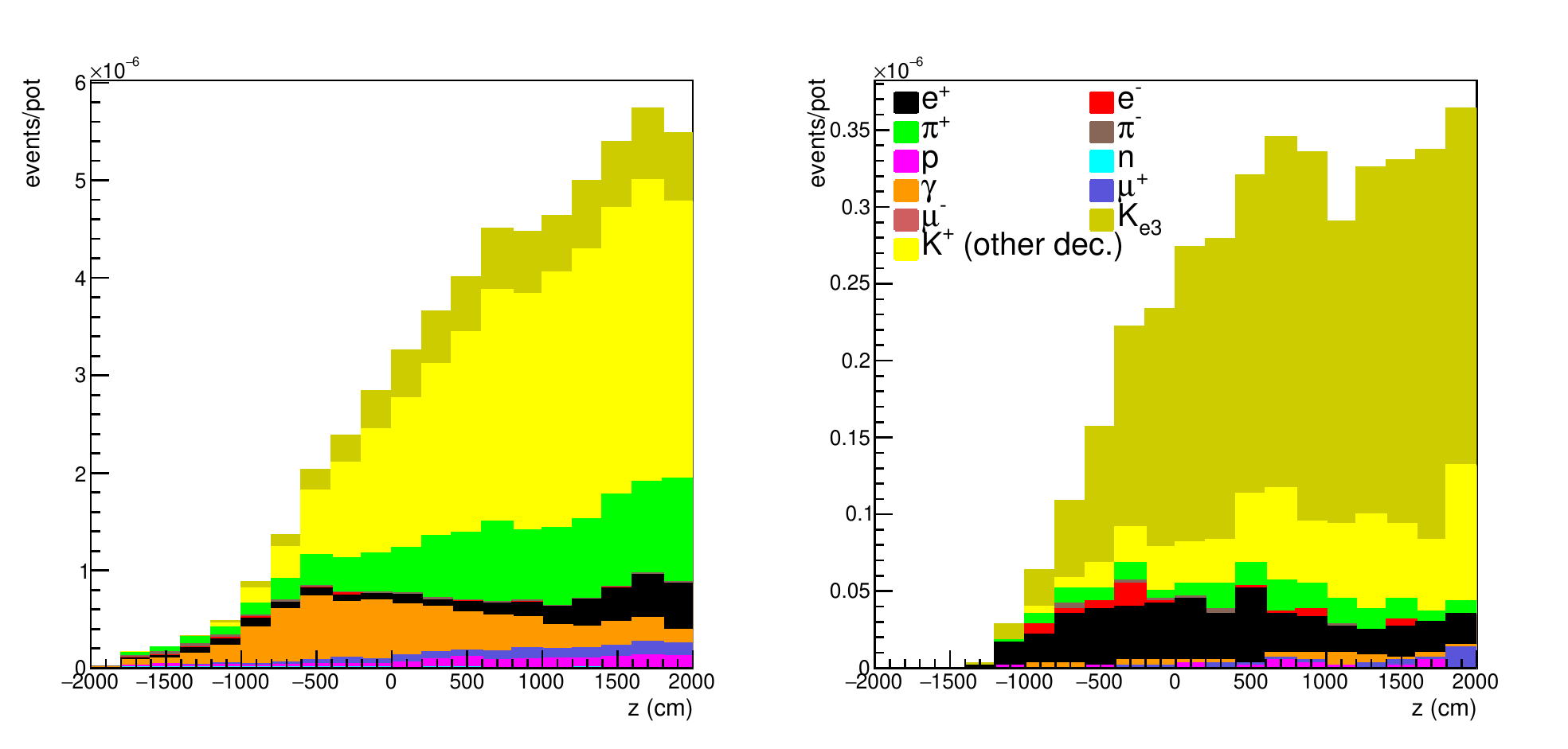}
\includegraphics[scale=0.65]{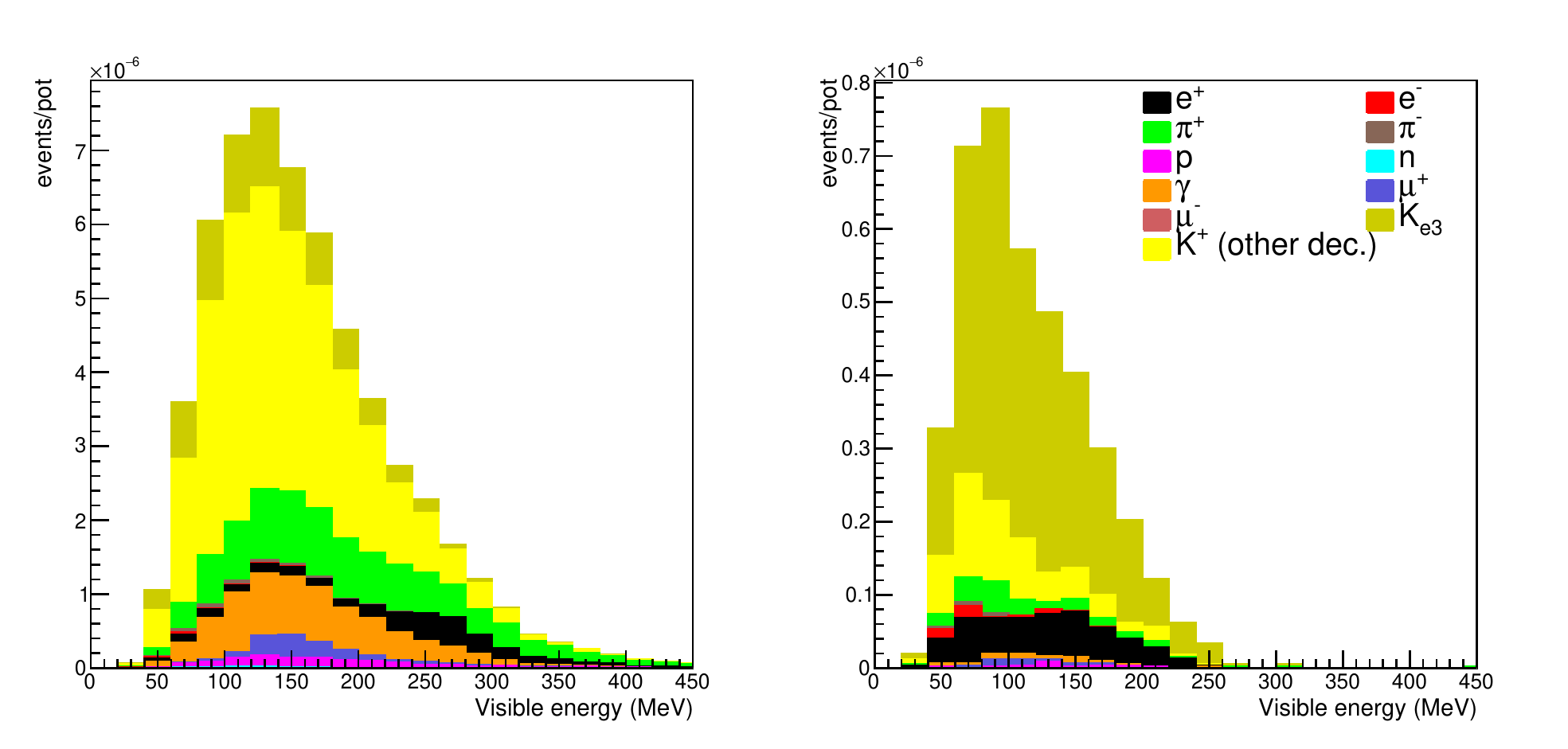}
\caption{Longitudinal position (top) and total energy (bottom) of selected events at different stages in the selection of $K_{e3}$ events: (left) after the event building; (right) after the NN discrimination.}
\label{fig:sel_z-ene}
\end{figure}

Fig.~\ref{fig:sel_z-ene} shows the distributions of two relevant observables, the longitudinal position and the total visible energy, for events at different stages of the selection: after the event building and after the NN discrimination.
% The performance of the NN in the hadron background rejection can be appreciated in particular in the reduction of the pion component (in green), while the $t_0$-layers are effective in the suppression of photons from the beam halo and of the other surviving kaon decay modes (in yellow), dominated by the ones with a $\pi^0$ production. The dominant background consists of beam induced positrons and electrons.

\subsection{Muon monitoring}
\label{sec:muon}
% Antonio, Thessaloniki
\subsubsection{Tagger}
\label{sec:mutagger}

A similar procedure to that applied for positron reconstruction and discrimination is used for muons from $K_{\mu2}$ and $K_{\mu3}$, produced at mean angles such that they cross the calorimetric tunnel. First of all, a clustering of the energy deposits in the LCMs, compatible in space and time with the expected track from a muon particle, is performed. This step, known as muon building, is unchanged since the last report~\cite{spsc_2020}: the clustering is triggered by the identification of a seed (likewise the $e^{+}$ clustering), an energy deposit in the innermost LCM layer compatible with a mip ($5 < E < 15$~MeV); the track length is required to be at least 3 LCMs in the first calorimetric layer and at least 2 LCMs in each one of the second and third layer, for a total of 7 LCMs all in different positions along the longitudinal direction. The cuts on the seed energy and track length have been fine tuned such that the product of efficiency times purity is maximized for the muon building. 

\begin{figure}[ht]
\begin{subfigure}{.5\textwidth}
  \centering
  % include first image
  \includegraphics[width=1.0\linewidth]{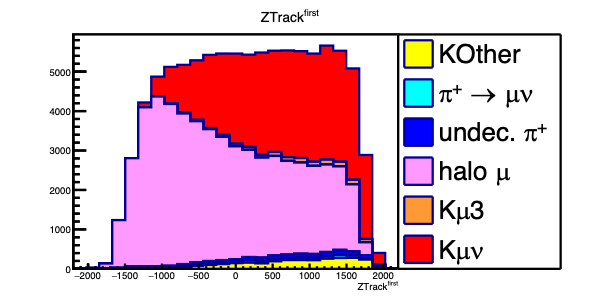}  
  \caption{}
  \label{fig:zMuBuilder_noNN}
\end{subfigure}
\begin{subfigure}{.5\textwidth}
  \centering
  % include second image
  \includegraphics[width=1.0\linewidth]{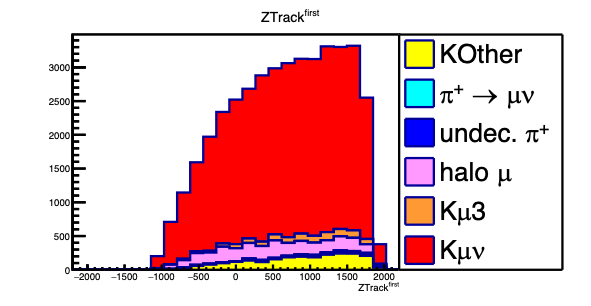} 
  \caption{}
  \label{fig:zMuBuilder_NNcut}
\end{subfigure}
\caption{(a) distribution of the impact point along the calorimeter for muons from kaon decays. Signal contribution in red, $K_{\mu2}$, and orange, $K_{\mu3}$. The main background contribution, in purple, is from halo muons. Other colors show subdominant background contributions. (b) same as in (a) but after applying the NN cut.}
\label{fig:zMuBuilder}
\end{figure}

Among the built events a signal-background discrimination is achieved by applying a NN. Different algorithms have been investigated, but like for $e^{+}$ also the $\mu^{+}$ signal identification shows best performance with the Multilayer perceptron. Concerning the background, the larger contribution comes from halo muons, while second order contributions are muons from pion decays emitted at large angles and pions either from the secondary beam or kaon decays. A set of variables has been identified to train the NN: they exploit the differences in the topology and in the amount of energy deposit between signal and background. In particular, a good discrimination variable for the halo muons is the muon impact point position along the calorimeter, shown in Fig~\ref{fig:zMuBuilder}. In fact, the signal tend to be more prominent in the forward part of the tunnel, whereas the halo muons hit the tunnel preferentially upstream. For pions the discrimination is helped by energy related variables, since muons (being mips) release less energy in the calorimeter. 

\begin{figure}
\centering
\includegraphics[width=0.47\linewidth]{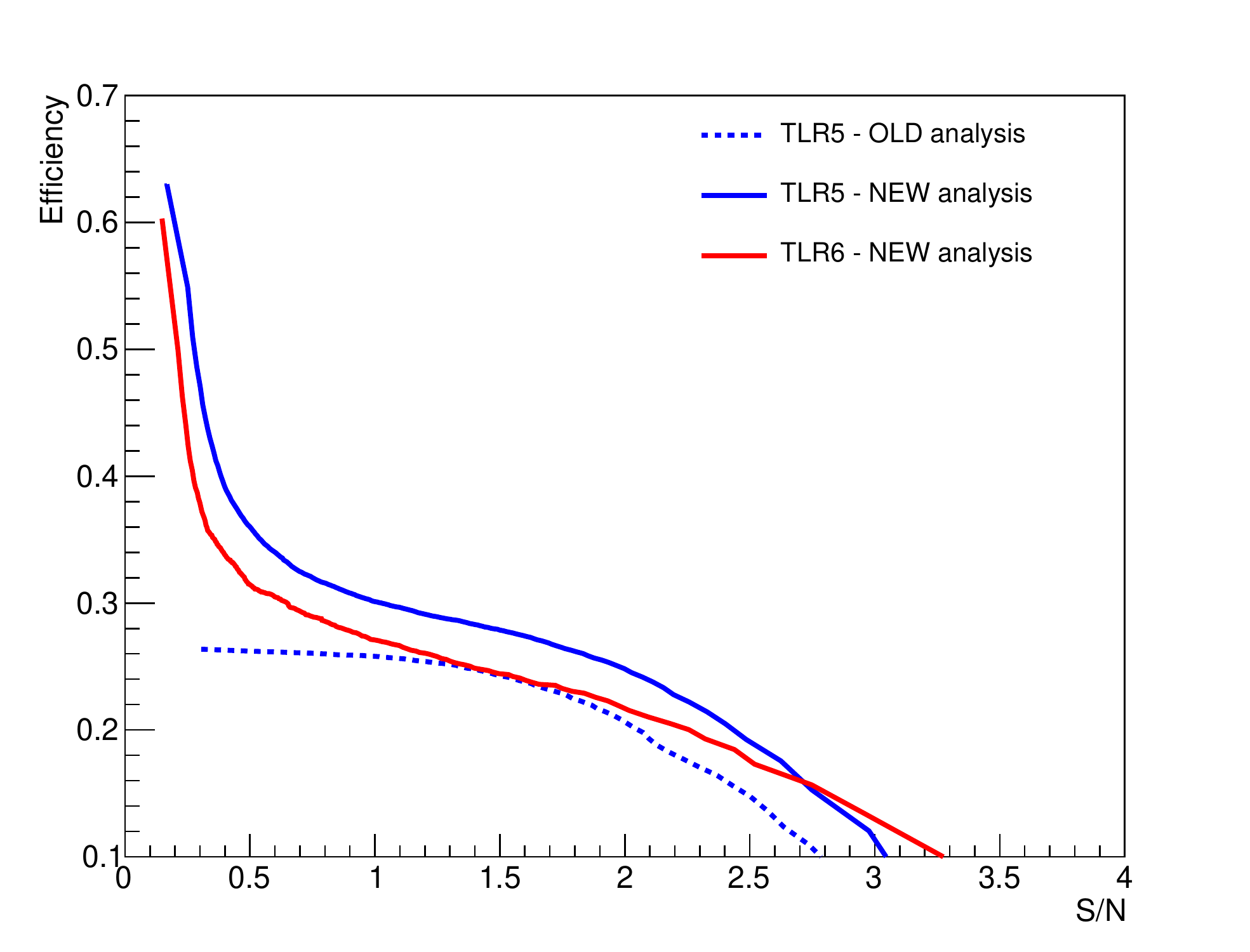}%
\includegraphics[width=0.53\linewidth]{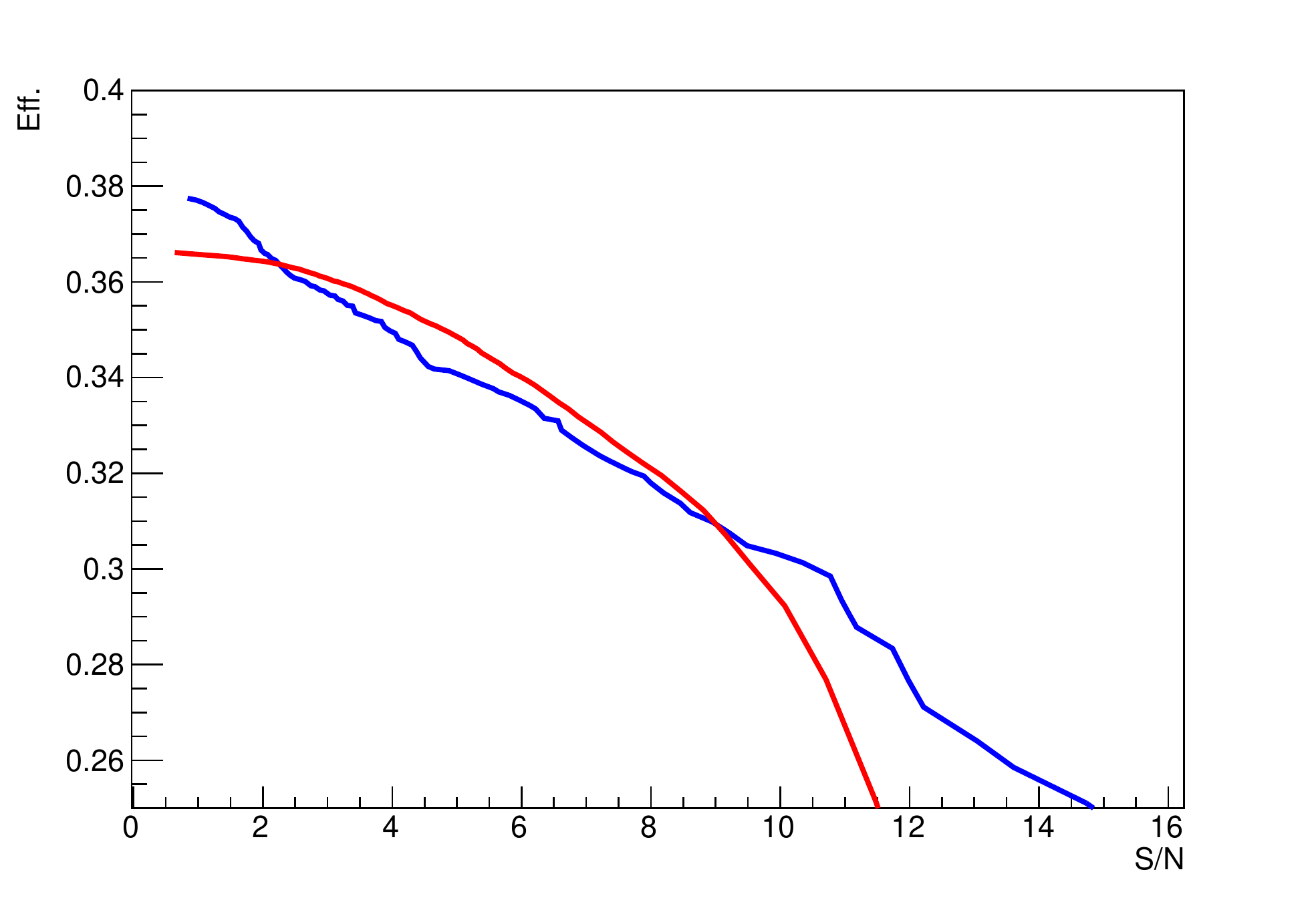}
\caption{Left: Efficiency as a function of the signal-to-noise ratio in the selection of $K_{e3}$ events. Blue lines refers to the previous version of the beamline (TLR5), the dashed one is obtained with the positron identification analysis described in the previous report~\cite{spsc_2020}, the solid one to the improved analysis described here (see text). The red line refers to the new beamline, TLR6 (Sec.~\ref{sec:tl}). Right: signal efficiency versus signal-to-noise ratio for $K_{\mu2}$ and $K_{\mu3}$ events selection. The result from the new transfer line, in red, is compared to the one from the previous report, in blue.
}
\label{fig:eff_vs_SN}
\end{figure}

With the new version of the transfer line, a factor of $\sim3-4$ more halo muons in the forward region of the calorimeter is observed. This could cause a worsening of the discrimination power for the muon impact point observable. Moreover, a reduction of the clustering performance could be caused by the increase in the rate of particles hitting the tunnel, inducing higher mixing effects between different events. Anyway, a worsening of the muon building performance has not been observed. In Fig.~\ref{fig:eff_vs_SN} (right) we present the signal selection efficiency versus the signal-to-noise ratio after applying the NN on the built muon candidates. The two curves show the comparison between the results in the previous report and the one updated with the new transfer line: performances are very similar, a part in the very low and high S/N regions. The very low S/N region corresponds to looser cuts on the NN (no cut in the lower S/N limit), the lower efficiency is then dominated by slightly lower event building performance, probably related to the larger mixing effects observed with the new transfer line. The very high S/N region corresponds to tighter cuts on the NN, and selected background events are very similar to the signal. Thus, the lower efficiency here could be related to the forward region of the calorimeter being polluted by a higher amount of halo muons with the new transfer line. Nevertheless, when cutting at the NN value maximizing the efficiency times purity we get a signal efficiency of $\sim$~34\% and S/N of $\sim$~6 (Fig.~\ref{fig:eff_vs_SN}, right).

\subsubsection{Muon chambers}
\label{sec:muchambers}

\begin{figure}[ht]
\begin{subfigure}{.5\textwidth}
  \centering
  % include first image
  \includegraphics[width=1.0\linewidth]{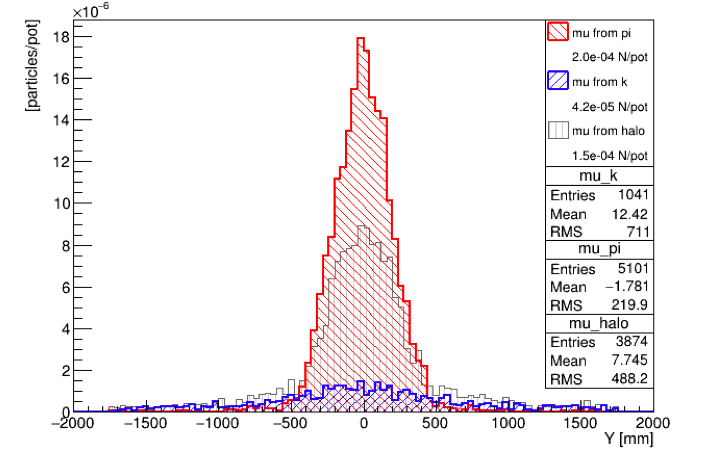}  
  \caption{}
  \label{fig:muSpatial}
\end{subfigure}
\begin{subfigure}{.5\textwidth}
  \centering
  % include second image
  \includegraphics[width=1.0\linewidth]{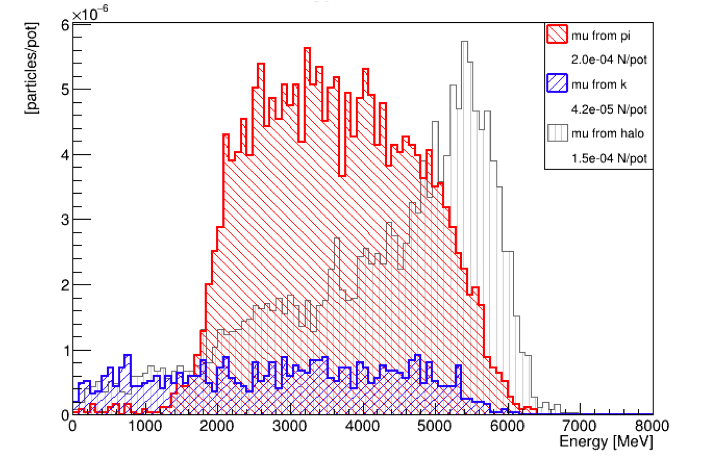} 
  \caption{}
  \label{fig:muE}
\end{subfigure}
\caption{Relevant distributions for muons in the instrumented hadron-dump. (a) spatial distribution along one of the two dimensions of a detector layer. (b) energy distribution. Red and blue are the signal contribution, from pion and kaon decays. Gray distribution are halo-muons.}
\label{fig:muDump}
\end{figure}

Low energy muon-neutrinos from pions can be constrained by monitoring the associated muons produced in the decays. These $\mu$ are emitted at low angle, and go through the decay tunnel crossing the final hadron-dump. A measurement of relevant physics observables can be performed by instrumenting the hadron-dump, using different detector layers installed at increasing depth and interleaved by absorbing material. The detector layout was presented in the last report~\cite{spsc_2020}. It allows to measure the spatial distribution of muons, shown in Fig.~\ref{fig:muSpatial}, and their energy, in Fig.~\ref{fig:muE}, by range-out. The differences in the distributions between the contributions can be exploited to disentangle the signal from the background, again dominated by halo-muons. 

In collaboration with the new group from University of Thessaloniki we are now in the process of determining the best detector technology to employ for the instrumentation of the hadron-dump. The detector needs to cope with a muon rate of 2~MHz/cm$^{2}$ and a neutron fluence of about $10^{12}~1~$MeV-{neq}$/$cm$^{2}$ in the layer closer to the end of the decay tunnel.

\subsection{Narrow band off-axis technique}
\label{sec:NBOA}
A slow proton extraction scheme allows for the direct measurement of muons from pion decays after the hadron dump: the combination of this information with the one from the tagger on large angle muons from kaons provides a complete monitoring of the $\nu_{\mu}$ flux in ENUBET (see Sec.~\ref{sec:muon}).
In addition, the narrow momentum width of the beam ($\mathcal{O}$(5-10\%)) can be exploited to provide the neutrino energy on an event by event basis, thanks to its correlation with the position of the interaction vertex in a detector placed at short distance.
%Inserire qui figura con correlazioni E-R?
The determination of the neutrino energy at source with enough precision could overcome the uncertainties related to its determination through the reconstruction of the final state, thus allowing for a high precision differential cross section measurement.

This method, named ``narrow band off-axis'', was already described in~\cite{enubet_proposal}. In this report we update the related results using the realistic beam provided by the latest version of the transfer line (Sec.~\ref{sec:tl}), thus including also muons and neutrinos from the beam halo and off-momentum mesons. Low energy neutrinos from the target station and the proton dump are not included in the present estimation, but they are expected to give a negligible contribution to the final interaction budget, as a result of their cross section suppression. A final re-evaluation of the technique capabilities will be performed after the fine tuning of the complete GEANT4 simulation of the ENUBET facility (Sec.~\ref{sec:simulation}).

\begin{figure}
\centering
\includegraphics[scale=0.41]{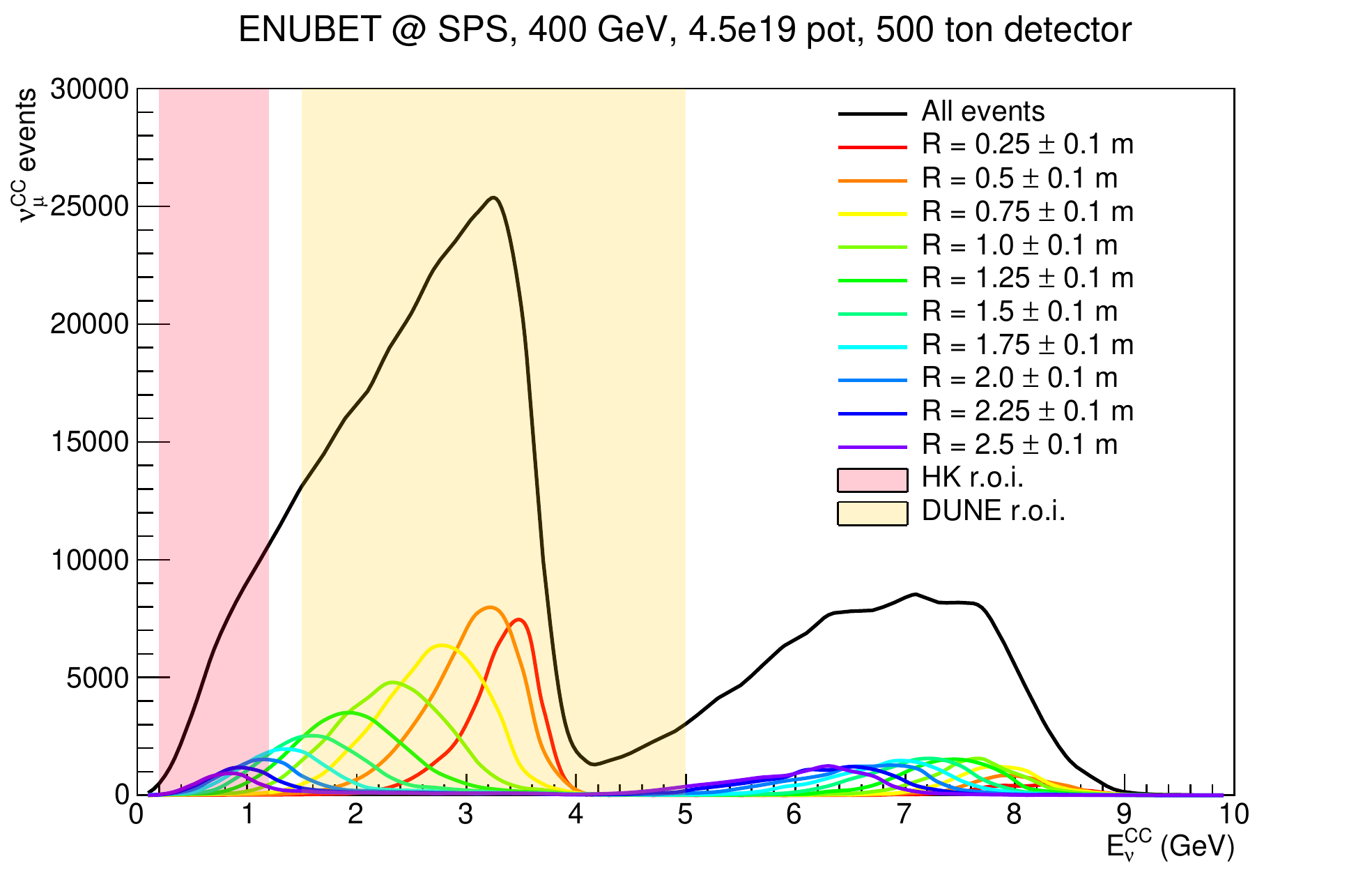}
\includegraphics[scale=0.41]{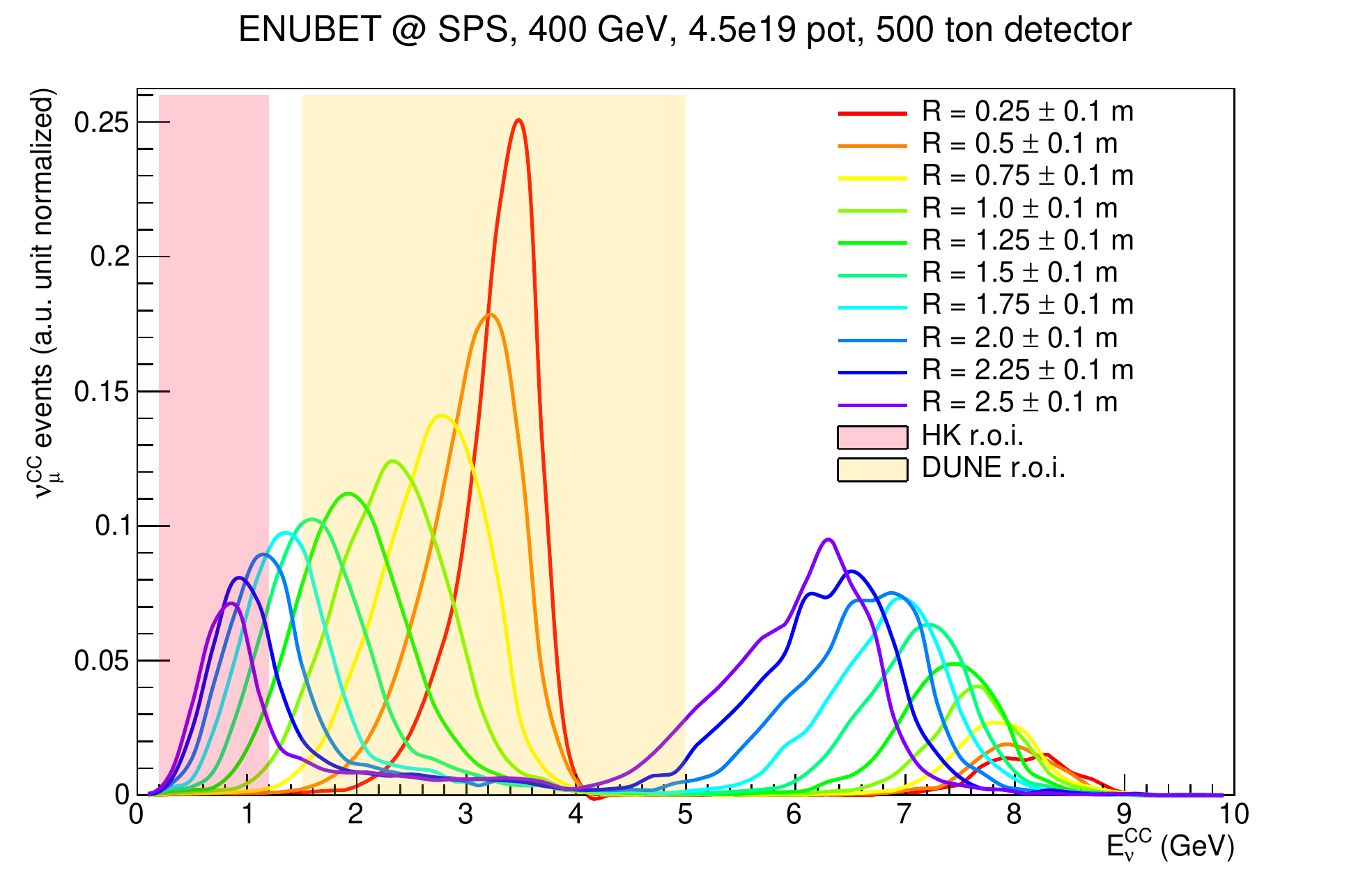}
\caption{(left) $\nu_{\mu}^{CC}$ interaction spectrum in black. The coloured lines show spectra of interactions at different radial distances (R) from the beam axis, higher frequencies of the line colour refer to larger R. (right) Unit normalized spectra. }
\label{fig:NBOA_spectra}
\end{figure}

In order to estimate the expected number of neutrino interactions a 500~ton liquid Argon detector (isoscalar target) placed 90~m from the entrance of the decay tunnel (50~m from the hadron dump) and with a cross sectional area of 6$\times$6~m$^2$ has been considered. The detailed detector response is not included and the interaction rate is the convolution of the neutrino flux with the CC cross section: assuming 4.5$\times$10$^{19}$~POT, about 4$\times$10$^5$ $\nu_{\mu}^{CC}$ interactions will be observed at the detector and their spectrum is shown by the black line in Fig.~\ref{fig:NBOA_spectra} (left). The colored lines describe the spectra obtained selecting neutrino interactions at different radial distances $R$ from the beam axis, in windows of $\pm$10~cm. The same spectra, but normalized to one, are reported in the right plot.

The low energy pion component of the spectra can be separated from the kaon one with a loose cut on the reconstructed energy. The width of the pion peaks at different $R$ can then be used as an estimator of the precision on the incoming neutrino energy.
%As shown in Fig.~\ref{fig:NBOA_Eres} it ranges from 8\% to 25\% in the core of the DUNE energy domain, for which the ENUBET beam is optimized. 
%Possible improvements on the performance of the technique in the HyperK energy region, where the energy resolution is limited to $\sim$30\%, will be
%evaluated once the the multi-momentum beamline (Sec.~\ref{sec:multimom}), that can enrich the low-energy part of the spectrum, will be finalized.

%\begin{figure}
%\centering
%\includegraphics[scale=0.75]{figs/rate1R-8.5.pdf}
%\caption{Spectra unit normalized}
%\label{fig:NBOA_spectra1}
%\end{figure}

%\begin{figure}
%\centering
%\includegraphics[scale=0.6]{figs/en_res-8.5.pdf}
%\caption{Beam energy spread (in black) and peak energy (in red) as a function of the distance R of the interaction vertex at the detector from the beam axis.}
%\label{fig:NBOA_Eres}
%\end{figure}

%\subsection{Tagged neutrino beams}
%\label{sec:tagged}
%Antonio Andrea Thessaloniki
%
%\section{Next steps}
% Andrea, Francesco

%\begin{figure}
%\centering
%\includegraphics[scale=0.6]{figs/en_res-8.5.pdf}
%\caption{Beam energy spread (in black) and peak energy (in red) as a function of the distance $R$ of the interaction vertex at the detector from the beam axis.}
%\label{fig:NBOA_Eres}
%\end{figure}

%\subsection{Tagged neutrino beams}
%\label{sec:tagged}
%Antonio Andrea Thessaloniki
%
%\section{Next steps}
% Andrea, Francesco

\subsection{Demonstrator: design, procurement, construction}
\label{sec:demonstrator}
% Andrea

\begin{figure}
    \centering
    \includegraphics[width=0.565\linewidth]{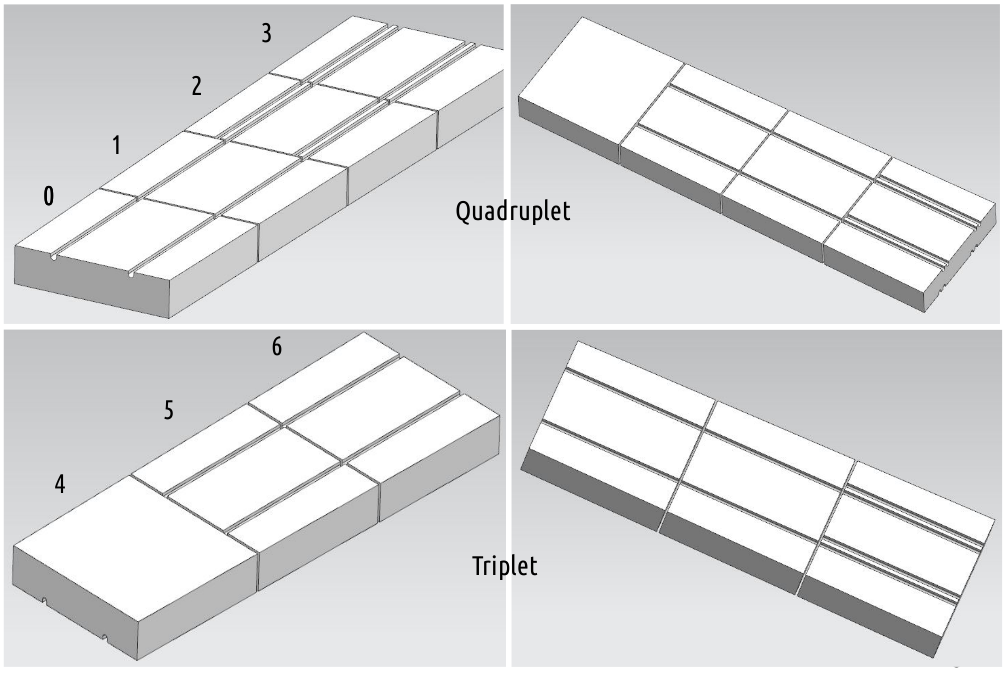}%
    \includegraphics[width=0.435\linewidth]{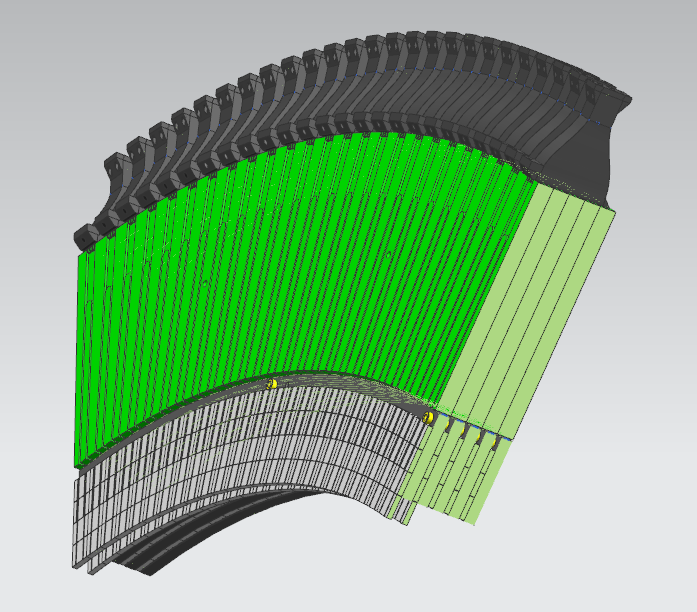}
    \caption{Left: layout of the frontal grooves readout scheme.
    Right: rendering of a section of the demonstrator corresponding to the extension along the beam axis of single module (LCM). Scintillators are shown in gray, borated polyethylene shielding in green and fiber concentrators in black.}
    \label{fig:frontalscint}
\end{figure}

%The design of demonstrator has undergone significant improvements in the last year. 
%The definition of the large scale production of the scintillators has driven the adoption of a new light readout scheme that employs frontal grooves in place of the lateral grooves that had been envisaged formerly. This allows a safer production with injection molding and also guarantees a more uniform light collection. 
%The new adopted scheme is outlined in 
The layout of scintillators is shown in Fig.~\ref{fig:frontalscint}, left. All tiles are trapezoidal in shape such that to minimize dead regions. The $t_0$-layer tile (innermost radial position) holds two grooves. They are 1 mm deep such that the full WLS diameter is enclosed in the scintillator. These two grooves are meant for light collection (``readout'' grooves) and they are propagated to the other three outer tiles where they only have the function of letting space to host the WLS fiber without letting the scintillation light to enter (``transit'' grooves). Transit grooves hence need to be covered with a chemical etching treatment to make them optically opaque. The second tile has readout grooves on the other side of the transit grooves. This pattern repeats as in the Figure. For tile 2 and 3 readout grooves are slightly offset-ed with respect to 0 and 1 to accommodate them with the transit ones.

The scintillators are in the production phase with UNIPLAST (Moskow) in collaboration with the INR group. The tiles will be first produced with the proper shape using injection molding. 
Afterwards transit grooves will be milled
and the tiles will be treated with chemical etching surface treatment to make them white and opaque. Finally the readout grooves will be milled.
The total number of scintillator tiles for the demonstrator will be of about 10000 pieces.
The scintillators will be interspersed with 15~mm thick iron slabs to form the sampling calorimeter. Their
position with respect to the other elements of the demonstrator is shown in the right part of Fig.~\ref{fig:frontalscint}.

%The new scheme has been validated with a GEANT4 optical simulation and the distance between pair fibers has been optimized to achieve the best possible light collection efficiency and uniformity. Overall, taking a 1.5~cm distance the number of collected photons is significantly higher in the new scheme (left top plot of Fig.~\ref{fig:frontalscintOS}) and more spatially uniform with respect to the former lateral readout  (top right of the same figure). We are also measuring efficiency maps of some tiles with a similar geometry (Fig.~\ref{fig:frontalscintOS}, bottom plots) at INFN-Bologna with a cosmic ray tracker.
%\begin{figure}
%    \centering
%    \includegraphics[width=\linewidth]{figs/comp_lo.png}
%    \includegraphics[width=0.41\linewidth]{figs/mich1.png}%
%    \includegraphics[width=0.5\linewidth]{figs/mich2.png}
%    \caption{Top: Comparison between the light collection using lateral  -left - or frontal WLS fibers with a 1.5~cm spacing -right- according to a GEANT4 optical simulation. Bottom: uniformity tests on tiles with frontal readout done with cosmic rays.}
%    \label{fig:frontalscintOS}
%\end{figure}

%\begin{figure}
%    \centering
%    \includegraphics[width=\linewidth]{figs/3Dsettore.png}
%    \caption{Rendering of a section of the demonstrator corresponding to the extension along the beam axis of single module (LCM). Scintillators are shown in gray, borated polyethylene shielding in green and fiber concentrators in black.}
%    \label{fig:demo3dx}
%\end{figure}

The procurement of Silicon Photomultipliers for the ENUBET demonstrator has been finalized and Hamamatsu models S14160-3050HS ($3\times 3$~mm$^2$) and S14160-4050HS ($4\times 4$~mm$^2$) were selected.

\begin{figure}
    \centering
    \includegraphics[width=0.4065\linewidth]{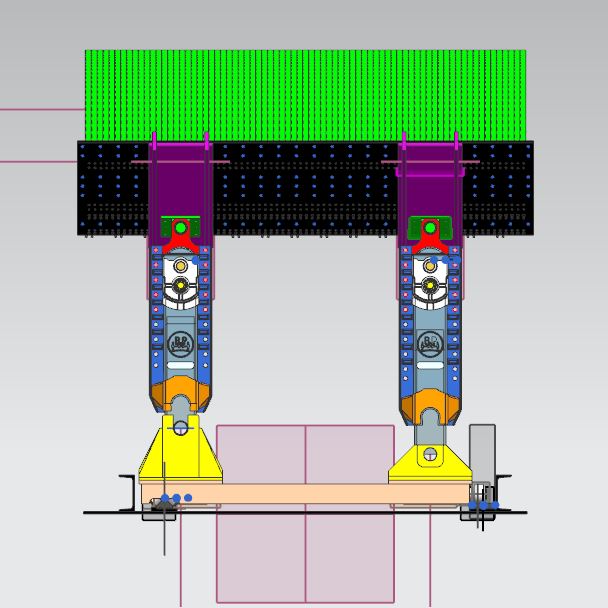}%
    \includegraphics[width=0.5935\linewidth]{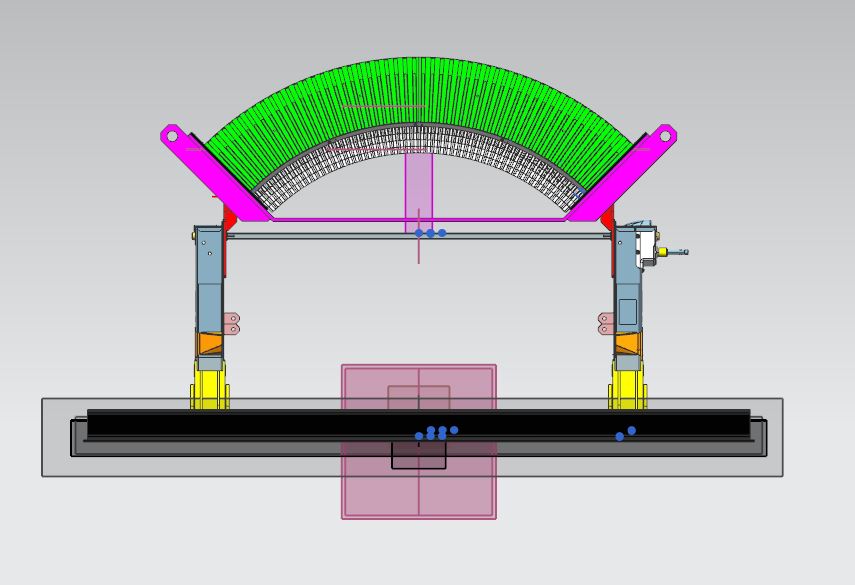}
    \caption{Left: rendering of the demonstrator.}
    \label{fig:demo2d}
\end{figure}

\begin{figure}
    \centering
    \includegraphics[width=0.42\linewidth]{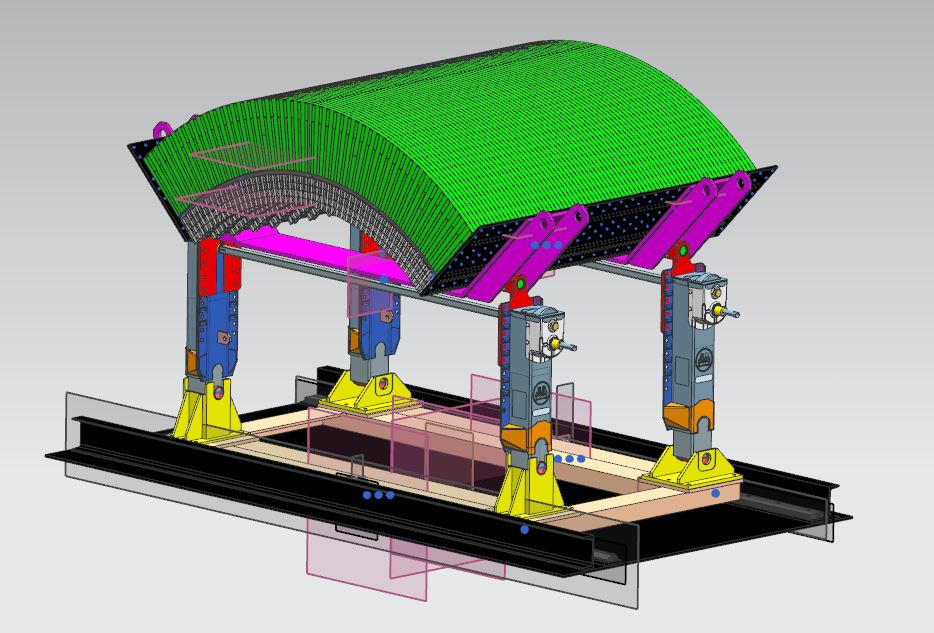}%
    {~}
    \includegraphics[width=0.58\linewidth]{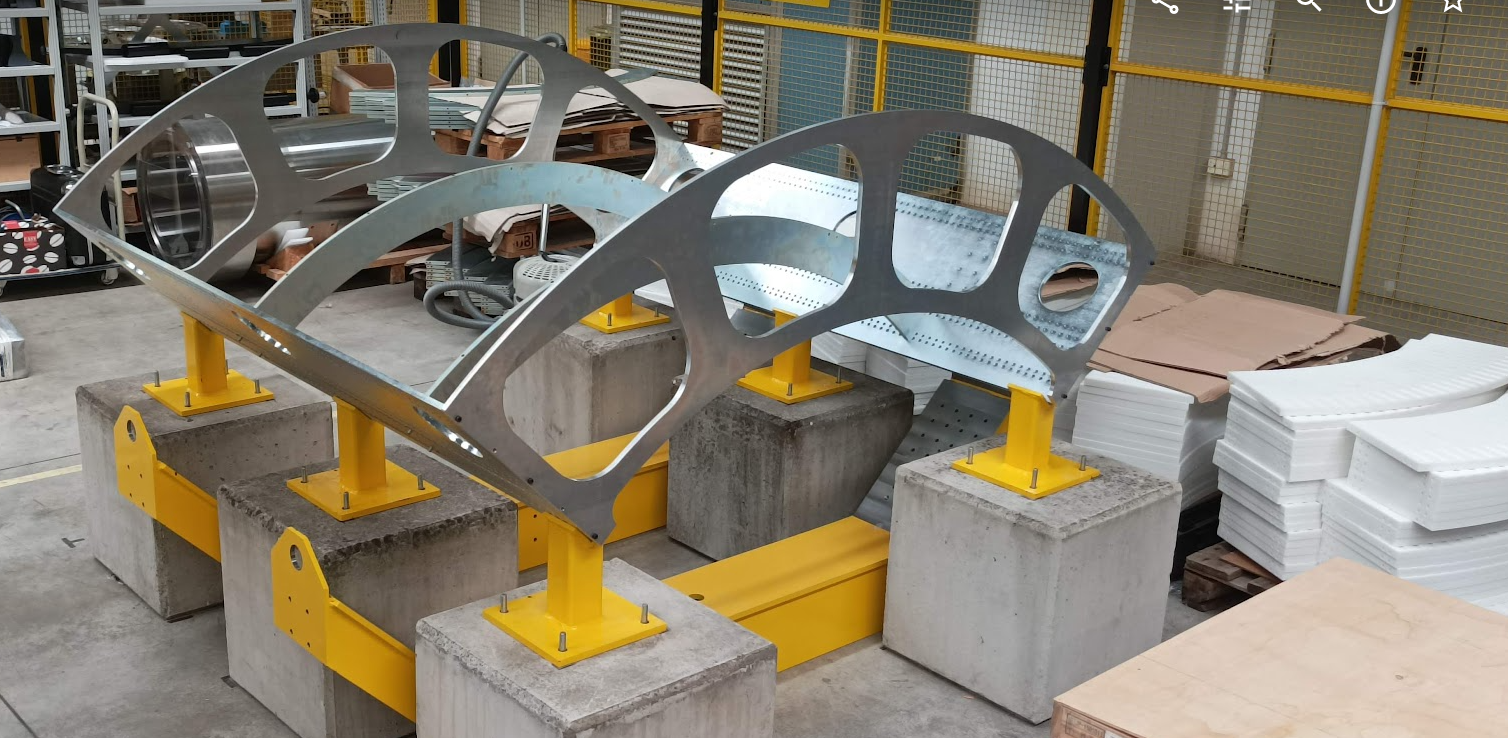}
    \caption{Left: 3D rendering of the demonstrator. Right: assembly at INFN-LNL.}
    \label{fig:demo3d}
\end{figure}

The final choice for the layout of the demonstrator to be exposed at CERN in October 2022 is shown in Fig.~\ref{fig:demo2d} with a beam-transverse (right) and a lateral view (left). The same is shown with a 3D rendering in Fig.~\ref{fig:demo3d}. The object is 1.65~m long in the longitudinal direction and it spans 90$^\circ$ in azimuth. It comprises 75 layers or iron (15 mm thick) and 75 layers of scintillators (7~mm thick) i.e. $12~\times 3$~LCM. We expect to instrument the central 45$^\circ$ central region while the rest has been kept for mechanical considerations. The structure is designed to be extendable to a full $2\pi$ object by joining four similar detectors with minimal dead regions. The iron part extends radially for 11~cm while the remaining 30~cm are filled with 22.5~mm thick with Borated P.E. slabs with 5\% Boron (green in the Figure). 

The structure is held by a mechanical crawl sitting on four extensible legs that are designed to allow tilting the calorimeter-beam angle in the vertical plane and adjusting the horizontal position by means of a wheel-rail mechanism.

\begin{figure}
    \centering
    {\includegraphics[width=0.6\linewidth,angle=90]{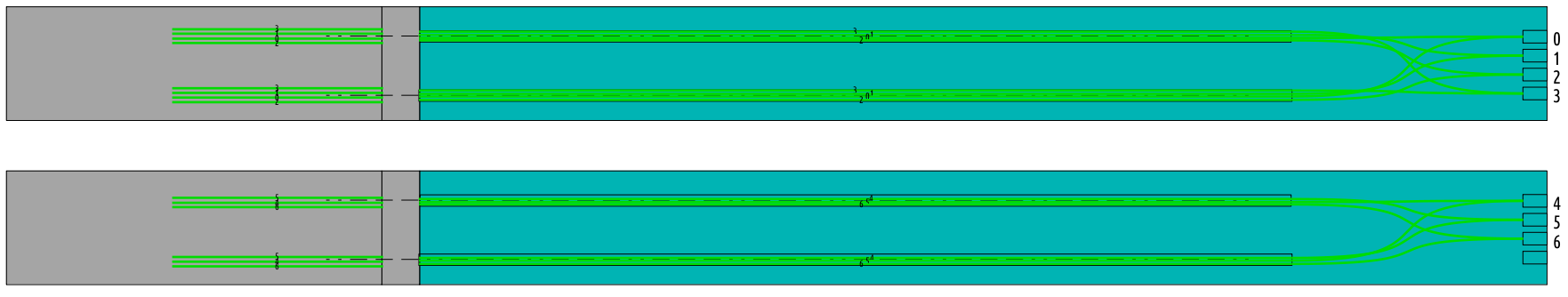}%
    \includegraphics[width=0.6\linewidth,angle=0]{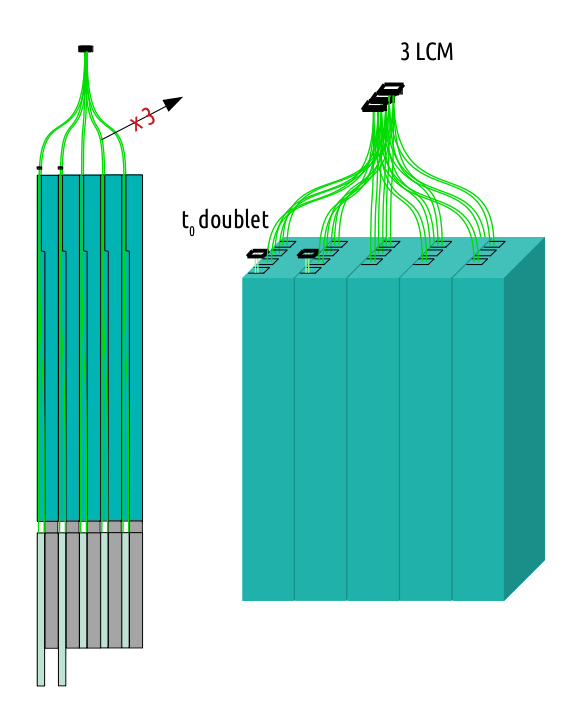}\\
    \includegraphics[width=0.7\linewidth]{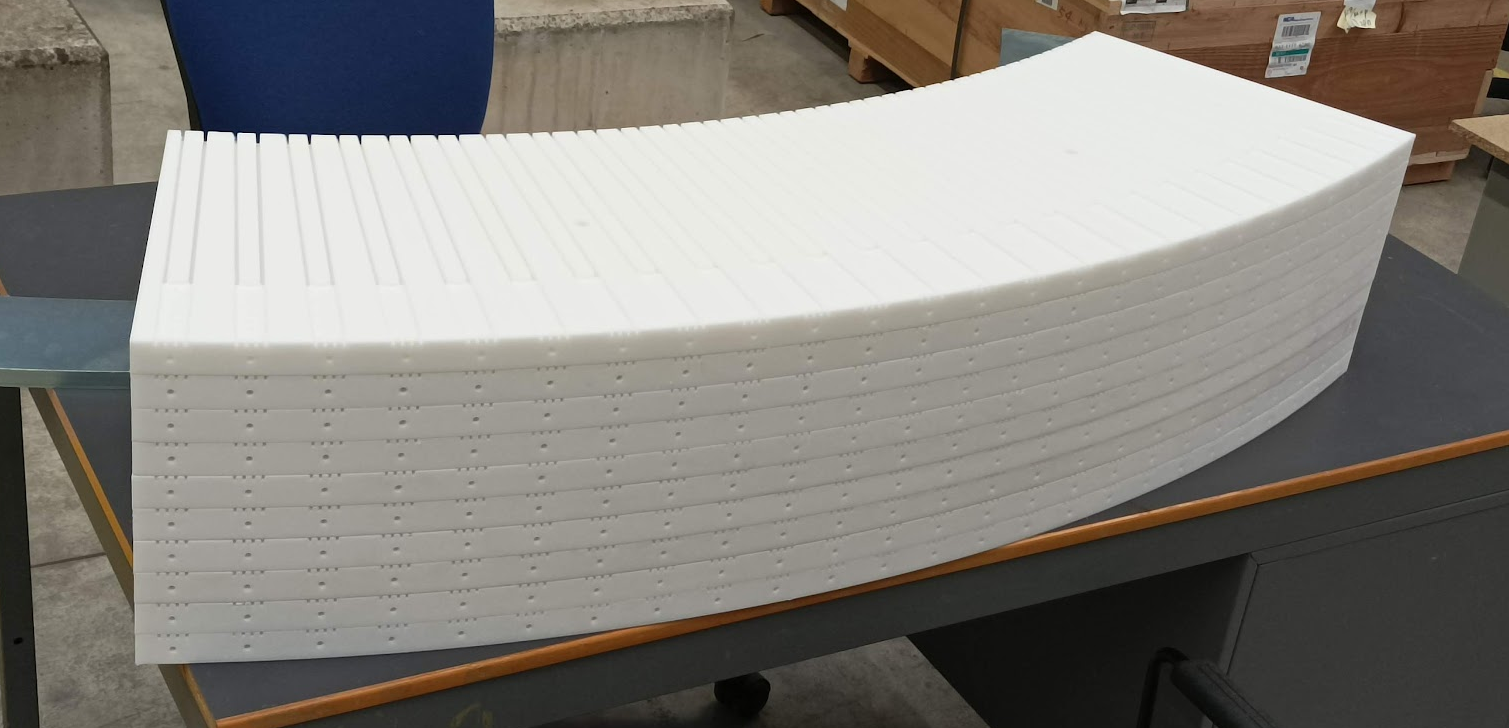}
    }
    \caption{Top: schematic view of the WLS routing thought the borated polyethylene shielding. Bottom: the borated polyethylene shielding slabs.}
    \label{fig:WLSSiPM}
\end{figure}

\begin{figure}
    \centering
    \includegraphics[width=\linewidth]{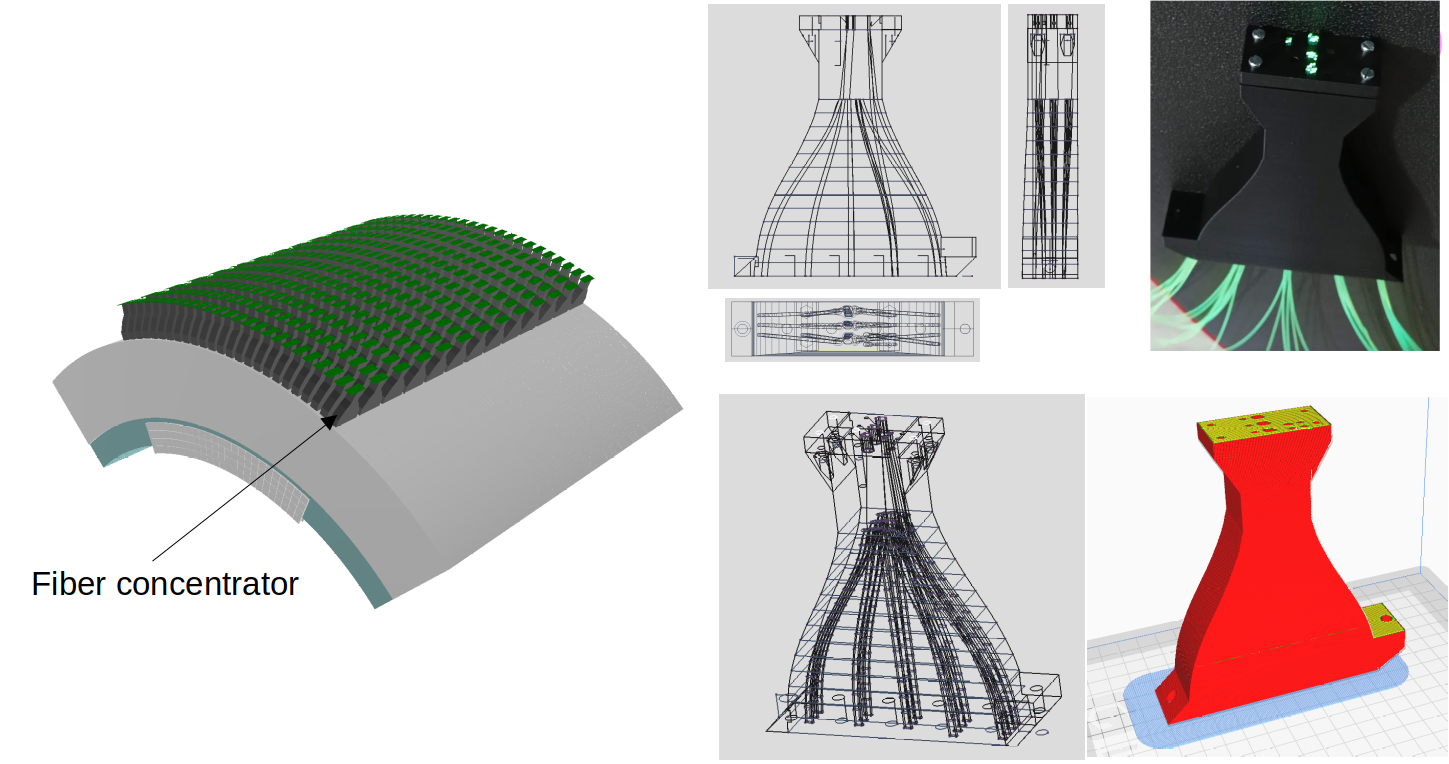}
    \caption{Fiber concentrators.}
    \label{fig:FC}
\end{figure}

The strategy for routing the WLS and matching them to the photo-sensors is sketched in Fig.~\ref{fig:WLSSiPM}. The ten WLS fibers belonging to the same LCM are routed by passing through grooves milled in the BPE and bundled to a $4\times 4$~mm$^2$ SiPM. The $t_0$-layer fibers will be read by two independent SiPMs.
The demonstrator will adopt WLS fiber ``concentrators'' that we developed using 3D-printing techniques. These objects (Fig.~\ref{fig:FC}) are essential to ease the process of fibers bundling. In such a way the coupling of WLS fibers to the SiPM can be accomplished in a neat and reproducible way allowing a good, homogeneous optical contact. After fiber polishing, five SiPM (three 4$\times$4 mm$^2$ and two 3$\times$3 mm$^2$ for the $t_0$-layer), soldered on a small PCB, will be fixed to the top of each fiber concentrator with screws.

\begin{figure}
    \centering
    \includegraphics[width=0.99\linewidth,angle=0]{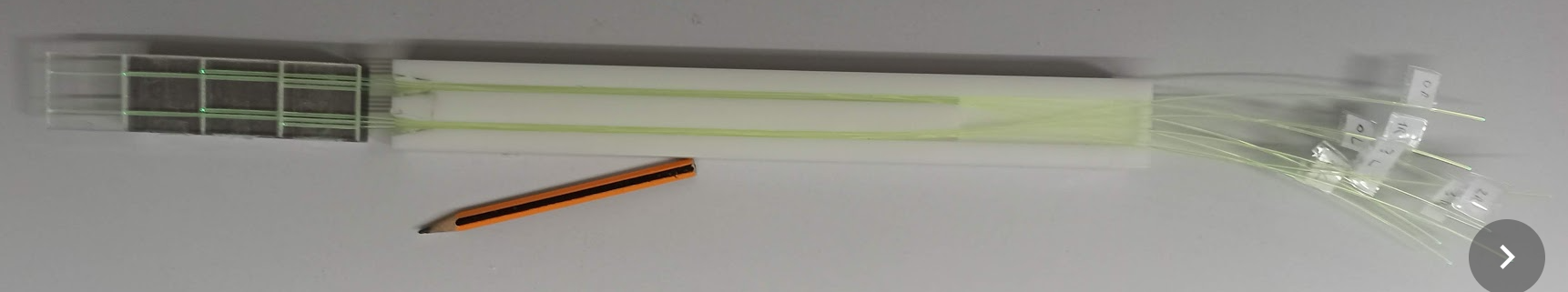}
    \includegraphics[width=0.99\linewidth,angle=0]{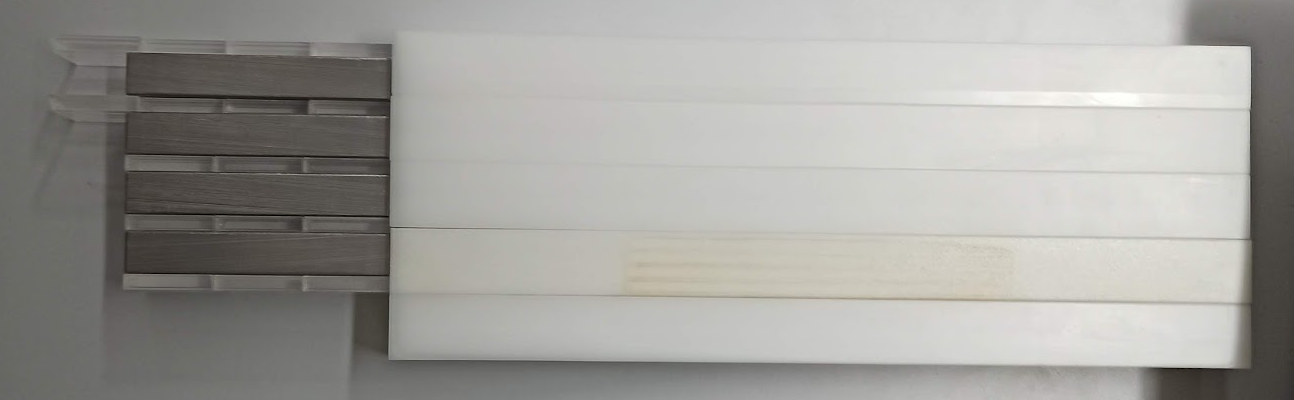}
    \caption{Prototype of a single azimuthal section of the demonstrator (ENUBINO) shown during its assembly to highlight the composing elements (scintillators, iron absorbers, WLS fibers and the BPE shielding.}
    \label{fig:mockup}
\end{figure}

\begin{figure}
    \centering
    \includegraphics[width=0.39\linewidth]{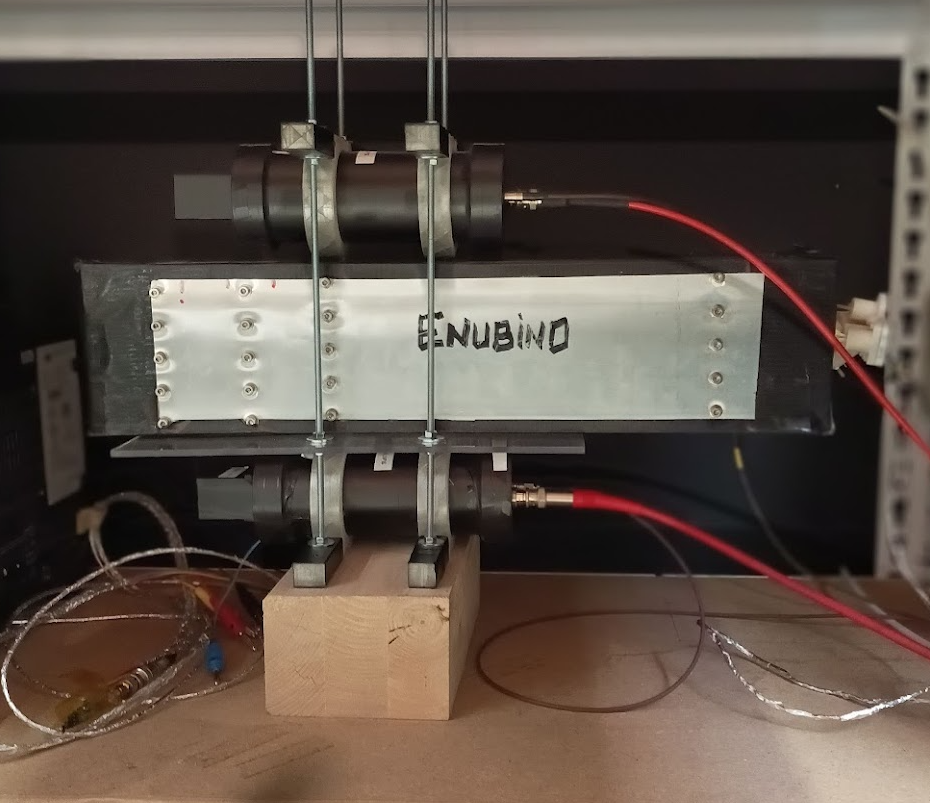}%
    {~~~}
    \includegraphics[width=0.61\linewidth]{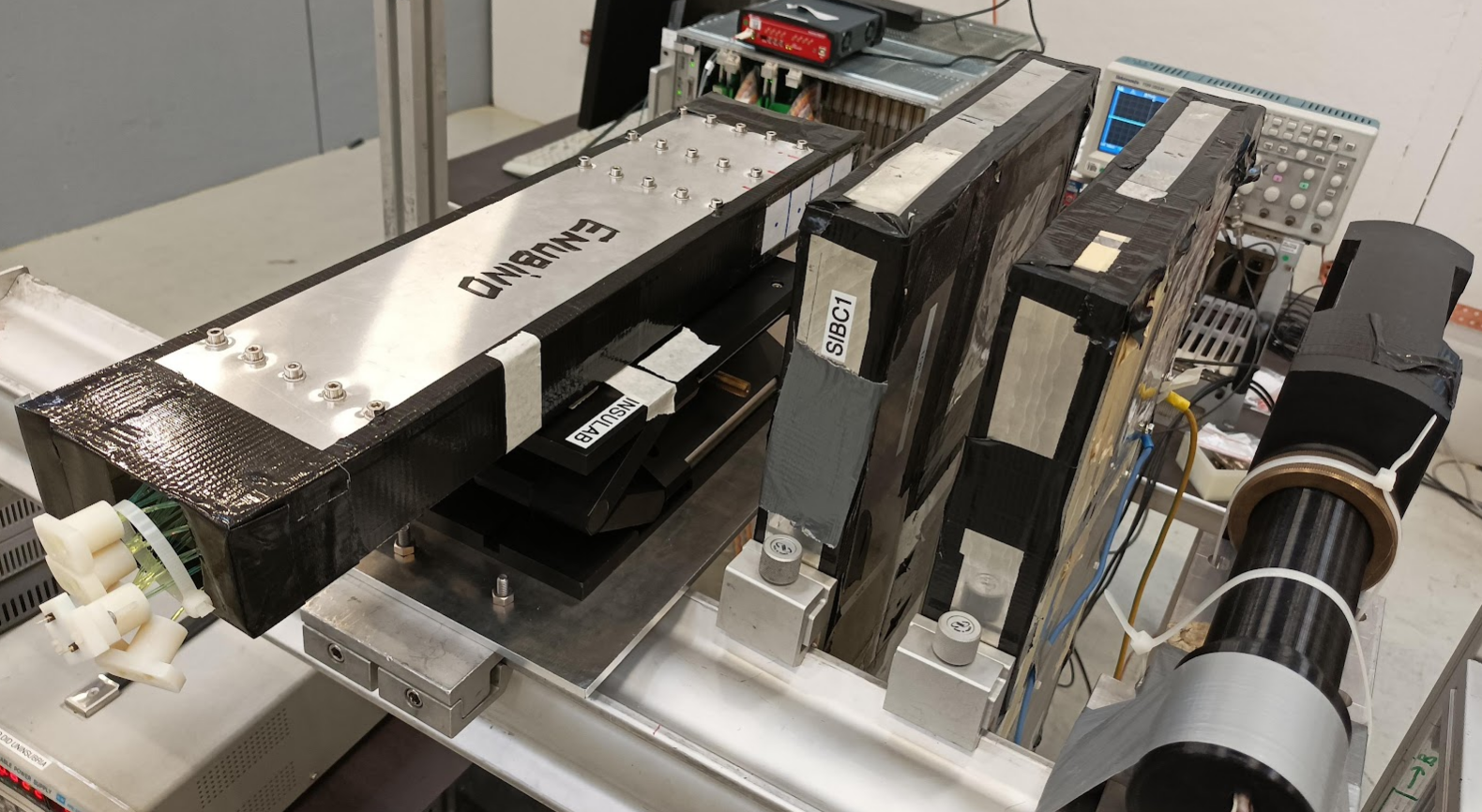}
    \caption{The ENUBINO prototype exposed to cosmic rays and beams at the CERN-PS.}
    \label{fig:enubino_exposure}
\end{figure}

\begin{figure}
    \centering
    \includegraphics[width=0.49\linewidth]{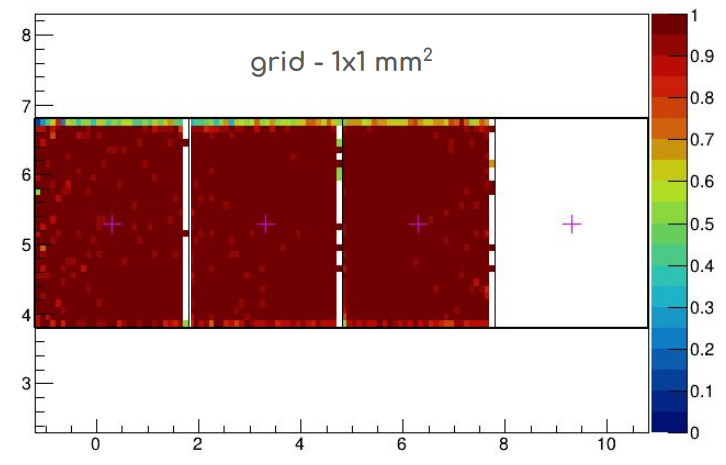}%
    \includegraphics[width=0.51\linewidth]{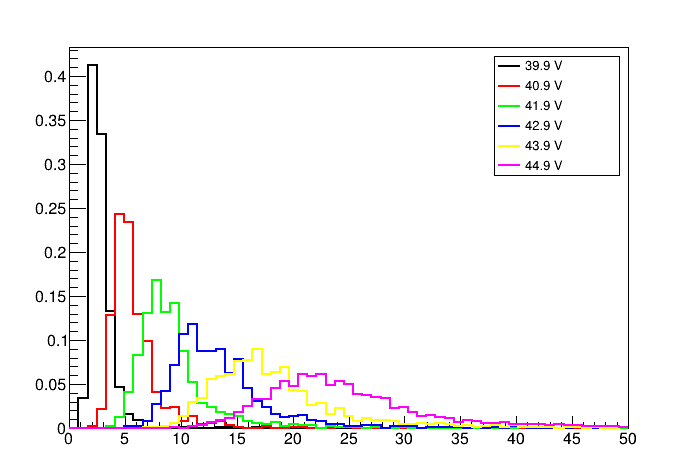}
    \caption{ENUBINO: efficiency maps (left) and m.i.p. signals vs over-voltage for the chosen Hamamatsu SiPMs, in mV (right).}
    \label{fig:unif_advan}
\end{figure}

A pre-demonstrator small prototype comprising 3 LCMs (ENUBINO) (Fig.~\ref{fig:mockup}) has been built and characterized with cosmic ray tracks at INFN-LNL laboratory and with particle beams at CERN in November 2021 (Fig.~\ref{fig:enubino_exposure}). It faithfully implements the chosen geometry and the solutions for the light readout both at the level of scintillators and of the neutron absorber layer. The prototype behaves according to expectations with a good uniformity and a tolerable level of optical cross talk between the scintillator tiles. 
Efficiency maps show high efficiency and good uniformity (Fig.~\ref{fig:unif_advan}, left). The response to m.i.p signals was also tested with the final selected Hamamatsu SiPM model as a function of the supply voltage (Fig.~\ref{fig:unif_advan}, right).

%We have performed some tests for the production of the iron slabs using calendering of straight slabs (Fig.~\ref{fig:iron}). The process introduces a thickening of the inner side
%of the circumference by $\sim 1$~mm. We have hence decided to proceed with high pressure water cutting of large slabs since the achievable precision is better, especially for the thickness. 
%\begin{figure}
%    \centering
%    \includegraphics[width=0.5\linewidth]{figs/calandratura.png}
%    \caption{Calendering tests for the iron absorbers slabs.}
%    \label{fig:iron}
%\end{figure}

\subsubsection{Assessment of the systematics and physics performance}
\label{sec:systematics}
% Antonio, Claudia D., Andrea, Francesco

In conventional beams the systematic uncertainties impacting on the neutrino flux are relatively large: they come from dedicated experiments to constrain the hadro-production processes and from the precision of the parameters characterizing the facility (for instance beamline related ones, like currents, alignment of the magnets, etc.). As it was already mentioned, typically, the largest systematic contribution comes from the hadro-production uncertainties, both at the level of the target and from the re-interactions of secondaries in the passive elements of the beamline. %A summary of the relevant contributions is reported in table \colorbox{BurntOrange}{XY}. 

Within NP06/ENUBET, the monitoring of the leptons produced together with the neutrinos, allows to set better constraints on the neutrino flux, factorizing-out these nuisance effects. Basically, the constraints on the parameters inferred a posteriori by exploiting the neutrino monitoring data, imply a reduction of the systematic uncertainties. The parameters constrained from data are then used to reweight the Monte Carlo simulation and get in turn a higher precision on the neutrino flux.

A model describing the measured physics observables is built from the distributions predicted by the simulation. The model Probability Density Function (PDF) is:

\begin{equation}\label{eq:pdf}
    PDF = N_{S}(\Vec{\alpha}, \Vec{\beta})\cdot S(\Vec{\alpha}, \Vec{\beta}) + N_{B}(\Vec{\alpha}, \Vec{\beta})\cdot B(\Vec{\alpha}, \Vec{\beta})
\end{equation}

It is given by the sum of the signal, $S$, and background, $B$, shape templates, normalized to the corresponding number of events $N_{S,B}$. The systematic effects are introduced as nuisance parameters in the model: the set of $\Vec{\alpha}$ and $\Vec{\beta}$ are the hadroproduction and beamline related nuisances, respectively. As shown in Eq.~\ref{eq:pdf}, systematics can affect both the normalization and shape of signal and background. An extended likelihood for the observed data can be written using the model PDF in Eq.~\ref{eq:pdf}, where the nuisance parameters are constrained by their own PDFs, $pdf(\Vec{\alpha}|0,1)$ and $pdf(\Vec{\beta}|0,1)$, representing the a priori knowledge on the hadroproduction and facility parameters discussed before. By maximizing the extended likelihood, a specific estimation of the nuisance for the NP06/ENUBET facility can be obtained.

This approach, has been tested using toy Monte Carlo data to study the level of improvement in the systematic uncertainties and the consequent gain on the neutrino flux precision. The implementation of the model PDF in Eq.~\ref{eq:pdf} and of the extended maximum likelihood on toy Monte Carlo data is done through the RooFit package from ROOT~\cite{root}. A \emph{RooRealSumPdf} is used to write the model from Eq.~\ref{eq:pdf}, where the signal and background templates are \emph{RooHistPdf} built from the distribution predicted by the simulation. Templates normalization and shape are allowed to change around their nominal values, following the PDFs of the nuisance parameters $\Vec{\alpha}$ and $\Vec{\beta}$:

\begin{equation}\label{eq:norm}
    N(\Vec{\alpha}, \Vec{\beta}) = N_{0}\cdot (1 + \vec{r}_{\alpha}\cdot\Vec{\alpha} + \vec{r}_{\beta}\cdot\Vec{\beta})
\end{equation}

\begin{equation}\label{eq:shape}
    T(\Vec{\alpha}, \Vec{\beta}) = T_{0}\cdot + \Vec{\alpha}\cdot\Delta \Vec{T}_{\alpha} + \Vec{\beta}\cdot\Delta \Vec{T}_{\beta}
\end{equation}
where $N_{0}$ and $T_{0}$ are the nominal normalization and shape, respectively, for either a signal or background template. $\Vec{r}_{\alpha, \beta}$ and $\Delta \vec{T}_{\alpha, \beta}$ are the relative changes in normalization and the variations in shape due to a one standard deviation change in $\Vec{\alpha}, \Vec{\beta}$, with respect to their nominal values. The final model is the product of the model PDF and the constraining PDFs for the nuisance parameters $\Vec{\alpha}$ and $\Vec{\beta}$. A set of toy Monte Carlo experiments are generated by setting the nuisance parameters to values extracted from their PDFs. A fit is then performed on each experiment to study the improvement on nuisance parameters constraint due to the generated pseudo-data. In order to do this the \emph{RooMCStudy} tool is exploited. This implementation, using the tools available from RooFit, is highly modular and flexible, allowing to add different systematic contributions relatively easily, without having to rewrite the model for each new parameter to be included. 

Preliminary tests to debug the algorithm have been performed by using a toy hadroproduction model. Mock physics observables for the fit have been sampled using an analytical parametrization of hadroproduction and basic decay kinematics. 

\begin{figure}[ht]
\begin{subfigure}{.33\textwidth}
  \centering
  % include first image
  \includegraphics[width=1.0\linewidth]{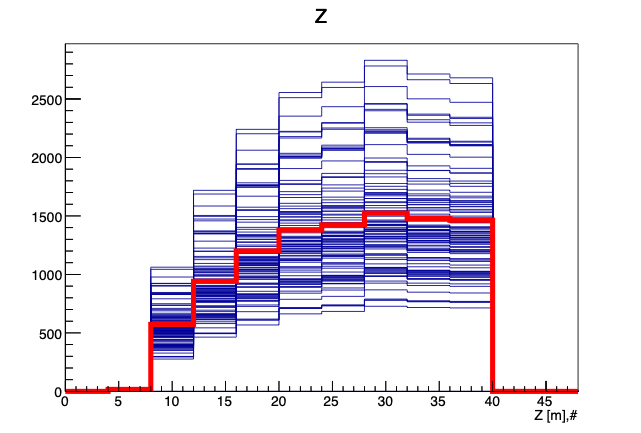}  
  \caption{}
  \label{fig:mockZmu-a}
\end{subfigure}
\begin{subfigure}{.33\textwidth}
  \centering
  % include second image
  \includegraphics[width=1.0\linewidth]{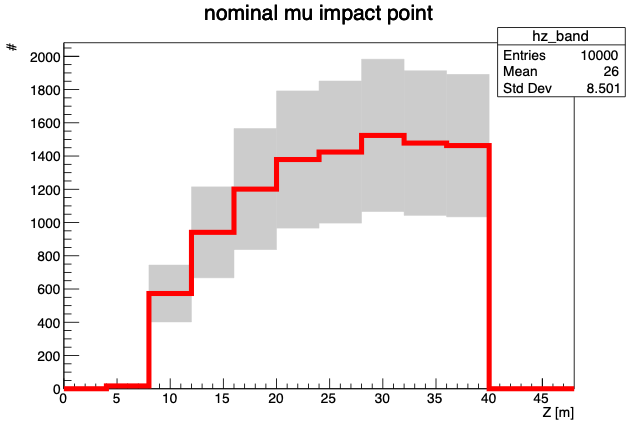}  
  \caption{}
  \label{fig:mockZmu-b}
\end{subfigure}
\begin{subfigure}{.33\textwidth}
  \centering
  % include third image
  \includegraphics[width=1.0\linewidth]{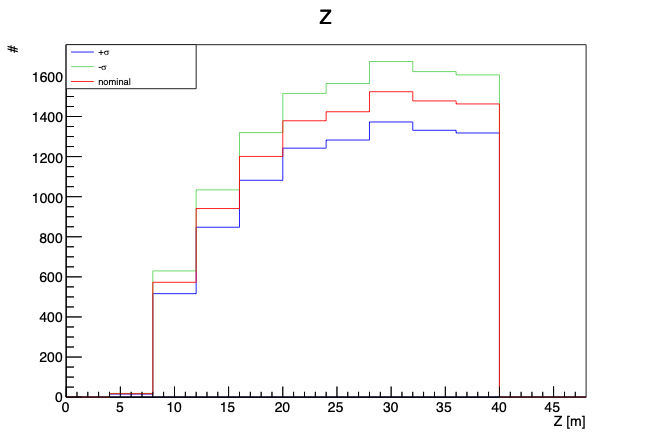}  
  \caption{}
  \label{fig:mockZmu-c}
\end{subfigure}
\caption{Example of mock observable distribution for the impact point along the tagger of muons from kaon decays. The red line is the nominal distribution, corresponding to the mean value of 100 varied histograms computed by sampling the toy hadroproduction parameters from their PDFs. (a) blue lines are the ensemble of the 100 varied histograms. (b) gray band represents the one standard deviation evaluated from the varied histograms. (c) blue and green lines correspond to $+1~\sigma$ and $-1~\sigma$ deviation from nominal, respectively, due to the variation of one of the toy hadro-production parameters.}
\label{fig:mockZmu}
\end{figure}

As an example, let us consider the impact point distribution along the tagger, $Z$, for muons from kaon decays. To understand how the observable changes with the variation of the hadroproduction parameters, a distribution for each set of the parameters value, sampled from their PDF, can be computed. This way, an ensemble of $N$ possible realization of the distribution is obtained, as shown in Fig.~\ref{fig:mockZmu-a}. This approach is dubbed multi-universe method~\cite{kordowsky}. The nominal distribution (red lines in Fig.~\ref{fig:mockZmu}) is obtained from the mean value of the ensemble, and is used as shape template in the model PDF. From the same ensemble, the $1~\sigma$ band of the distribution is also evaluated, as reported in Fig.~\ref{fig:mockZmu-b}. In principle, with the multi-universe method the $1~\sigma$ variation of the distribution can be obtained from the sampling of each hadroproduction parameter independently, taking into account correlations with other parameters. In Fig.~\ref{fig:mockZmu-c} are shown the $\pm1~\sigma$ distributions of $Z$ corresponding to the variation of one such parameter. These varied templates can be used in the model PDF to describe the systematic effect of that parameter: say that the parameter is $\alpha$, then the model allows a variation of the observable around the nominal distribution constrained by PDF$(\alpha|0,1)$, with $\pm1~\sigma$ values described by the templates in Fig~\ref{fig:mockZmu-c}.  

\begin{figure}[ht]
\begin{subfigure}{.5\textwidth}
  \centering
  % include first image
  \includegraphics[width=1.0\linewidth]{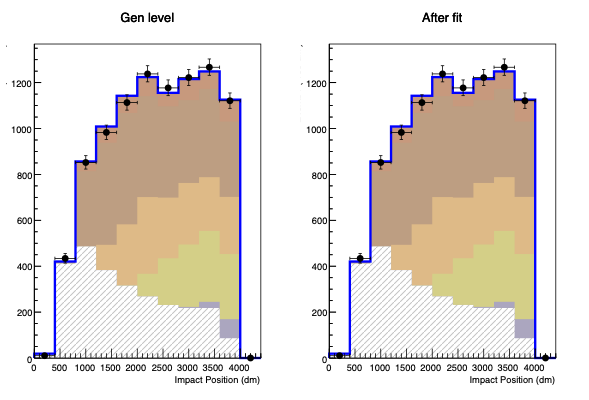}  
  \caption{}
  \label{fig:toyMCZ}
\end{subfigure}
\begin{subfigure}{.5\textwidth}
  \centering
  % include second image
  \includegraphics[width=1.0\linewidth]{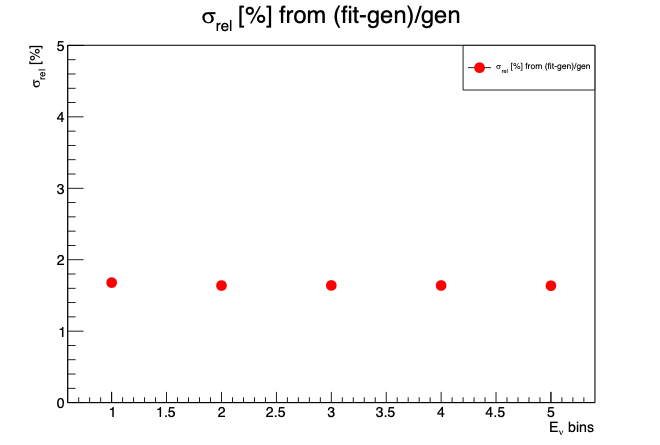}  
  \caption{}
  \label{fig:postFitErr}
\end{subfigure}
\caption{Outcome from a 500 toy Monte Carlo test. (a) distribution of the impact point along the calorimeter for muons from kaon decays, from one of the toy experiments. Black points are the toy Monte Carlo data. The model is built from the signal templates for different neutrino energies, coloured histograms, and the background template, dashed histogram. Left: model with parameters values set to that of data generation. Right: model after performing the extended maximum likelihood fit. (b) relative error on the neutrino flux after the fit.}
\label{fig:toyMC}
\end{figure}

With the model PDF available, a toy Monte Carlo study can be pursued to establish the fit stability and performance. In the model, as already pointed out, the nuisance parameters are allowed to vary around their nominal values, following their PDF constraints. For each toy experiment, data are generated from the model by fixing the nuisance parameters to values sampled from their constraints. After data generation, the same model with nominal nuisances is used to perform an extended maximum likelihood fit. Fig.~\ref{fig:toyMC} reports the outcome from a study done by generating 500 toy Monte Carlo experiments. In this test, the model is built by considering a signal template for each energy interval in the neutrino spectrum (in Eq.~\ref{eq:pdf} the first term is a sum of more than one signal template). The output from one of these toys is shown in Fig.~\ref{fig:toyMCZ} for the $Z$ observable.
Being the normalization of each signal template the number of the neutrinos in a given energy interval, the fit provides a constraint on the neutrino spectrum. Fig.~\ref{fig:postFitErr} reports the relative error for the neutrino spectrum, after performing the 500 toy Monte Carlo fits: it amounts to $\sim 1.8\%$, taking into account an initial systematic uncertainty of $\sim 15\%$.

\begin{figure}
    \centering
    \includegraphics[width=0.99\linewidth]{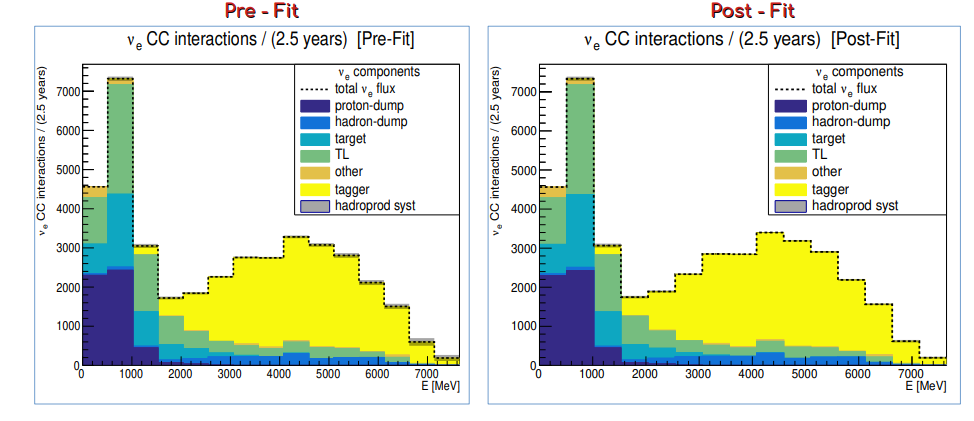}\\
    \includegraphics[width=0.5\linewidth]{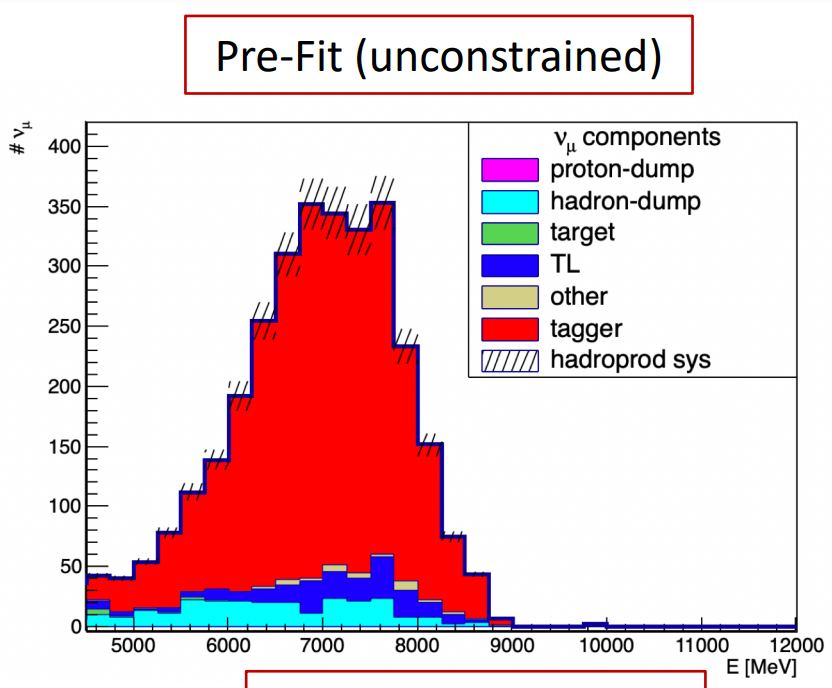}%
    \includegraphics[width=0.5\linewidth]{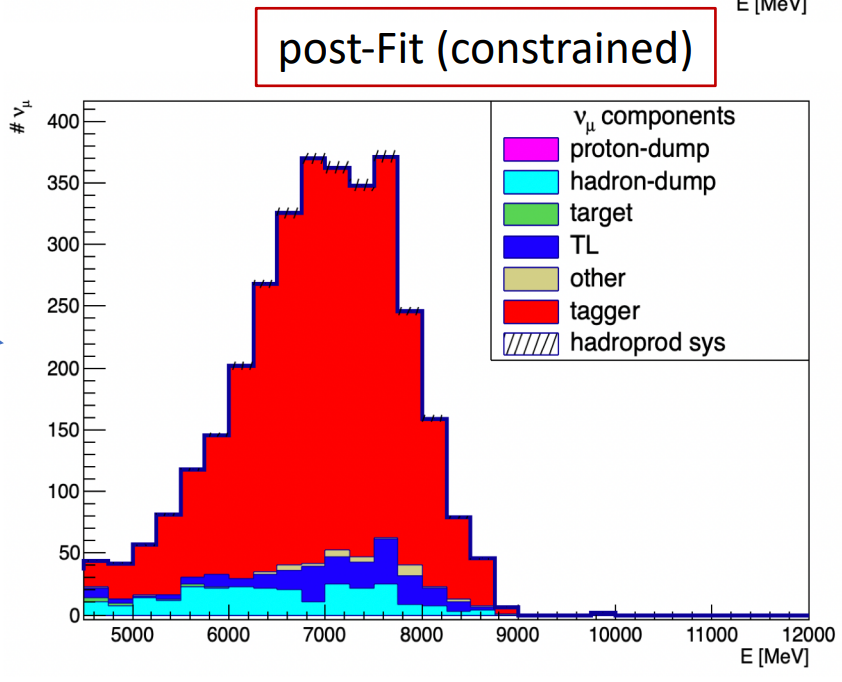}
    \caption{Example of the reduction of hadroproduction uncertainties on $\nu_e$ and $\nu_\mu$ spectra after the constraint given by fitting the lepton spectra reconstructed in the tagger.}
    \label{fig:sys1}
\end{figure}

More recently a model based on real hadroproduction data and related experimental systematic uncertainties
is being worked out. Work in progress is ongoing to get a re-weight of the simulation from these data: this will provide templates for the nominal distributions of the observables. Furthermore, the uncertainty on the data can be propagated to the re-weighting, and by means of the multi-universe method will provide the $\pm1~\sigma$ variation of the nominal templates. Basically, distributions analogous to the one shown in Fig.~\ref{fig:mockZmu} will be computed by correcting the Monte Carlo simulation with real data. The full GEANT4 simulation of NP06/ENUBET contains all the required information to build the re-weighted distribution, since the events producing a detectable signal are linked to the hadrons produced in the target. The hadroproduction data under consideration are the one from the NA56/SPY experiment~\cite{Ambrosini1999}, taken with primary proton beam energy of 400 GeV, the same as the one we are considering for our facility. With the model obtained in this manner a realistic assessment of the hadro-production systematics will be possible. We are also working to characterize the a priori uncertainty of the hadroproduction by employing different generators (i.e. hadronic physics list available in GEANT4).
Figure~\ref{fig:sys1} shows preliminary results in which the reduction of the systematics uncertainties
after the fit constraint on lepton variables is visible by comparing the uncertainty envelope before
and after the fit. The machinery gas been developed for both the $\nu_e$ and $\nu_\mu$ spectra.
We are presently assessing the impact of the facility parameters (magnetic fields, alignments) following the same procedure. 

\section{Cross section measurements with ENUBET}

\begin{figure}[ht]
\centering
\includegraphics[scale=0.3]{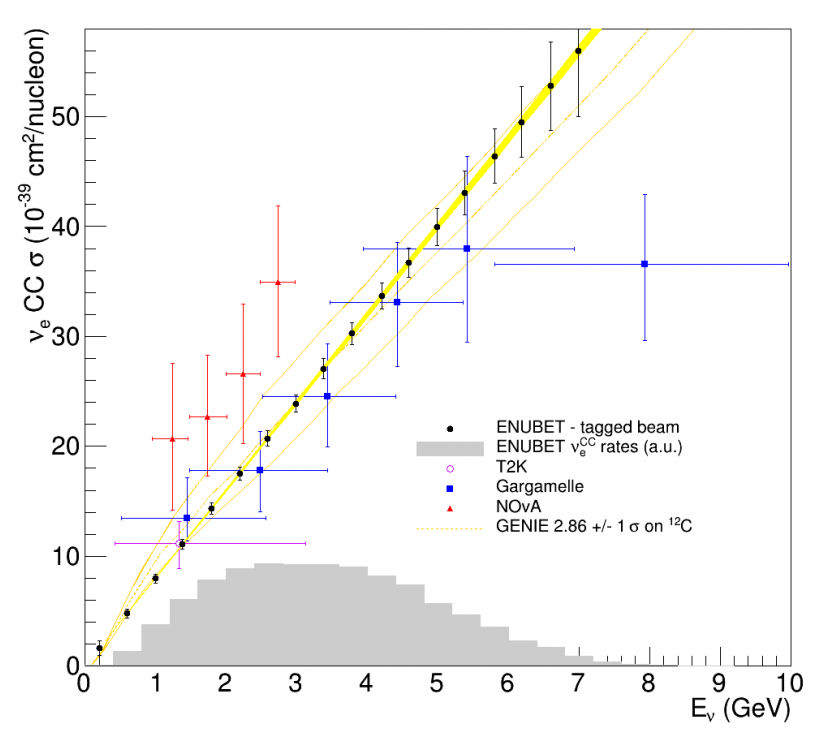}
    \caption{Impact of the ENUBET data on existing $\nu_{e}$ cross-section measurements assuming 1\% flux precision.}
    \label{fig:xsecenubet}
\end{figure}

The impact of the ENUBET data on the $\nu_{e}$ cross-section measurement assuming 1\% flux precision is shown in Figure~\ref{fig:xsecenubet} and compared with current data.

The results obtained by ENUBET fostered the possibility to constrain the low-energy $\nu_{\mu}$ flux from pion decays by monitoring associated muons emitted at low angles by instrumenting the hadron dump. Muons that exit the decay tunnel and go through the hadron dump can be measured by detectors placed between absorber layers. With a static focusing beamline and fast muon detectors, it is possible to reconstruct muons on an event-by-event basis and measure their momentum from their range in the detector stations. The instrumented hadron dump allows to separate the three different components of the muon spectrum: the measured momentum can be exploited to separate muons from pions, and beam-halo muons from muons of $K_{\mu 2}$ ($K^+ \rightarrow \mu^+ \nu_\mu$) and $K_{\mu 3}$ decays that, in turn, can be used to determine the $\nu_{\mu}$ flux. The energy of $\nu_{\mu}$ can be studied in bins corresponding to muon momenta reconstructed by range as they show a clear anti-correlation due to the 2-body decay kinematics, allowing to constrain the $\nu_{\mu}$ shape and normalization.
The ENUBET collaboration is studying a  system of 8 muon stations to be placed right after the hadron dump with iron absorbers from 2 m (upstream) to 0.5 m (downstream) depth. Muon and neutron fluences have already  been estimated: the most upstream detector needs to cope with a muon rate of 2 MHz/cm$^{2}$ and the total neutron fluence integrated over the experiment lifetime is 10$^{12}$ n(1-MeV-eq)/cm$^{2}$. The radiation damage expected in a monitored beam is much smaller compared to beams where the dump is used to stop non-interacting primary protons. This is the reason while instrumented dumps are not an option in conventional beams like the ones used for long-baseline experiments.

It is worth mentioning that ENUBET is working also on a site-dependent multi-momentum beamline to cover the lower energy region, which is of interest for HK: the current design of the multi-momentum beamline uses existing CERN magnets. The beamline optics is flexible enough to select hadron momenta down to 4 GeV or less. The design consists of quadrupoles for focusing and two dipoles for momentum selection and a large total bending to ensure the separation of the different $\nu_{e}$ components at the neutrino detector. 

%The study  will be finalized with background reduction studies to provide a large sample of $\nu_\mu$ with a flux known at per-cent level and
The possibility of measuring the flux of $\nu_\mu$ with a precision comparable with $\nu_e$ is a significant asset for ENUBET. Since ENUBET is a narrow band beam and the pions produce muon-neutrinos in a two-body decay, there is a strong correlation between the neutrino angle and its energy. As a consequence, the measurement of the interaction vertex of the neutrino in the detector (and, hence, its angle) provides a measurement of the energy with an average precision of 10\%. It is worth stressing that this indirect energy measurement (\textit{narrow-band off-axis technique}) can be performed at the level of a single neutrino without resorting to the energy reconstruction at the detector. This is the main tool to remove all reconstruction biases in the neutrino detector and suppress the systematics on neutrino interaction modeling that plague the Near-Detector measurements without any tuning with data.

%\section{DUNE and HyperKamiokande}

\section{Other opportunities}
%sterili, NSI, possibile implementazione in ESS, nustorm
%FT
Even if we gave special emphasis on cross section measurement, readers should be aware that ENUBET is a full-fledged short baseline facility and can address many items currently addressed by the US short baseline program. Thanks to an unprecedented control of the flux, ENUBET provides a substantial improvement in the study of Non Standard Interactions (NSI) and is a mandatory step in case the SBN experiments at FNAL support the LSND and MiniBoone anomaly. SBN results will be arriving in the course of the Snowmass process and beyond and are not addressed in this document. Still they could potentially change the neutrino program strategy  in a few years from now. For a detailed discussion of the NSI potential of ENUBET we refer to \cite{Delgadillo:2020uvm}.

Finally, the common R\&D and engineering studies pursued in coordination with nuSTORM and the Muon Collider Collaborations are described here \cite{snowmass_nustorm}.

\section{Conclusions}

The Snowmass 2021 DPF Community Planning Exercise is particularly timely for the consideration of monitored neutrino beams and the global strategy for the study of neutrino physics. In 2019-2022, the ENUBET Collaboration has devised the first end-to-end simulation of a monitored beam with a static focusing system addressing all technology and physics (systematics) challenges. The technology of a monitored neutrino beam has been proven to be feasible and cost-effective, and the complexity does not exceed significantly the one of a conventional short baseline beam. The ENUBET Collaboration is in the process of evaluating site specific implementations at CERN, FNAL, and J-PARC, and the CERN design is rather mature thank to NP06/ENUBET and the contributions of the CERN PBC group. 

Having ENUBET in data taking in a timescale comparable with DUNE is an outstanding opportunity both to enhance the DUNE physics reach and to advance substantially the study of neutrino interaction with matter and weak nuclear physics. In particular, a neutrino facility as ENUBET complemented by a dedicated suite of moderate-mass detectors represent the ideal playground for the next generation of cross section experiments.

\tableofcontents

\end{document}